\documentclass[12pt]{article}

\usepackage{draft} 

\begin{document}

\unitlength = .8mm

\begin{titlepage}

\begin{center}

\hfill \\
\hfill \\
\vskip 1cm

\title{Little String Amplitudes \\ \Large (and the Unreasonable Effectiveness of 6D SYM)}

\author{Chi-Ming Chang$^\heartsuit$, Ying-Hsuan Lin$^\heartsuit$, Shu-Heng Shao$^\heartsuit$, Yifan Wang$^\clubsuit$, Xi Yin$^\heartsuit$}

\address{
$^\heartsuit$Jefferson Physical Laboratory, Harvard University, \\
Cambridge, MA 02138 USA
\\
$^\clubsuit$Center for Theoretical Physics, Massachusetts Institute of Technology, \\
Cambridge, MA 02139 USA}

\email{cmchang@physics.harvard.edu, yhlin@physics.harvard.edu, \\ shshao@physics.harvard.edu, yifanw@mit.edu,
xiyin@fas.harvard.edu}

\end{center}

\abstract{ We study tree level scattering amplitudes of four massless states in the double scaled little string theory, and compare them to perturbative loop amplitudes in six-dimensional super-Yang-Mills theory. The little string amplitudes are computed from correlators in the cigar coset CFT and in ${\cal N}=2$ minimal models. The results are expressed in terms of integrals of conformal blocks and evaluated numerically in the $\alpha'$ expansion. We find striking agreements with up to 2-loop scattering amplitudes of massless gluons in 6D $SU(k)$ SYM at a $\mathbb{Z}_k$ invariant point on the Coulomb branch. We comment on the issue of UV divergence at higher loop orders in the gauge theory and discuss the implication of our results. }

\vfill

\end{titlepage}

\eject

\tableofcontents

\section{Introduction}

There are two interesting maximally supersymmetric quantum field theories in six dimensions: the $(2,0)$ superconformal field theory and the $(1,1)$ super-Yang-Mills theory. They are believed to be the low energy limits of $(2,0)$ or $(1,1)$ little string theories (LST) \cite{Berkooz:1997cq,Seiberg:1997zk,Aharony:1998ub,Giveon:1999zm,Aharony:1999ks,Kutasov:lecturenote}. While the LST is a strongly coupled string theory that is difficult to get a handle on, it admits a 1-parameter deformation known as the double scaled little string theory (DSLST) \cite{Giveon:1999px,Giveon:1999tq}, the parameter being an effective string coupling. In the weak coupling limit, DSLST can be studied perturbatively. In particular, one can compute the spectrum and in principle perturbative scattering amplitudes to all orders in $\alpha'$.

In this paper, we will focus on the $(1,1)$ theories. In this case, the LST reduces to 6D SYM at low energies. More precisely, the interactions of massless modes of LST are expected to be described by an effective theory that is 6D SYM deformed by higher dimensional operators (such operators are highly constrained by supersymmetry and we will return to this point later). $\alpha'$ will be mapped to the 6D Yang-Mills coupling via the relation
\ie\label{idaa}
{1\over 2\pi\alpha'} = {8\pi^2\over g_{YM}^2}.
\fe
This can be understood by identifying the little string with instanton strings in the SYM. The DSLST, on the other hand, is related to 6D SYM on its Coulomb branch. The relation between the string coupling and the Coloumb branch parameter takes the form
\ie\label{idbb}
g_s \sim {1\over g_{YM} m_W}.
\fe
 Thus one may anticipate a relation between scattering amplitudes of massless modes in the DSLST and those of the Cartan gluons in the 6D SYM of the schematic form
\ie
{\cal A}_{DSLST}(g_s, \A', E) = {\cal A}_{SYM}(g_{YM}, m_W, E),
\fe
provided the identifications (\ref{idaa}) and (\ref{idbb}).
Note however that the weak coupling limit of DSLST corresponds to the regime far from the origin on the Coloumb branch. In the 6D gauge theory, a priori one would expect higher dimensional operators to become important in this limit, and it is not a priori clear whether the gluon scattering amplitudes in the SYM, expanded in $1/m_W^2$, should agree with that of the perturbative amplitudes of massless modes of DSLST. 

In fact, the 6D SYM diverges at three-loops, and one might be led to think that such an agreement is impossible with just the SYM perturbative amplitudes (without including any higher dimensional operators). However, a more careful inspection indicates that the scattering amplitude of gluons in the Cartan $U(1)^{k-1}$ on the Coulomb branch of $SU(k)$ gauge theory is free of UV divergence at 3-loops. This is because there are only two candidate counter terms that are dimensional 10 operators allowed by sixteen supersymmetries; one of them is a double trace operator known to be 1/4 BPS and finite in 6D SYM according to \cite{Bossard:2010pk}, the other is a single trace operator that is non-BPS but vanishes when restricted to the Cartan subalgebra of the $SU(k)$. (We will explicitly verify the absence of this UV divergence in Appendix \ref{3ldiv}.) This suggests that a certain non-renormalization theorem is at play, and one could hope that the 6D SYM by itself already captures some of the dynamics of massless gluons on the Coulomb branch of the $(1,1)$ little string theory.

Remarkably, {\bf we find agreement} between up to two-loop amplitudes of the massless gluons on the Coulomb branch of the 6D SYM, expanded to leading order in $1/m_W^2$, and the tree level amplitude of the corresponding massless string modes in DSLST, expanded up to second order in $\A'$. This is achieved through explicit computation of amplitudes in the $SU(k)$ SYM, based on previously known two-loop results derived from unitarity cut methods \cite{Dixon:1996wi,Dixon:2010gz,Bern:1996je,Brandhuber:2010mm,Bern:1997nh,Bern:2010qa}, and in the DSLST by a worldsheet computation that involves expressing $SL(2)/U(1)$ coset CFT correlators in terms of Liouville correlators, which are then expressed as integrals of Virasoro conformal blocks. The computations on the two sides utilize completely different techniques and the results are rather involved. Nonetheless, we will evaluate the results numerically for $k$ up to 5, and the answers on the two sides strikingly agree.

While our result may be viewed as a strong check of the duality between DSLST and 6D SYM, it also hints that the higher dimensional operators correcting the 6D SYM are under control. Put differently, the perturbative 6D SYM seems to know a lot more about its UV completion and particularly the instanton strings than one might have naively expected!


The paper is organized as the following. After reviewing the construction of DSLST and its spectrum, we will study the tree level four-point amplitude of massless RR vertex operators in six dimensions. In computing such amplitudes, we will need the singular limits of certain correlators in the $SL(2)/U(1)$ coset (``cigar") CFT, which are further related to Liouville correlators by Ribault and Teschner's relations \cite{Ribault:2005wp}. This will allow us to express the DSLST amplitudes as integrals of Liouville/Virasoro conformal blocks.\footnote{The integration of conformal blocks was circumvented in \cite{Aharony:2003vk} as the authors of \cite{Aharony:2003vk} were only concerned with the leading order answer in the $\A'\to 0$ limit. We find it necessary to compute the full conformal blocks in order to extract the next order terms in the $\A'$ expansion, and numerical integration of conformal blocks will be carried out.} Order by order in the $\alpha'$ expansion, these integrals can be computed numerically, thereby producing an explicit $\alpha'$ expansion of the massless amplitudes in DSLST. The DSLST results are listed in Table \ref{tab:lstresults} and the 6D SYM results are summarized in Table \ref{tab:2loop}.

These amplitudes are then compared with gluon scattering amplitudes of 6D SYM on its Coulomb branch, expanded in $1/m_W^2$ to the leading order (if these amplitudes were the full story, higher order terms in the $1/m_W^2$ expansion would be mapped to higher genus double scaled little string amplitudes). 
The 1-loop and 2-loop amplitudes are perfectly finite and can be obtained using unitarity cut method \cite{Bern:1996je,Brandhuber:2010mm,Bern:1997nh}, and the 3-loop amplitudes have been reduced to scalar integrals as well \cite{Bern:2007hh} (as already asserted, they are free of UV divergences when the external lines are restricted to the Cartan subalgebra). 
An agreement of the 1-loop amplitude at order $1/m_W^2$ with the low energy limit of four gluon amplitude in DSLST was found by \cite{Aharony:2003vk}. 
We will  carry out the $\alpha'$ expansion of DSLST amplitude to the next order in $\alpha'$ and compare it with the 2-loop $SU(k)$ SYM amplitudes (which involve highly nontrivial group theory factors). While the expressions on both sides are quite complicated, we will evaluate them numerically for $k=2,3,4,5$, and remarkably, they agree. (In the $k=5$ case, for instance, the four-point functions of nontrivial primaries in the three-state Potts model went into the DSLST amplitude computation!)
We conlude with a discussion on the implication of these results. Details on the numerical integration of conformal blocks, based on Zamolodchikov's recursion relations, are described in Appendix \ref{numerics}.

\section{Double scaled little string theory}

In Section~2.1, we review holographic descriptions of DSLST. In Sections~2.2-2.6, we construct the normalizable vertex operators of the supersymmetric $( SL(2)_k/U(1) \times SU(2)_k/U(1) ) / \bZ_k$ theory.  We will find that these vertex operators $V^{sl (\eta, \bar\eta)}_{j,m,\bar m} V^{su (\eta',\bar\eta')}_{j',m',\bar m'}$ have quantum numbers satisfying (2.56) subject to the identification (2.62) and (2.64).
Our main result is in Section~2.7, where we identity the massless bosonic vertex operators of type IIA string theory on $\bR^{1,5} \times ( SL(2)_k/U(1) \times SU(2)_k/U(1) ) / \bZ_k$, which is the T-dual description of IIB (1,1) DSLST.  These massless string states correspond to the scalars and gluons in the 6D SYM.

\subsection{The holographic description}

The physical definition of double scaled little string theory is the decoupled theory on $k$ NS5-branes in type IIA or IIB string theory, spread out on a circle of radius $r_0$ in a transverse plane $\mathbb{R}^2\subset \mathbb{R}^4$, in the double scaling limit $r_0, g_s\to 0$, with $r_0/g_s$ held finite in string units. Here one may consider an energy scale of interactions comparable to the string scale (without taking a low energy limit). The configuration or the decoupled theory has a residual global symmetry $\mathbb{Z}_k\times U(1)$.

In the case of IIB NS5-branes, D1-branes extending between NS5-branes correspond to W-bosons in the six-dimensional $SU(k)$ gauge theory. They become tensionless in the $r_0\rightarrow 0$ limit, where we recover the strongly coupled (1,1) LST.

On the other hand, one may directly take the decoupling limit of the string worldsheet theory in the NS5-brane background. Recall that (1,1) LST is described by IIB string theory on the linear dilaton background from the NS5-brane near horizon geometry {\cite{Callan:1991at,Aharony:1998ub}. Its deformed cousin (1,1) DSLST is described by $\cN=2$ Liouville theory in IIB which is T-dual to the ${SL(2)_k \over U(1)}$ cigar CFT in IIA, similarly for (2,0) DSLST \cite{Giveon:1999px,Ooguri:1995wj, Kutasov:1995te} (see also \cite{Sfetsos:1998xd}).

This leads to the ``holographic" description of DSLST, as IIA (in the $(1,1)$ case) or type IIB (in the $(2,0)$ case) string theory with the target space given by
\ie
{\mathbb R}^{5,1} \times \left( {SL(2)_k\over U(1)}\times {SU(2)_k\over U(1)}\right)/\bZ_k
\fe
in the NSR formalism. The $\bZ_k$ orbifolding acts simultaneously on the two (supersymmetric) coset models.

The supersymmetric coset model $SL(2)_k/U(1)$ can be constructed from the bosonic $SL(2)_{k+2}$ plus three free fermions $\lambda^a$, $a=1,2,3$, and gauging the $U(1)$ supercurrent that contains $\lambda^3$ and the total bosonic $U(1)$ current at level $k$ (combining a $U(1)$ from the bosonic $SL(2)_{k+2}$ and a $U(1)$ current from the fermions at level $-2$). The supersymmetric $SU(2)_k/U(1)$, constructed in a similar manner from the bosonic $SU(2)_{k-2}$ and three fermions, is the same as the $k$-th ${\cal N}=2$ minimal model. The total worldsheet matter central charge is
\ie
c^m = 9+{3(k+2)\over k}+{3(k-2)\over k} = 15,
\fe
as required for critical string theory.

\subsection{$SL(2)_k/U(1)$}

The ${\cal N}=2$ supersymmetric $SL(2)_k/U(1)$ coset model can be constructed as follows \cite{Giveon:2003wn}. One starts with the bosonic $SL(2)_{k+2}$ WZW model, governed by the current algebra
\ie
j^a(z) j^b(0) \sim {(k+2)\eta^{ab} \over 2z^2} + i\epsilon^{abc} {j_c(0)\over z},
\fe
where $\eta_{ab}={\rm diag}(1,1,-1)$. We will lower and raise the index $a$ by $\eta_{ab}$ and its inverse $\eta^{ab}$. The anti-holomorphic currents will be denoted as $\bar j^a$. The extension to supersymmetric $SL(2)_k$ WZW model simply requires adding three fermions $\lambda^a$, $a=1,2,3$, that obey a slightly non-standard reality condition. Namely, the OPEs of the $\lambda$'s are given by
\ie
\lambda^a(z) \lambda^b(0) \sim {\eta^{ab}\over z}.
\fe
The current $- {i\over 2} \epsilon_{abc}\lambda^b\lambda^c$ gives rise to an $SL(2)$ current algebra at level $-2$. Altogether, we have a level $k$ $SL(2)$ current
\ie
J_a = j_a - {i\over 2}\epsilon_{abc} \lambda^b \lambda^c\label{totalcurrent}
\fe
for the supersymmetric $SL(2)_k$ WZW model.
This theory has ${\cal N}=1$ superconformal symmetry, with the supercurrent
\ie
G = \sqrt{2\over k} (\eta_{ab} \lambda^a j^b - {i\over 6}\epsilon_{abc} \lambda^a\lambda^b\lambda^c).
\fe
The three fermions $\lambda^a$ are superconformal primaries with respect to $G$, while the $J^a$'s are superconformal descendants of $\lambda^a$.

The supersymmetric $SL(2)_k/U(1)$ coset model is defined by gauging the $U(1)$ ${\cal N}=1$ supermultiplet that contains the primary $\lambda^3$ and $J_3$. The coset model $SL(2)_k/U(1)$ has $\mathcal{N}=2$ superconformal symmetry with superconformal currents $G^\pm$ and $R$-current $J_R$ given by
\ie
& G^\pm = \sqrt{2\over k} j^\mp \lambda^\pm,
\\
& J_R = {k+2\over k}\lambda^+\lambda^-+ {2\over k} j^3 = \lambda^+\lambda^- + {2\over k} J^3,\label{Rcurrent}
\fe
where we have defined
\ie
\lambda^\pm \equiv {\lambda^1\pm i\lambda^2\over \sqrt{2}},~~~~
j^\pm \equiv j^1\pm i j^2.
\fe


Let us summarize the relations between the various currents in this construction. We have the bosonic $SL(2)_{k+2}$ currents $j^a$ and the fermonic currents $-{i\over2}\epsilon_{abc}\lambda^b\lambda^c$, with their sum being the total $SL(2)_k$ current $J^a$ for the supersymmetric $SL(2)_k$ WZW model. The coset model $SL(2)_k/U(1)$, as an $\mathcal{N}=2$ superconformal field theory, has the $R$-current $J_R$ defined above. Let $X$, $x$, $X_R$, and $H$ be the bosonzations of $J^3$, $j^3$, $J_R$, and $\LL^+\LL^-$ respectively,
\ie
& J^3 = - \sqrt{k\over 2} \partial X,\\
&j^3 = - \sqrt{k+2\over2} \partial x,\\
&J_R = i \sqrt{ k+2\over k} \partial X_R,\\
&\LL^+ \LL^- = i \partial H.\label{slbos}
\fe 
The (chiral) bosons are normalized with the standard OPE $X(z) X(0)\sim - \log z$, etc. 
It follows from \eqref{totalcurrent} and \eqref{Rcurrent} that the following relations hold among the bosonization scalars:
\ie
&i H = \sqrt{2\over k} X + i \sqrt{k+2\over k} X_R,\\
& x= \sqrt{k+2\over k}X + i \sqrt{2\over k} X_R.
\fe
Note that $H(z)x(0)\sim 0$ by definition, and it follows that $X(z)X_R(0)\sim 0$. This is necessary for $J_R$ to be well defined in the coset theory.

The NS superconformal primaries of the coset model can be constructed starting from the primaries of the supersymmetric $SL(2)_k$ WZW model, which can be taken to be the primaries of the bosonic $SL(2)_{k+2}$ WZW model, $\Phi^{sl}_{j,m,\bar m}$, with conformal weight
\ie
 -{j(j+1)\over k}
\fe
and charge $m$ and $\bar m$ with respect to $j^3$ and $\bar j^3$. The range of $j$ and $m,\bar m$ will be discussed later. The $SL(2)/U(1)$ primary $V_{j,m,\bar m}^{sl}$ is then obtained by factoring out a $U(1)$ primary\footnote{As we are focusing on the massless string modes in this paper, we only need to take into account those coset primaries that come directly from current primaries.},
\ie
\Phi_{j,m,\bar m}^{sl} = V_{j,m,\bar m}^{sl} e^{\sqrt{2\over k} (m X-\bar m \bar X)},
\fe
where $X$ and $\bar X$ are related to the $U(1)$ currents by the bosonization dictionary $J^3=-\sqrt{k\over 2}\partial X$, $\bar J^3=+\sqrt{k\over 2}\bar\partial\bar X$. The plus sign convention for $\bar X$ in $\bar J^3$ is chosen for later convenience. The coset model primary $V_{j,m,\bar m}^{sl}$ has conformal weights
\ie
\Delta_{j,m} = {-j(j+1)+m^2\over k}, ~~~\bar \Delta_{j,\bar m} = {-j(j+1)+\bar m^2\over k}.
\fe
The ${\cal N}=2$ superconformal $R$-charge can be determined as follows. First recall that $J_R = i\partial H  + {2\over k}J^3$ and $H$ does not have singular OPE with $\Phi_{j,m,\bar m}^{sl}$. It follows that $R(\Phi_{j,m,\bar m}^{sl}) = 2m/k$. Next, since $X_R(z)X(0)\sim 0$, the operator $e^{ m\sqrt{2\over k} X}$ is uncharged with respect to $J_R$. Combining the above two observations, we obtain the $R$-charges for the coset primary $V_{j,m,\bar m}^{sl}$,
\ie
R(V_{j,m,\bar m}^{sl}) = {2m\over k},~~~\bar R(V_{j,m,\bar m}^{sl}) = {2\bar m\over k}.
\fe

\subsubsection{Non-normalizable, delta function normalizable, and normalizable primaries}

Depending on the value of $j$, the primary operator $V_{j,m,\bar m}^{sl}$ can either be non-normalizable, delta function normalizable, or normalizable along the radial direction of the cigar. The non-normalizable primaries correspond to generic real $j$, whereas the delta function normalizable primaries are given by $j\in -{1\over 2} + i\mathbb{R}$ (the imaginary part of $j$ being the asymptotic momentum along the radial direction of the cigar). Note that the conformal weight formula is invariant under $j\to -j-1$. In fact, the operators $V_{j,m,\bar m}^{sl}$ and $V_{-j-1,m,\bar m}^{sl}$ are related by \cite{Aharony:2004xn}
\ie
& V_{j,m,\bar m}^{sl} = R(j,m,\bar m;k) V_{-j-1,m,\bar m}^{sl},
\\
& R(j,m,\bar m;k) = \nu(k)^{2j+1} {\Gamma(1-{2j+1\over k}) \Gamma(j+m+1) \Gamma(j-\bar m+1)\Gamma(-2j-1)\over \Gamma(1+{2j+1\over k}) \Gamma(m-j) \Gamma(-\bar m-j) \Gamma(2j+1) },
\\
& \nu(k) = {1\over \pi} {\Gamma(1+{1\over k}) \over \Gamma(1-{1\over k}) }.
\fe
In the above formula $m\geq \bar m$ is assumed, i.e. the momentum $n=m-\bar m$ (defined below) is non-negative. If $n$ is negative we simply exchange the role of $m$ and $\bar m$ in the formula for the reflection coefficient.
Using this reflection relation we can restrict the non-normalizable operators to real $j>-{1\over 2}$. In constructing string vertex operators, we must also impose an upper bound $j<{k-1\over 2}$ to ensure the two point function of $V_{j,m,\bar m}^{sl}$ is nonsingular \cite{Giveon:1999px,Giveon:1999tq}.\footnote{This bound is slightly stronger than that imposed by the no-ghost theorem in string theory on $SL(2)$ \cite{Evans:1998wq,Maldacena:2000hw}.}  

The $j^3$, $\bar j^3$ charges $m$ and $\bar m$ are related to momentum and winding on the cigar (along the circle direction),
\ie
m - \bar m = n\in \mathbb{Z},~~~~ m+\bar m=kw\in k\mathbb{Z},
\fe
or
\ie
m = {n+wk\over 2},~~~~\bar m = {-n+wk\over 2}.
\fe
So far, for either the non-normalizable or the delta function normalizable primary operators, there are no constraining relations between $m,\bar m$ and $j$.

Importantly, there are also {\it normalizable} primary operators at special values of real $j$ \cite{Aharony:2003vk,Aharony:2004xn}, corresponding to principal discrete series of $SL(2)$ \cite{Maldacena:2000hw}, namely
\ie
j= j_* = m_0-1, m_0-2,\cdots,~~~j_*>-{1\over 2}
\fe
where $m_0$ is given by
\ie\label{m0102}
m_0 =
\begin{cases}
{\rm min}\{ |m|,|\bar m|\}, & m,\bar m<-{1\over 2}. \\
{\rm min}\{ m,\bar m\}, & m,\bar m>{1\over 2}.
\end{cases}
\fe

Only delta function normalizable primaries and normalizable primaries are being used to construct the vertex operators of DSLST. The delta function normalizable primaries of the $SL(2)/U(1)$ will lead to a continuum of string modes that propagate down the cigar, whereas the normalizable primaries will give rise to string modes localized at the tip of the cigar, thus effectively living in six dimensions \cite{Giveon:1999px,Aharony:2004xn}. The scattering amplitudes of the latter is the subject of this paper.

\subsection{$SU(2)_k/U(1)$ and ${\cal N}=2$ minimal model}

The supersymmetric coset model $SU(2)_k/U(1)$ can be constructed similarly to the $SL(2)_k/U(1)$ model. We will denote the primary operators and charges in the $SU(2)_k/U(1)$ model with primes in order to distinguish them from those in the $SL(2)_k/U(1)$ model. Let $j'_i$, $i=1,2,3$, be bosonic $SU(2)_{k-2}$ currents and let $\lambda'_i$ be three free fermions. The OPEs are (in this case there is no distinction between upper and lower indices)
\ie
&j'_i(z)j_j'(0)\sim{(k-2)\delta_{ij}\over 2z^2}+i\epsilon_{ijk}{{j'}^k(0)\over z},\\
&\lambda'_i(z)\lambda'_j(z)\sim{\delta_{ij}\over z}.
\fe
The overall $SU(2)_k$ current of the supersymmetric WZW model is given by
\ie
J_i'=j_i'-{i\over 2}\epsilon_{ijk}{\lambda'}^j{\lambda'}^k.\label{totalcurrentp}
\fe
The supersymmetric $SU(2)_k/U(1)$ coset has $\mathcal{N}=2$ superconformal currents ${G'}^\pm$ and $U(1)$ $R$-current $J_R'$,
\ie
{G'}^\pm = \sqrt{2\over k} {j'}^\mp {\lambda'}^\pm,~~~J_R' = - {2\over k} j_3' + {k-2\over k}{\lambda'}^+ {\lambda'}^-\label{Rcurrentp}.
\fe

Let $X'$, $x'$, $X_R'$, $H'$ be the bosonization of $J_3'$, $j_3'$, $J_R'$, and ${\lambda'}^+{\lambda'}^-$, respectively:
\ie
&J_3' = i \sqrt{k\over 2}\partial X',\\
&j_3' = i \sqrt{ k-2\over 2} \partial x',\\
&J_R' = i \sqrt{k-2\over k}\partial X_R',\\
&{\lambda'}^+ {\lambda'}^- = i \partial H'.
\fe
From \eqref{totalcurrentp} and \eqref{Rcurrentp}, we can read off the relations between the bosonization scalars
\ie
&X' = \sqrt{k-2\over k} x' + \sqrt{2\over k} H',\\
&X_R '= - \sqrt{2\over k} x'+\sqrt{k-2\over k} H'.\label{relation}
\fe
In particular, it follows that $X'(z)X_R'(0)\sim 0$, as is required for $J_R'$ to survive the coset construction.

%

The NS superconformal primaries of the $SU(2)_k/U(1)$ coset model (which is the same as the $k$-th ${\cal N}=2$ minimal model), denoted by $V^{su}_{j',m',\bar m'}$, can be constructed starting from the bosonic $SU(2)_{k-2}$ primary $\Phi^{su}_{j',m',\bar m'}$ and factoring out the $U(1)$ part. Their conformal weights and $R$-charges are 
\ie
\Delta_{ j', m'} = { j'(j'+1) -  m'^2\over k},~~~R = - {2m'\over k},
\fe
where $j'$ is half-integer valued, in the range $j'=0,{1\over 2},1,\cdots,{k\over 2}-1$, and $m'=-j',-j'+1,\cdots,  j'$. 

For later application we will recall below two examples of supersymmetric $SU(2)_k/U(1)$ coset models where the primary operators and correlators can be easily written down. The $k=2$ case has zero central charge and is a trivial theory. The $k=3$ model will be described as follows.

\subsubsection{Supersymmetric $SU(2)_3/U(1)$ and the compact boson}\label{compact}

In the $k=3$ case, the supersymmetric coset theory $SU(2)_3/U(1)$ is a compact boson\footnote{We add a prime to the field to distinguish from the linear dilaton $\phi$.} $\phi'$ of radius $1/\sqrt{3}$ \cite{Lerche:1989uy}. This compact boson CFT enjoys the $\mathcal{N}=2$ superconformal symmetry with superconformal currents
\ie
&G'^\pm (z) = \sqrt{2\over3} \exp \left[ \pm i \sqrt{3} \phi'(z)\right],\\
&\bar G'^\pm (\bar z) = \sqrt{2\over3} \exp \left[ \mp i \sqrt{3} \bar\phi'(\bar z)\right],
\fe
with dimension ${3\over 2}$. The coefficient $\sqrt{2\over3}$ is fixed by the OPE for $G'^+(z)G'^-(0)= {2c\over 3 z^3} +\cdots$.

The $R$-current can be determined by looking at the OPE for $
G'^+ (z) G'^-(0)$,
\ie
G'^+(z) G'^-(0) &\sim {2c\over 3z^3 } + {2\over z^2 } J_R'(0) + {2\over z} T'(z) +{1\over z} \partial J_R'(0) \\
& \sim {2\over3 z^3}
+ { 2 \over  z^2} {i\over \sqrt{3}}\partial \phi'(0)   + \cdots,
\fe
hence
\ie
J_R'(z) = {i\over \sqrt{3} } \partial \phi'(z).
\fe
The antiholomorphic $R$-current $\bar J_R'(\bar z) =-{i\over \sqrt{3} } \partial \bar\phi'(\bar z)$ has an extra sign because of our convention for $\bar G'^\pm(\bar z)$.

The Virasoro primary operators are determined as usual
\ie\label{virasoro}
\exp\left[ i  \left(   { \sqrt{3} n} + {w\over2\sqrt{3}}\right) \phi'(z)\right]
\exp\left[ i  \left(   { \sqrt{3}n  } - {w\over2\sqrt{3}}\right) \bar \phi'(\bar z)\right],~~~~n,w\in \bZ. 
\fe

In the following we will determine the superconformal primaries. The $\cN=2$ superconformal primary states $|\Phi\ra$ are Virasoro primaries which in addition satisfy
\ie\label{superprimary}
&G'^\pm_{r} | \Phi\ra= \oint {dz \over 2\pi i} ~z^{r+{1\over2} } G'^\pm (z) \Phi(0)= 0,~~~r>0,\\
&J'_{R,n} | \Phi \ra = \oint {dz\over 2\pi i} z^n J'_R(z) \Phi(0)=0,~~n>0.
\fe
The second condition implies that the OPE for $J'_R(z)\Phi(0)$ can't be more singular than $1/z$, which is always true for the Virasoro primaries \eqref{virasoro}. 
In the NS-sector, this implies that the OPE for $G'^\pm (z)\Phi(0)$ can't be more singular than $1/z$. This imposes constraints on the momentum $n$ and the winding number $w$,
\ie
& -1 \le 3n+ {w\over 2} \le 1,\\
& -1 \le 3n-{w\over 2} \le 1,
\fe
and $3n\pm{w\over 2}$ being integers. 
It follows that $(n,w)=(0, \pm2)$ or $(n,w)=(0,0)$. The latter is the identity operator. The former ones are
\ie
\text{NS}:~&\exp\left[ i   { 1 \over \sqrt{3} } \phi'(z) \right] \exp\left[ - i   { 1 \over \sqrt{3} } \bar \phi'(\bar z)\right],~~\Delta=\bar \Delta ={1\over 6},~~R = \bar R= {1\over 3},\\
&\exp\left[ -i   { 1 \over \sqrt{3} } \phi'(z) \right] \exp\left[  i   { 1 \over \sqrt{3} } \bar\phi'(\bar z)\right],~~\Delta=\bar \Delta ={1\over 6},~~R = \bar R=-{1\over 3}.
\fe

Moving on to the R-sector, the superconformal primary condition \eqref{superprimary} implies that the OPE for $G'^\pm (z)\Phi(0)$ can't be more singular than $1/z^{3/2}$. This imposes constraints on $n,w$ in \eqref{virasoro}:
\ie
& -{3\over 2} \le 3n+{w\over 2} \le {3\over2},\\
& -{3\over 2} \le 3n-{w\over 2} \le {3\over2},
\fe
and $3n\pm{w\over 2}$ being half integers. The possible solutions are $
(n,w) = (0 ,\pm1 ),~(0,\pm3)$, corresponding to the R-sector primary operators 
\ie
\text{R}:~~&\exp\left[ \pm i   { 1 \over 2\sqrt{3} } \phi'(z)\right] \exp\left[ \mp i   { 1 \over 2\sqrt{3} } \bar \phi'(\bar z)\right]  ,~~~\Delta=\bar\Delta = {1\over 24},~~R=\bar R  = \pm {1\over 6},\\
&\exp\left[ \pm i   { 3 \over 2\sqrt{3} } \phi'(z)\right] \exp\left[ \mp i   { 3 \over 2\sqrt{3} } \bar\phi'(\bar z)\right]  ,~~~\Delta=\bar\Delta = {3\over 8},~~R=\bar R  = \pm {1\over 2}.
\fe

\subsection{Spectral flow}


In an ${\cal N}=2$ superconformal theory of central charge $c$, the spectral flow automorphism \cite{Schwimmer:1986mf,Lerche:1989uy}, labeled by a real parameter $\eta$, takes an operator $\cal O$ of weight $\Delta$ and $R$-charge $R$ to another operator ${\cal O}^\eta$ of weight $\Delta^\eta$ and $R$-charge $R^\eta$, related by
\ie
\Delta^\eta=\Delta+\eta R+\eta^2{c\over 6},~~~R^\eta=R+\eta{c\over 3}.
\fe
In the $SL(2)_k/U(1)$ superconformal coset theory, the spectral flowed operator $V^{sl,\eta}_{j,m}$ has weight and $R$-charge 
\ie
\Delta={-j(j+1)+(m+\eta)^2\over k}+{\eta^2\over 2},~~~R={2(m+\eta)\over k}+\eta.\label{sfsl}
\fe
On the other hand, in the $SU(2)_k/U(1)$ superconformal coset, the spectral flowed operator $V^{su,\eta'}_{j',m'}$ has
\ie
\Delta={j'(j'+1)-(m'+\eta')^2\over k}+{\eta'^2\over 2},~~~R=-{2(m'+\eta')\over k}+\eta'.\label{sfsu}
\fe
In particular, when $\eta,\eta'$ are $\pm {1\over 2}$, $V^{sl,\pm{1\over 2}}_{j,m}$ and $V^{su,\pm{1\over 2}}_{j',m'}$ are R-sector vertex operators.

\subsection{$\bZ_k$ orbifold}

As already mentioned, the worldsheet CFT in the holographic description of DSLST (either type IIA or type IIB case) is
\ie
\bR^{1,5}\times\left({SL(2)_k\over U(1)}\times {SU(2)_k\over U(1)}\right)/\bZ_k,
\fe
The $\bZ_k$ orbifold  is inherited from the holographic dual of the (non-doubly-scaled) LST, with worldsheet description
\ie
\bR^{1,5}\times\bR_\phi\times SU(2)_k.
\fe
Here $\bR_\phi$ is a linear dilaton direction, coming from the radial direction transverse to the stack of NS5-branes.
It is a standard fact that the supersymmetric $SU(2)_k$ WZW model can be written as the $\bZ_k$ orbifold of the product of a supersymmetric $U(1)_k$ WZW model and a supersymmetric coset model $SU(2)_k/U(1)$ \cite{DiFrancesco:1988xz},
\ie
SU(2)_k=\left(U(1)_k\times {SU(2)_k\over U(1)}\right)/\bZ_k. \label{rewriting}
\fe
Before proceeding to the $\bZ_k$ orbifolding in DSLST, let us recall how \eqref{rewriting} works.

\subsubsection{$SU(2)_k$ as a $\bZ_k$ orbifold of $U(1)_k \times {SU(2)_k\over U(1)} $}

We will write down primary operators of the supersymmetric $SU(2)_k$ WZW model in the language of $\left( U(1)_k \times { SU(2)_k\over U(1)} \right)/\bZ_k$. By comparing with the general primary operators in the  unorbifolded $ U(1)_k \times { SU(2)_k\over U(1)}$, we will be able to identify the action of the $\bZ_k$ orbifold.

Let $\Phi_{j',m'}$ be a primary of the bosonic $SU(2)_{k-2}$ current algebra. In the supersymmetric $SU(2)_k$ WZW model, we can adjoin $\Phi_{j',m'}$ with $e^{ i \eta' H'}$ (where $i\partial H' = {\lambda'}^+{\lambda'}^-$). We will see shortly that $\eta'$ may be identified as the spectral flow parameter. 
Now factor out the $J_3'$ charge,
\ie
\Phi_{j',m'}e^{i\eta'H'}=e^{i\sqrt{2\over k}(m'+\eta')X'}V^{su,\eta'}_{j',m'},\label{idsu2}
\fe
where recall that $X'$ is the bosonization of the $J_3'$ current via $J_3'=  i\sqrt{k\over 2} \partial X'$. In the language of $\left( U(1)_k \times {SU(2)_k\over U(1)}\right)/\bZ_k$, $X'$ is the (holomorphic part of the) compact boson at radius $\sqrt{2k}$, that represents the $U(1)_k$. The operator $V^{su,\eta'}_{j',m'}$ is indeed the spectral flowed operator \eqref{sfsu} in the coset model $SU(2)_k/U(1)$, as can be checked by comparing the weights and $R$-charges on both sides. 

To complete the identification of vertex operators on the two sides of \eqref{rewriting}, we need to include the antiholomorphic part as well. The vertex operators 
coming from (\ref{idsu2}) are
\ie\label{primc}
e^{i\sqrt{2\over k}(m'+\eta')X'-i\sqrt{2\over k}(\bar m'+\bar\eta')\bar X'}V^{su,(\eta',\bar \eta')}_{j',m',\bar m'}.
\fe
On the other hand, if we were considering the {unorbifolded} $U(1)_k\times {SU(2)_k\over U(1)}$, the vertex operators would take the form
\ie\label{prima}
e^{i \sqrt{2\over k} M X' -i \sqrt{2\over k} \bar M \bar X' } V^{su ,(\eta',\bar \eta')}_{j' , m',\bar m'},
\fe
where the quantum numbers $M$ and $\bar M$ satisfy $M-\bar M \in \bZ, ~M+\bar M \in k\bZ$. Under the $\mathbb{Z}_k$ action, the $U(1)$ and $SU(2)/U(1)$ parts of (\ref{prima}) carry charge
$-{M-\bar M\over k}$ and ${m'+\eta' - \bar m'-\bar \eta'\over k}$ (mod 1) respectively.

(\ref{primc}) would be reproduced from (\ref{prima}) with the identification
\ie\label{ida}
M= m' +\eta' ,~~\bar M = \bar m' +\bar \eta'.
\fe
However, the quantization condition on $M$ and $\bar M$ in the $U(1)_k$ are different from that on $m', \bar m'$ of vertex operators in the supersymmetric $SU(2)_k$ via this identification.
The condition $M-\bar M\in \bZ$ translates into
\ie
m' + \eta' - \bar m ' -\bar\eta' \in \bZ. \label{U(1)1}
\fe
The other condition $M+\bar M\in k\bZ$, however, needs to be relaxed. This is achieved by taking the $\mathbb{Z}_k$ orbifold of $U(1)_k\times {SU(2)_k\over U(1)}$. Including twisted sectors now allows for $M+\bar M \in \bZ$, or
\ie
m'+\eta' +\bar m' +\bar\eta'\in \bZ,\label{U(1)2}
\fe
as desired. The special case $m'+\eta' +\bar m' +\bar\eta'\in k\bZ$ gives operators in the untwisted sector.
Note that the orbifold projection demands that the total $\mathbb{Z}_k$ charge vanishes, and this is in particular obeyed by (\ref{ida}).

In LST, the $\bZ_k$ orbifold does not act on the linear dilaton $\bR_\phi$. After the deformation to DSLST, the linear dilaton $\bR_\phi$ combined with the $U(1)_k$ will be deformed to the $SL(2)_k/U(1)$ coset theory, on which the $\bZ_k$ orbifolding acts nontrivially. In preparation for the deformed case, let us introduce some notations in the linear dilaton theory.

Let $\phi$ be the linear dilation with background charge $Q^{LST}=\sqrt{2\over k}$ with the standard OPE $\phi(z)\phi(0)\sim-\log z$. Let $\psi_\phi$ be the supersymmetric partner to $\phi$. The linear dilaton fermion $\psi_\phi$ and the third component $\lambda_3'$ of the fermion in the supersymmetric $SU(2)_k$ WZW model pair up to give the current $\psi\psi^*$, where
 \ie
 \psi = {1\over \sqrt{2} } (\psi_\phi + i \lambda'_3), ~~~\psi^* = {1\over \sqrt{2} } (\psi_\phi - i \lambda'_3).
 \fe
After the deformation to DSLST, the current $\psi\psi^*$ turns into $\lambda^+\lambda^-$ in \eqref{slbos} of the $SL(2)_k/U(1)$ coset model. Hence we will use the same symbol $H$ for the bosonization of $\psi\psi^*$,
\ie
\psi\psi^* = i\partial H.
\fe
The total superconformal $R$-current for the internal CFT $\bR_\phi\times \left( U(1)_k \times{ SU(2)_k\over U(1)}\right)/\bZ_k$ can be written in the $\mathbb{R}_\phi\times SU(2)_k$ language as
\ie
J^{tot}_R = i\partial H + i \partial H',\label{totalR}
\fe
where we recall that $i\partial H'=\lambda'^+\lambda'^-$ in the $SU(2)_k$.

Including the linear dilaton $\phi$ and the current $i\partial H = \psi \psi^\ast$, we may consider the more general vertex operator $e^{\sqrt{2\over k} j\phi}e^{i\eta H+i\eta'H'}\Phi_{j',m'}$ in $\bR_\phi\times SU(2)_k$. As before, we can factor out its $J_3'$ charge and write it in the language of $\bR_\phi\times \left( U(1)_k \times{ SU(2)_k\over U(1)}\right)/\bZ_k$ as
\ie
e^{\sqrt{2\over k} j\phi}e^{i\eta H+i\eta 'H'}\Phi_{j',m'}&=e^{\sqrt{2\over k} j\phi}e^{i\eta H}e^{i\sqrt{2\over k}(m'+\eta')X'}V^{su,\eta'}_{j',m'}.\label{defoperator}
\fe

\subsubsection{$\left( {SL(2)_k\over U(1)}\times {SU(2)_k\over U(1)}\right)/\bZ_k$}

Now we will consider the deformation from (the internal part of) the worldsheet theory of LST to that of DSLST \cite{Giveon:1999px,Giveon:1999tq}, namely
\ie
 \bR_\phi \times  \left(U(1)_k\times {SU(2)_k\over U(1)}\right)/\bZ_k\rightarrow
\left({SL(2)_k\over U(1)}\times {SU(2)_k\over U(1)}\right)/\bZ_k.
\fe
We would like to see how the vertex operator \eqref{defoperator} is deformed. 
The weight and the $R$-charge (determined from \eqref{totalR}) of the vertex operator \eqref{defoperator} are
\ie
&\Delta= - {j(j+1)\over k} + {\eta^2+\eta'^2\over2} + {j'(j'+1)\over k},\\
&R = \eta+\eta'.
\fe
In the deformed theory, \eqref{defoperator} maps to the following spectral flowed operator in $\left({SL(2)_k\over U(1)}\times {SU(2)_k\over U(1)}\right)/\bZ_k$
\ie
V^{sl,\eta}_{ j,m'+\eta'-\eta}V^{su,\eta'}_{j',m'}\label{VV},
\fe
which indeed has the same weight and $R$-charge (by \eqref{sfsl} and \eqref{sfsu}).
In other words, the vertex operators $V^{sl,\eta}_{ j,m}V^{su,\eta'}_{j',m'}$ in ${SL(2)_k\over U(1)}\times {SU(2)_k\over U(1)}$ that obey
\ie
m+\eta=m'+\eta',
\fe 
as well as the constraints \eqref{U(1)1} and \eqref{U(1)2}, will survive the orbifold.

Combining the holomorphic and antiholomorphic part, we see that a class of vertex operators in the deformed theory $\left({SL(2)_k\over U(1)}\times {SU(2)_k\over U(1)}\right)/\bZ_k$ can be written as
\ie
 V^{sl,(\eta,\bar\eta)}_{ j,m,\bar m}V^{su,(\eta',\bar \eta')}_{j',m',\bar m'},
\fe
with the quantum numbers satisfying
\ie
 &m+\eta=m'+\eta',~~ \bar m+\bar\eta=\bar m'+\bar \eta',\\
 &m'+ \eta' - \bar m' -\bar \eta'\in \bZ,\\
&m'+\eta'+\bar m'+\bar\eta'\in\bZ.\label{Zk}
\fe 
The vertex operators with $m'+\eta'+\bar m'+\bar\eta'\in k\bZ$ are in the untwisted sector. 


\subsection{Identifications among vertex operators in the coset theories}

%

There are nontrivial identifications between vertex operators $V^{\eta}_{j,m}$ with different $\eta,~j,~m$ in both the $SL(2)/U(1)$ and $SU(2)/U(1)$ coset theories. These identifications can be traced back to the ones in the bosonic coset theories. Let us start with the bosonic $SU(2)_{k_{bos}}/U(1)$ at level $k_{bos}$, with primary operators $V^{bos}_{j',m'}$. The quantum number $j'$ lies in the range $0\le j' \le {k_{bos}\over 2}$, whereas $m'$, unlike in the bosonic $SU(2)_{k_{bos}}$ WZW model, is \textit{a priori} unconstrained. There is the following identification among primaries labeled by different quantum numbers \cite{Fateev:1985mm,Aharony:2003vk}\footnote{Note that the conformal weight formula $h_{j',m'}={j'(j'+1)\over k_{bos}+2}-{m'^2\over k_{bos}}$ only applies for $|m'|\leq j'$ which labels the primaries obtained directly from factoring out $U(1)$. Within this range, the identification is only nontrivial for $m'=j'$. Otherwise it may be regarded as a way to extend the definition of $V^{bos}_{j',m'}$.}:
\ie
&V^{bos}_{j',m'}=V^{bos}_{{k_{bos}\over 2}-j',-{k_{bos}\over 2}+m'}.
\label{id}
\fe
The identification in particular implies $V^{bos}_{j',m'}=V^{bos}_{j',m'+k_{bos}}$. This is consistent with the statement that ${m'-\bar m'\over k_{bos}}$ (mod 1) is the charge with respect to the residual $\mathbb{Z}_{k_{bos}}$ action on the coset theory. 
We now describe the identifications among the vertex operators $V^{su,\eta'}_{j',m'}$ in the supersymmetric $SU(2)_k/U(1)$. 
Note the following relation between the primary operator $\Phi_{j',m'}$ for the bosonic $SU(2)_{k-2}$ WZW model, the vertex operator $V^{su,\eta'}_{j',m'}$ for the supersymmetric coset $SU(2)_k/U(1)$, and the primary operator $V^{bos}_{j',m'}$ for the bosonic coset $SU(2)_{k-2}/U(1)$,
\ie
\Phi_{j',m'} e^{i{\eta'}  H'} = V_{j',m'}^{su,{\eta'}  } e^{ i \sqrt{2\over k } ( {\eta'}  +m') X'}
=V^{bos}_{j',m'} e^{ i \sqrt{ 2\over k-2} m' x'} e^{i{\eta'}  H'}.
\fe
Using (\ref{relation}), we have
\ie
V^{bos}_{j',m'} = V^{su,{\eta'}  } _{j',m'} e^{ - i {\eta'}   \sqrt{k-2\over k} X_R '+ i m' {2\over\sqrt{ k(k-2)} }X_R'} 
\fe
From (\ref{id}) we have $V_{j',m'}^{bos} = V_{{k-2\over 2} -j', m' -{k-2\over 2}}^{bos}$ (recall that $k_{bos}= k-2$), hence
\ie
V^{su,{\eta'}  }_{j',m'} = V^{su,{\eta'}  }_{ {k-2\over 2} -j ', m'- {k-2\over 2}} e^{ -i \sqrt{k-2\over k}X_R'} = V^{su,{\eta'}  -1}_{{k-2\over2}-j',m'-{k-2\over2}}.
\fe
Note that even though $V^{bos}_{j',m'} = V^{bos}_{j',m' + (k-2)}$, the vertex operator for the supersymmetric $SU(2)_k/U(1)$ is not invariant under a shift on $m'$ alone. Rather, we have
\ie
V^{su,\eta'}_{j',m'
 +(k-2)} = V^{su,\eta'-2}_{j',m'}.
 \fe
This is consistent with the interpretation of ${m'+\eta'-\bar m'-\bar\eta'\over k}$ (mod 1) as the charge with respect
to the $\mathbb{Z}_k$ symmetry. We can write the above identifications in a more compact form,
 \ie
 V^{su,{\eta'}  }_{j',m'}  = V^{su,{\eta'}  -1}_{{k-2\over2}-j',m'-{k-2\over2}}= V^{su,{\eta'}  +1}_{{k-2\over2}-j',m'+{k-2\over2}}.\label{id1}
 \fe

Similarly, primary operators in the bosonic $SL(2)_{k+2}/U(1)$ are subject the following identification\footnote{We are considering here coset primaries that come directly from the lowest weight (resp. highest weight) principal discrete representations of $SL(2)$ \cite{Maldacena:2000hw}, i.e $  D^+_j=\{|j;m\ra:m\in j+{\mathbb N}\}$ (resp. $ D^-_j=\{|j;m\ra:m\in -j-{\mathbb N}\}$) with $-{1\over 2}<j<{k-2\over 2}$, to which the conformal weight formula $h_{j,m}=-{j(j+1)\over k_{bos}-2}+{m^2\over k_{bos}}$ applies. When restricted to this subset of primaries, the identification (2.63) is meaningful only for $j=-m-1$, i.e. it maps the highest weight state of $D^-_j$ to the lowest weight state of $D^+_{{k_{bos}\over 2}-j-2}$. The rest of the relations may be thought of as extending the definition of $V^{bos}_{j,m}$ (see \cite{Parnachev:2001gw} for an explanation of the origin of this identification).} \cite{Parnachev:2001gw,Aharony:2003vk}
\ie\label{idsl}
V^{bos}_{j,m} = V^{bos}_{  {k-2\over2}  - j, {k+2\over 2}+m},
\fe
from which we obtain the identification for the supersymmetric $SL(2)_k/U(1)$ theory
\ie
V^{sl,{\eta } }_{j,m}  =V^{sl,{\eta } +1}_{{k-2\over2}-j,m-{k+2\over2}}= V^{sl,{\eta } -1}_{{k-2\over2}-j,m+{k+2\over2}}.\label{id2}
\fe
Note the sign difference in $\eta\pm1$ and $m\mp {k+2\over2}$ when compared with $SU(2)_k/U(1)$. Once again, this is consistent with the interpretation of $-{m+\eta-\bar m-\bar\eta\over k}$ (mod 1) as the charge of $V^{sl,(\eta,\bar\eta)}_{j,m,\bar m}$ with respect to the $\mathbb{Z}_k$ symmetry.

\subsection{Massless string states}

Now we discuss the construction of physical vertex operators in type IIA string theory on $\mathbb{R}^{1,5}\times \left({SL(2)_k\over U(1)}\times {SU(2)_k\over U(1)}\right)/\bZ_k$, which is the T-dual description of IIB $(1,1)$ DSLST \cite{Giveon:1999px,Giveon:1999tq}. We will focus on the explicit description of massless string modes in the $\mathbb{R}^{1,5}$, localized at the tip of the cigar. We will further restrict our attention to bosonic string modes, and discuss the (NS,NS) sector and (R,R) sector separately.

\subsubsection{(NS,NS)-sector}
Consider the (NS,NS)-sector vertex operators of the form \cite{Aharony:2003vk,Aharony:2004xn}
\ie\label{vns}
{\cal V}_{NS} = e^{-\varphi-\tilde\varphi} e^{ip_\mu X^\mu} V_{j,m,\bar m}^{sl,{(\eta,\bar\eta) } } V_{j',m',\bar m'}^{su,{(\eta',\bar\eta')}  }
\fe
where $\varphi,\tilde\varphi$ are the bosonized superconformal ghosts, $V^{sl,(\eta,\bar\eta)} _{j,m,\bar m}$ and $V^{su,(\eta',\bar\eta')} _{j',m',\bar m'}$ are vertex operators of $SL(2)_k/U(1)$ and $SU(2)_k/U(1)$ respectively, as described earlier. $X^\mu$ with $\mu=0,\cdots,5$ are the bosons for $\bR^{1,5}$ (not to be confused with the bosonization of $J^3 = - \sqrt{k\over2} \partial X$ in the $SL(2)$). The spectral flow parameters $\eta,\bar\eta,\eta',\bar\eta'$ are integer valued in the (NS,NS)-sector. Since $\cV_{NS}$ has no $\bR^{1,5}$ spacetime index, it should be the vertex operator for the six-dimensional scalar fields.

The mass shell condition is
\ie
{1\over 2}p^2+ {(m+{\eta}  )^2-j(j+1)\over k}+{{\eta } ^2\over 2} + {j'(j'+1)-(m'+{\eta'}  )^2\over k} +{{\eta'}  ^2\over 2}= {1\over 2}.
\fe
We will focus on the massless case $p^2=0$. We also demand the quantum numbers to obey the $\bZ_k$ orbifold condition \eqref{Zk}. The on-shell condition for massless states then reduces to
\ie
{ j' (j'+1)- j (j+1)\over k } + {{\eta} ^2\over 2} + {{\eta'}  ^2\over 2} ={1\over2}.\label{onshell}
\fe

Next let us examine the chiral GSO projection condition,
\ie
F_L + R \in 2\bZ+1,
\fe
where $F_L$ is the holomorphic worldsheet fermion number (and similarly in the anti-holomorphic sector). 
The total $R$-charge can be computed using spectral flow \eqref{sfsl} and \eqref{sfsu},
\ie
R = {2(m+\eta)\over k } + {\eta } 
-{2(m'+\eta')\over k} + {\eta'} = \eta + {\eta'},
\fe
where we have used the $\bZ_k$ orbifold condition \eqref{Zk} $m+\eta= m'+\eta'$. 
Altogether, we need $F_L+{\eta }  +{\eta'}  \in 2\bZ+1$ for the GSO condition to be met.

It is straightforward to verify that in the case $F_L=0$, a {\it normalizable} (NS,NS) massless vertex operator of the form (\ref{vns}) that obeys GSO projection condition must satisfy (see Appendix~\ref{n1}) 
\ie
\eta^2+{\eta'}  ^2=1.
\fe
In other words, one of $\eta$ and $\eta'$ must be zero and the other equal to $\pm1$.

Recall that a normalizable vertex operator must have $j=m_0-u$ with $u\in \mathbb{N}$, where $m_0={\rm min}\{|m|, |\bar m|\}$.  
Consider first the case $\eta=1,~{\eta'}  =0$, and so $m=m'-1$. In this case, the normalizable vertex operators must have $j=j'$. 
We observe that $|m+1| = |m'| \le j' = j \leq |m|-u$. Hence we must require $m\le -1$ and $j=|m|-1$. The analysis for $\eta=0, ~{\eta'}  =-1$ is identical. The allowed values of the quantum numbers for the normalizable massless vertex operators in these two cases are 
\ie
&j={\ell\over 2},~~m = - {\ell+2\over 2},~~~~j' = {\ell\over 2} ,~~m' = - {\ell\over 2},\\
&\text{for}~~~\ell = 0 ,1,\cdots, k-2,
~~~~~(\eta=1,~{\eta'}  =0~~{\rm or}~~\eta=0,~{\eta'}  =-1).
\fe 
The upper bound on $\ell$ comes from the constraint $j' \le {k-2\over 2}$. Note that $m,m'$ are both negative in these cases.

Similarly, for $\eta=0,~{\eta'}  =1$, we have $j=j'$ and $m=m'+1$, which implies that $|m-1|\le |m|-u$ for some $u\in \mathbb{N}$. The analysis for $\eta=-1,~{\eta'}  =0$ is identical. We end up with the solutions
\ie
&j={\ell\over 2} ,~~ m = {\ell+2\over 2},~~~~j'={\ell\over2},~~ m'={\ell\over2},\\
&\text{for}~~~\ell=0,1,\cdots,k-2,~~~(\eta=0,~{\eta'}  =1~~{\rm or}~~\eta=-1,~{\eta'}  =0).
\fe
Note that $m,m'$ are both positive in these cases.

To summarize, the normalizable vertex operators from the internal CFT have, in their holomorphic part,
\ie
V_{{\ell\over 2},-{\ell+2\over 2}}^{sl,1} V_{{\ell\over 2},-{\ell\over 2}}^{su,0},~~ V_{{\ell\over 2},{\ell+2\over 2}}^{sl,0} V_{{\ell\over 2},{\ell\over 2}}^{su,1},~~ V_{{\ell\over 2},-{\ell+2\over 2}}^{sl,0} V_{{\ell\over 2},-{\ell\over 2}}^{su,-1},~~ V_{{\ell\over 2},{\ell+2\over 2}}^{sl,-1} V_{{\ell\over 2},{\ell\over 2}}^{su,0},~~~\ell=0,1,\dots,k-2.\label{nsvertex0}
\fe
The operators in (\ref{nsvertex0}) are not all independent, however. Recall that we have the identifications \eqref{id1} and \eqref{id2}, and therefore,
\ie
 V_{{\ell\over 2},-{\ell+2\over 2}}^{sl,1} V_{{\ell\over 2},-{\ell\over 2}}^{su,0}&=V_{{k-2-\ell\over 2},{k -\ell\over 2}}^{sl,0} V_{{k-2-\ell\over 2},{k-2-\ell\over 2}}^{su,1},
\\
 V_{{\ell\over 2},-{\ell+2\over 2}}^{sl,0} V_{{\ell\over 2},-{\ell\over 2}}^{su,-1}&=V_{{k-2-\ell\over 2},{k -\ell\over 2}}^{sl,-1} V_{{k-2-\ell\over 2},{k-2-\ell\over 2}}^{su,0}.\label{idns}
\fe
In particular, the identification flips the sign of $m$ and $m'$. This will be important when we include the anti-holomorphic part of the vertex operators.

Combining with the anti-holomorphic part, normalizability requires in addition that either $m,\bar m<-{1\over 2}$ or $m,\bar m>{1\over 2}$ (see \eqref{m0102}). This constrains the possible pairings between $\eta,\eta'$ and $\bar\eta,\bar \eta'$. For example, $\eta=1,~\eta'=0$ cannot pair up with $\bar\eta=0, ~\bar \eta'=1$ since in this case $m<0$ but $\bar m>0$. On the other hand, from the identifications \eqref{idns}, some of the pairings are the same as others. For example, $\eta=1,~\eta'=0,~\bar\eta=1,~\bar\eta'=0$ would be identified as $\eta=0,~\eta'=1,~\bar\eta=0,~\bar\eta'=1$.

In the end, there are four inequivalent pairings between the holomorphic and anti-holomorphic quantum numbers that are allowed in the massless vertex operator $V^{sl,(\eta,\bar\eta)}_{j,m,\bar m}V^{su,(\eta',\bar\eta')}_{j',m',\bar m'}$
\ie
\left.\begin{array}{|c|c|c|c|}\hline \eta & \bar\eta & \eta' & \bar \eta' \\\hline ~1~ &~ 1~ & 0 & 0 \\\hline 0 & 0 & -1 & -1 \\\hline ~1~ & 0 & 0 & -1 \\\hline 0 & ~1~ & -1 & 0 \\\hline \end{array}\right.
\fe
In fact, here the normalizability condition relating $j$ to $m$ and $\bar m$  implies that $m=\bar m$, and the orbifold projection condition further implies $m'=\bar m'$. Note that even though $m,\bar m,m',\bar m'$ are all negative for the vertex operators pairing this way (see \eqref{nsvertex0}), we can use the identifications (\ref{idns}) to have them all positive by changing $\eta,\bar \eta,\eta',\bar\eta'$.

The explicit forms of the four sets of normalizable vertex operators are 
\ie
&\cV^-_{NS1,\ell}=e^{-\varphi-\tilde\varphi} e^{ip_\mu X^\mu}  V_{{\ell\over 2},-{\ell+2\over 2},-{\ell+2\over 2}}^{sl,(1,1)} 
V_{{\ell\over 2},-{\ell\over 2},-{\ell\over 2}}^{su,(0,0)}
,
~~\cV^-_{NS2,\ell}=e^{-\varphi-\tilde\varphi} e^{ip_\mu X^\mu}  
V_{{\ell\over 2},-{\ell+2\over 2},-{\ell+2\over 2}}^{sl,(0,0)} V_{{\ell\over 2},-{\ell\over 2},-{\ell\over 2}}^{su,(-1,-1)},
\\
&\cV^-_{NS3,\ell}=e^{-\varphi-\tilde\varphi} e^{ip_\mu X^\mu}  V_{{\ell\over 2},-{\ell+2\over 2},-{\ell+2\over 2}}^{sl,(1,0)} 
V_{{\ell\over 2},-{\ell\over 2},-{\ell\over 2}}^{su,(0,-1)}
,
~\cV^-_{NS4,\ell}=e^{-\varphi-\tilde\varphi} e^{ip_\mu X^\mu}  
V_{{\ell\over 2},-{\ell+2\over 2},-{\ell+2\over 2}}^{sl,(0,1)} V_{{\ell\over 2},-{\ell\over 2},-{\ell\over 2}}^{su,(-1,0)},
\fe
or, equivalently, using the identifications \eqref{idns}, we can rewrite them as
\ie
&\cV^+_{NS1,\ell}=e^{-\varphi-\tilde\varphi} e^{ip_\mu X^\mu}  V_{{\ell\over 2},{\ell+2\over 2},{\ell+2\over 2}}^{sl,(0,0)} 
V_{{\ell\over 2},{\ell\over 2},{\ell\over 2}}^{su,(1,1)}
,
~~\cV^+_{NS2,\ell}=e^{-\varphi-\tilde\varphi} e^{ip_\mu X^\mu}  
V_{{\ell\over 2},{\ell+2\over 2},{\ell+2\over 2}}^{sl,(-1,-1)} V_{{\ell\over 2},{\ell\over 2},{\ell\over 2}}^{su,(0,0)},
\\
&\cV^+_{NS3,\ell}=e^{-\varphi-\tilde\varphi} e^{ip_\mu X^\mu}  V_{{\ell\over 2},{\ell+2\over 2},{\ell+2\over 2}}^{sl,(0,-1)} 
V_{{\ell\over 2},{\ell\over 2},{\ell\over 2}}^{su,(1,0)}
,
~\cV^+_{NS4,\ell}=e^{-\varphi-\tilde\varphi} e^{ip_\mu X^\mu}  
V_{{\ell\over 2},{\ell+2\over 2},{\ell+2\over 2}}^{sl,(-1,0)} V_{{\ell\over 2},{\ell\over 2},{\ell\over 2}}^{su,(0,1)},
\fe
with $\ell=0,1,\cdots,k-2$. $\cV^-_{NSi,\ell}$ is related to $\cV^+_{NSi,\ell}$ by \eqref{idns},
\ie
\cV^-_{NSi, \,\ell}  =\cV^+_{NSi, \,k-2-\ell},~~~~i=1,2,3,4.
\fe
These $4(k-1)$ vertex operators correspond to $2(k-1)$ complex scalars in the low energy supersymmetric Yang-Mills theory. Only two of the $4(k-1)$ normalizable vertex operators $\cV^-_{NS1,0},\cV^-_{NS2,k-2},$ or equivalently $\cV^+_{NS1,k-2},\cV^+_{NS2,0}$, are in the untwisted sector, i.e. they satisfy the condition $m+\bar m+\eta+\bar\eta \in k\bZ$. The rest are in the twisted sectors. 

Let us compare this with the NS5-brane or six dimensional gauge theory description. We are at the point in the Coulomb branch moduli space where the $k$ NS5-branes are spread on a circle in $\mathbb{R}^2\subset \mathbb{R}^4$, with a $\mathbb{Z}_k$ symmetry that permutes the NS5-branes cyclically and at the same time rotates the circle of spread. The center of mass mode decouples, and the relative motion of the NS5-branes gives rise to $4(k-1)$ massless real scalars. We can denote them by complex scalars $ Z_0,  Z_1,\cdots,  Z_{k-2}$ and $\widetilde Z_1,\widetilde Z_2,\cdots,\widetilde Z_{k-1}$. Here $ Z_i$ are scalars that are linear combinations of collective coordinates of the NS5-brane in the $\mathbb{R}^2$ that contains the circle of spread, while $\widetilde Z_j$ are scalars associated with the remaining traverse $\mathbb{R}^2$. The $\mathbb{Z}_k$ symmetry rotates $ Z_m$ and $\widetilde Z_m$ by the phase $e^{2\pi i m/k}$. Note that $\widetilde Z_0$ is a center of mass mode and decouples, thus absent from the DSLST spectrum. The center of mass mode in the $\phi$ direction, on the other hand, transforms with phase $e^{-2\pi i/k}$ under the $\mathbb{Z}_k$ (the sign in the exponent is a convention). This corresponds to $ Z_{k-1}$ which is absent from the spectrum. So indeed there is only a single massless {\it complex} scalar $ Z_0$ that is uncharged under the $\bZ_k$ symmetry.

\subsubsection{(R,R)-sector}

In the (R,R)-sector, we consider the vertex operators \cite{Aharony:2003vk,Aharony:2004xn}
\ie\label{RVO}
{\cal V}_R=\xi_{a,\dot a}e^{-{\varphi\over 2}-{\tilde\varphi\over 2}}e^{ip_\mu X^\mu}S_{a}\widetilde S_{\dot a}V^{sl,(\eta,\bar\eta)}_{j,m,\bar m}V^{su,(\eta',\bar\eta')}_{j',m',\bar m'},
\fe
where $S_a,~\widetilde S_{\dot a}$ are the spin fields in the $\mathbb{R}^{1,5}$, and $a$ and $\dot a$ are the indices in the $\bf 4$ and $\bf \bar 4$ of $SO(1,5)$ respectively. $\xi_{a,\dot a}$ is the polarization for a six-dimensinoal two-form field strength in the $\bf 15$. In the massless case, $\cV_R$ will give the vertex operators for the field strength of the $U(1)^{k-1}$ gauge bosons.

The on-shell condition for the vertex operator \eqref{RVO} is
\ie
{1\over 2}p^2+ {(m+{\eta}  )^2-j(j+1)\over k}+{{\eta } ^2\over 2} + {j'(j'+1)-(m'+{\eta'}  )^2\over k} +{{\eta'}  ^2\over 2}= {1\over 4}.
\fe
We will focus on the massless case $p^2=0$, and impose the $\bZ_k$ orbifold condition $m+\eta=m'+\eta'$ \eqref{Zk}. The on-shell condition for massless states is then
\ie
{ j' (j'+1)- j (j+1)\over k } + {{\eta} ^2\over 2} + {{\eta'}  ^2\over 2} ={1\over4}.
\fe
It's straightforward to derive that (see Appendix \ref{n2}) physical vertex operators surviving the GSO projection $F_L+R\in 2\bZ+{1\over 2}$ must have the following combinations of spectral flow parameters 
\ie
\eta=\pm{1\over 2},~~\eta'=\mp{1\over 2}.
\fe
Let us first consider the case $\eta={1\over 2},~\eta'= -{1\over2}$. We must have $j=j'$ and $m=m'-1$. The solutions for the normalizable states are
\ie
&j={\ell\over 2},~~m = - {\ell+2\over 2},~~~~j' = {\ell\over 2} ,~~m' = - {\ell\over 2},\\
&\text{for}~~~~\ell = 0 ,1,\cdots, k-2
~~~~~(\eta={1\over2},~{\eta'}  =-{1\over2}).
\fe 
 Note that $m,m'$ are both negative in this case.

Similarly, in the case $\eta=-{1\over 2},~\eta' = {1\over 2}$, we must have $j=j'$ and $m=m'+1$. The solutions for the normalizable states are
\ie
&j={\ell\over 2} ,~~ m = {\ell+2\over 2},~~~~j'={\ell\over2},~~ m'={\ell\over2},\\
&\text{for}~~~~\ell=0,1,\cdots,k-2,~~~(\eta=-{1\over2},~{\eta'}  ={1\over 2}).
\fe
 Note that $m,m'$ are both positive in this case.

However, these two sets of vertex operators with $\eta={1\over2},~\eta'=-{1\over2}$ and $\eta=-{1\over2},~\eta'= {1\over2}$ are in fact identified by \eqref{id1} and \eqref{id2},
\ie
V^{sl, 1/2}_{{\ell\over2}  ,  - {\ell+2\over2} } V^{su, -1/2}_{ {\ell\over2} , - {\ell\over2}}
=V^{sl, -1/2}_{{k-2-\ell\over2}  ,   {k -\ell\over2} } V^{su, 1/2}_{ {k-2-\ell\over2} , {k-2-\ell\over2}}.\label{idr}
\fe


Combining with the anti-holomorphic part, as before, normalizability of the vertex operator demands either $m,\bar m< -{1\over2}$ or $m,\bar m>{1\over2}$. The fact that $m$ and $\bar m$ must take the same sign, for instance, rules out the pairing between $\eta={1\over2},~\eta'=-{1\over2}$ in the holomorphic sector with $\bar\eta=-{1\over2},~\bar\eta'={1\over2}$ in the anti-holomorphic sector. 
In the end, the gauge boson vertex operators are
\ie
{\cal V}^-_{R,\ell}=\xi_{a,\dot a}e^{-{\varphi\over 2}-{\tilde\varphi\over 2}}e^{ip_\mu X^\mu}S_{a}\widetilde S_{\dot a}
V^{sl,(1/2,1/2)}_{{\ell\over2},-{\ell+2\over 2},-{\ell+2\over 2}}V^{su,(-1/2,-1/2)}_{{\ell\over2},-{\ell\over2},-{\ell\over2}},~~\ell=0,1,\cdots,k-2,\label{gaugevertex-}
\fe
or, equivalently,
\ie
{\cal V}^+_{R,\ell}=\xi_{a,\dot a}e^{-{\varphi\over 2}-{\tilde\varphi\over 2}}e^{ip_\mu X^\mu}S_{a}\widetilde S_{\dot a}
V^{sl,(-1/2,-1/2)}_{{\ell\over2},{\ell+2\over 2},{\ell+2\over 2}}V^{su,(1/2,1/2)}_{{\ell\over2},{\ell\over2},{\ell\over2}},~~\ell=0,1,\cdots,k-2.\label{gaugevertex+}
\fe
The two are related by the reflection \eqref{idr},
\ie
\cV^-_{R,\,\ell} = \cV^+_{R,\, k-2-\ell}.
\label{reflection}
\fe
These $k-1$ vertex operators $\cV^-_{R,\ell}$, $\ell=0,1,\cdots,k-2$, correspond to the $U(1)^{k-1}$ field strengths on the Coulomb branch of the six-dimensional gauge theory. Note that there are no untwisted sector massless RR vertex operators because $m'+\eta'+\bar m' + \bar \eta'= - \ell -1$ is never a multiple of $k$. This is consistent with the fact that there are no $\bZ_k$ invariant gauge bosons at this point on the Coulomb branch of the $SU(k)$ gauge theory. In fact, under the $\mathbb{Z}_k$ action that cyclically permutes the NS5-branes, $\cV^\pm_{R,\,\ell}$ rotates with the phase $e^{\pm {2\pi i\over k}(\ell+1)}$. Namely, the $\mathbb{Z}_k$ momentum for $\cV^\pm_{R,\,\ell}$ is $\pm(\ell+1)$.

\section{Correlators and amplitudes}
%
%
%
%
%
%

The goal of this section is to compute the string tree level amplitude of four gauge bosons in DSLST.  This amplitude is expressed in terms of correlators in $\bR^{1,5}$, $SL(2)_k/U(1)$, and $SU(2)_k/U(1)$.  The nontrivial part of the $SL(2)_k/U(1)$ correlator is the $SL(2)$ four-point function, which can be related to a correlator in Liouville theory via Ribault and Teschner's dictionary \cite{Ribault:2005wp}.  The $SU(2)_k/U(1)$ correlator, while unknown for general $k$, can be written in terms of correlators of a free boson and parafermions for $k = 2, 3, 4, 5$.  We write the final scattering amplitude as a series expansion in $\A'/2$:\footnote{The $1/2$ is conventional.}
\ie
\cA_{DSLST} = \cA^{(1)}_{DSLST} + {\A' \over 2} \cA^{(2)}_{DSLST}+ \cdots.
\fe
The main results of this section are the ratios ${\A'\over2}\cA^{(2)}_{DSLST} / \cA^{(1)}_{DSLST}$ (we do not fix the overall normalization of $\cA_{DSLST}$, so only the ratio is unambiguously computed) for different $k$'s presented in Table~\ref{tab:lstresults}. Remarkably, these ratios agree with the ratios between \textit{loop} amplitudes $g_{YM}^2\cA^{2-loop}/\cA^{1-loop}$ in the 6D $SU(k)$ SYM  computed in the next section.

\subsection{Winding number conserving correlators}

To begin with, consider the four-point CFT correlator of RR vertex operators ${\cal V}_{R,\ell}^\pm$ on the sphere, of the form
\ie
&\vev{{\cal V}^+_{R,\ell_1}{\cal V}^+_{R,\ell_2}{\cal V}^+_{R,\ell_3}{\cal V}^+_{R,\ell_4} },~~~\vev{{\cal V}^+_{R,\ell_1}{\cal V}^+_{R,\ell_2}{\cal V}^+_{R,\ell_3}{\cal V}^-_{R,\ell_4} },~~~\vev{{\cal V}^+_{R,\ell_1}{\cal V}^+_{R,\ell_2}{\cal V}^-_{R,\ell_3}{\cal V}^-_{R,\ell_4} },~~~{\rm etc}.
\fe
Let us examine the restrictions on the quantum numbers $\ell_i$ and the numbers of $\cV^+$'s versus $\cV^-$'s in a  nontrivial correlator due to conservation laws. Recall that the $SL(2)_k/U(1)$ part $V^{sl, \pm 1/2}_{j,m}$ of the R-sector vertex operator \eqref{gaugevertex-} and \eqref{gaugevertex+} is related to the primary operator $\Phi_{j,m}$ for the bosonic $SL(2)_{k+2}$ WZW model by
\ie
\Phi_{j,m}e^{\pm i{1\over 2}H}=V^{sl,\pm1/2}_{j,m}e^{\sqrt{2\over k}(m\pm{1\over 2})X},
\fe
where recall that $i\partial H=\lambda^+\lambda^-$ is the $U(1)$ current constructed out of the free fermions in the supersymmetric $SL(2)_k$, and $J^3=-\sqrt{k\over 2}\partial X$ is a component of the overall $SL(2)$ current. We will restrict ourselves to correlators with conserved $m+\eta$ and $\bar m+\bar\eta$ quantum numbers (loosely referred to as ``winding numbers"), since such correlators in the coset theory can be computed straightforwardly from correlators of primaries of the $SL(2)$ WZW model by factoring out the $U(1)$ part. In particular we will consider winding number conserving correlators of the form $\vev{{\cal V}^+_{R,\ell_1}{\cal V}^+_{R,\ell_2}{\cal V}^-_{R,\ell_3}{\cal V}^-_{R,\ell_4} }$, so that the $\ell_i$'s are subject to the constraint
\ie
\ell_1+\ell_2-\ell_3-\ell_4=0.
\fe
Recall that $\ell_i = 2j_i$ for the RR vertex operators \eqref{gaugevertex-} and \eqref{gaugevertex+}.

For the explicit computation below, we will focus on the special case $\ell_1 = \ell_2 = \ell_3= \ell_4\equiv \ell= 0,1,\cdots, k-2$, corresponding to the assignment of $SL(2)$ quantum numbers
\ie
&j_1= j_2 =j_3 = j_4 = {\ell\over 2},\\
&m_1= m_2 =- m_3 = - m_4 = {\ell+2\over 2},~~\bar m_i =m_i.\label{jonshell}
\fe
In fact, such a correlator would be well defined for the non-normalizable vertex operators with generic values of $j_i$'s as well. The correlator $\la \cV_{R,\ell_1}^+ \cV_{R,\ell_2}^+\cV_{R,\ell_3}^- \cV_{R,\ell_4}^-\ra$ of non-normalizable vertex operators has poles in the $j_i$'s at the values corresponding to normalizable vertex operators. The correlator of normalizable vertex operators is, after integration over the worldsheet,  the scattering amplitude $\cA_{DSLST}$ for the corresponding states. It is extracted from the residue of the four-point function of non-normalizable vertex operators, schematically in an LSZ form
\ie\label{vvvv}
\int_{\bC}d^2z\la \cV_{R,\ell_1}^+ \cV_{R,\ell_2}^+\cV_{R,\ell_3}^- \cV_{R,\ell_4}^-\ra \xrightarrow{ j_i \rightarrow {\ell\over 2} }
{ \cA_{DSLST}\over \prod_{i=1}^4( j_i - {\ell\over 2}) }.
\fe
Note that the DSLST amplitude has $\ell\rightarrow k-2-\ell$ reflection symmetry. This is obvious for the particular quantum number assignment due to the identification \eqref{reflection}. More generally, this is a consequence of flipping $\bZ_k$ momenta of the scattering states. 

The nontrivial part of (\ref{vvvv}) is the $SL(2)_k/U(1)$ coset CFT correlator 
\ie
\left\langle V^{sl, - 1/2}_{j_1, m_1}(z_1)
V^{sl, - 1/2}_{j_2 ,m_2}(z_2)
 V^{sl, 1/2}_{j_3,m_3}(z_3)
   V^{sl, 1/2}_{j_4,m_4}(z_4)
 \right\rangle.
\fe
This is related to the correlator of bosonic $SL(2)$ primaries $\Phi_{j,m,\bar m}$ by
\ie\label{winding}
&\left\la \prod_{i=1}^4 \Phi_{j_i,m_i,\bar m_i}(z_i)\right\ra
\left\la e^{- i {1\over 2} H(z_1)}e^{- i {1\over 2} H(z_2)}e^{ i {1\over 2} H(z_3)}e^{i {1\over 2} H(z_4)}\right\ra\\
& =\left\langle V^{sl, - 1/2}_{j_1, m_1}(z_1)
V^{sl, - 1/2}_{j_2 ,m_2}(z_2)
 V^{sl, 1/2}_{j_3,m_3}(z_3)
   V^{sl, 1/2}_{j_4,m_4}(z_4)
 \right\rangle\\
 &~~~~\times
\left\la e^{ \sqrt{2\over k}{\ell+1 \over2} X(z_1)}e^{ \sqrt{2\over k}{\ell+1\over2} X(z_2)}e^{ -\sqrt{2\over k}{\ell+1\over2} X(z_3)}e^{ -\sqrt{2\over k}{\ell+1\over2} X(z_4)}\right\ra.
\fe
The problem is thus reduced to computing the sphere four-point function of $SL(2)$ primaries,
$
\left\la \prod_{i=1}^4 \Phi_{j_i,m_i,\bar m_i}(z_i)\right\ra
$.
As stressed above, this four-point function has poles in $j_i$ as $j_i\rightarrow {\ell\over 2}$. We are only interested in extracting the residue of $\la \prod_{i=1}^4 \Phi_{j_i,m_i,\bar m_i}(z_i)\ra$. We will see in the next subsection that the pole structure is manifest after we rewrite the bosonic $SL(2)_{k+2}$ correlators in terms of Liouville correlators.

\subsection{Bosonic $SL(2)_{k+2}$ correlators and Liouville correlators}

%



In \cite{Ribault:2005wp}, a relation was established between an $n$-point function of primaries in the bosonic $SL(2)_{k+2}$ WZW model on the sphere and a $(2n-2)$-point function in Liouville theory on the sphere. The Liouville background charge $Q=b+1/b$ is related to the $SL(2)$ level $k$ by
\ie
b^2 = {1\over k}.
\fe
The Liouville cosmological constant $\mu$ is chosen to be $\mu={1\over \pi^2k}$ \cite{Ribault:2005wp}.

Before describing this relation between the correlators, we need to specify the normalization convention on the operators of question. Let $\Phi_{j,m,\bar m}$ be the $SL(2)$ WZW primaries, and $V_\A = e^{2\A\phi}$ be Liouville primaries of conformal weight $\Delta_\A=\A(Q-\A)$. In the convention of \cite{Ribault:2005wp}, the two-point function of Liouville primaries take the form
\ie
\langle V_{\A_2} (z_2) V_{\A_1} (z_1) \rangle = 2\pi \left[ \delta(Q-\A_1-\A_2)  + R^L(\A_1) \delta(\A_2-\A_1) \right] \cdot |z_{12}|^{-4\Delta_{\A_1}},
\fe
where $R^L(\A)$ is a reflection coefficient. The $SL(2)$ primaries $\Phi_{j,m,\bar m}(z)$ on the other hand are often conveniently packaged in terms of $\Phi_j(x|z)$, where $x$ is a complex auxiliary variable, such that
\ie
\Phi_{j,m,\bar m}(z) = \int d^2x \, x^{j+m} \bar x^{j+\bar m} \Phi_j(x|z).
\fe
$\Phi_j(x|z)$ are normalized such that their two-point functions take the form\footnote{It was argued in \cite{Aharony:2003vk} that the corresponding string theory two-point function has a slightly different normalization,
\ie
\langle \Phi_j \Phi_j \rangle_{string} = {1\over 2\pi^2} {2j+1\over k}\cdot {R^H(j)\over \pi \gamma(-2j-1)} .
\fe
}
\ie
\langle \Phi_{j_2}(x_2|z_2) \Phi_{j_1}(x_1|z_1) \rangle = \left[ \delta^2(x_{12}) \delta(j_1+j_2+1) + {R^H(j_1)\over \pi \gamma(-2j_1-1)} |x_{12}|^{-4j_1-4} \delta(j_1-j_2) \right] \cdot |z_{12}|^{-4\Delta_j},
\fe
where $\gamma(x)\equiv \Gamma(x)/\Gamma(1-x)$. The function $R^H(j)$ and $R^L(\A)$ coincide under the identification
\ie
\A = -b j + {1\over 2b}.
\fe

Now we can state Ribault and Teschner's relation connecting the $SL(2)$ and Liouville correlators. It takes the form\footnote{We use a different convention for $j$ than in \cite{Ribault:2005wp}: $j_{there} = - ( j_{here}+1)$. Our convention is consistent with that of \cite{Aharony:2003vk}.} 
\ie
& \left\langle \prod_{i=1}^n \Phi_{j_i,m_i,\bar m_i}(z_i,\bar z_i) \right\rangle \\
 &= {\pi^{2-2n} b\over (n-2)!} {\rm vol}\cdot\delta_{\sum m_i}\delta_{\sum \bar m_i} \prod_{i=1}^n {\Gamma(j_i+m_i+1)\over \Gamma(-j_i-\bar m_i)}
\prod_{1\leq r<s\leq n} z_{rs}^{m_r+m_s + {k\over 2}+1} \bar z_{rs}^{\bar m_r+\bar m_s + {k\over 2}+1}
\\
&~~~~\times 
\int \prod_{a=1}^{n-2} d^2 y_a \prod_{1\leq a<b\leq n-2} |y_{ab}|^{k+2} \prod_{r=1}^n \prod_{a=1}^{n-2} (z_r-y_a)^{-m_r-{k\over 2}-1} (\bar z_r-\bar y_a)^{-\bar m_r-{k\over 2}-1}
\\
&~~~~\times \left\langle \prod_{i=1}^n V_{\A_i}(z_i,\bar z_i) \prod_{a=1}^{n-2} V_{-{1\over 2b}}(y_a,\bar y_a) \right\rangle_{Liouville},\label{connection}
\fe
where $z_{rs} = z_r -z_s$ and $y_{ab}= y_a-y_b$;
vol is the volume factor coming from the integration
\ie
{1\over 2\pi} \int {d^2x\over |x|^2}.
\fe
So strictly speaking the correlators of $\Phi_{j,m,\bar m}$'s in the $SL(2)$ WZW model are divergent. The divergent volume factor will cancel against the correlator of the $U(1)$ part of the vertex operators, and the corresponding (winding number conserving) correlator in the $SL(2)/U(1)$ coset theory will be finite.
The Liouville momenta $\A_i$ of $V_{\A_i}$ are mapped to the $SL(2)$ spins $j_i$ via 
\ie\label{alphaj}
\A_i 	 = -bj_i+{1\over 2b}.
\fe

Let us now study the pole structure in \eqref{connection} for the case of interest, namely $n=4$,  $j_i\rightarrow {\ell\over 2}$, and $m_1=m_2=-m_3= -m_4= {\ell+2\over 2}$. Already the prefactors $\Gamma(j_i + m_i +1)$ give rise to poles in $j_3$, $j_4$ as
\ie
& j_3, \, j_4 \rightarrow {\ell\over 2} 
\fe
while $m_3$ and $m_4$ are kept fixed at $m_3=m_4 = - {\ell+2\over 2}$. 
The poles in $j_1$ and $j_2$ with positive (and fixed) $m_1$ and $m_2$, on the other hand, are expected to come from the $y_1,y_2$ integral, of the form
\ie\label{yint}
&\int d^2y_1 d^2 y_2 | y_{12}|^{k+2} \prod_{r,a=1}^2 |z_r-y_a|^{- (\ell +k+4)}
\prod_{r=3}^4 \prod_{a=1}^2 |z_r- y_a|^{\ell- k }\\
&
\times\left\la
V_{{k-\ell\over2} b +b\epsilon_1}(z_1)V_{{k-\ell\over2} b +b\epsilon_2}(z_2)
V_{{k-\ell\over2} b}(z_3)
V_{{k-\ell\over2} b }(z_4)
V_{-{1\over 2b}} (y_1)
V_{-{1\over 2b}}(y_2) 
\right\ra_{Liouville}
\fe
Here we have already taken the limit  $j_3, j_4\to {\ell\over 2} $, while writing 
\ie
j_i={\ell\over 2}-{\epsilon_i}~~ \text{for}~~ i=1,2.
\fe
 Evidently, the poles in $\epsilon_1, \epsilon_2$ must come from the integration over $y_1$, $y_2$ approaching either $z_1$ or $z_2$. Note that there is no singular contribution in the limits $y_1, y_2\to z_3, z_4$, due to the structure of the OPEs with $V_{-{1\over 2b}}$.
 
%

In Liouville theory, degenerate primaries have weights
\ie
\Delta_{m,n} = {Q^{2}\over 4} - {(n'b+m'/b)^2\over 4},~~~~n',m'\geq 1.
\fe
The first null descendant state occurs at level $n'm'$.
The primaries are given by $V_\A = e^{2\A\phi}$ with 
\ie
& 2\A = {nb+{m/ b}} = b(n+mk),
\\
& (n,m) = (n'+1,m'+1)~~{\rm or}~~(-n'+1,-m'+1),
\fe
where we need either $n,m\leq 0$, or $n,m\geq 2$. 

In particular, the operator $V_{-{1\over 2b}}$ is the degenerate primary labeled by $(n,m)=(0,-1)$.
It has the following OPE with $V_{b+\epsilon b} (z)$,
\ie\label{degOPE}
V_{-{1\over 2b}} (y) V_{ {k-\ell\over2} b +b\epsilon}(z)
&\sim |y-z|^{k-\ell +2 \epsilon}V_{- b{\ell\over 2} + \epsilon b}(z)\\&
+C_-({k-\ell\over2} b +b\epsilon) |y-z|^{k+\ell +2 -2\epsilon} V_{- b{\ell\over 2} +{1\over b} +\epsilon b}(z).
\fe
where the coefficient $C_-(\A)$ is given by 
\ie\label{opec}
C_-(\A) = \widetilde\mu^k k^{k+1} {\gamma( {2\A\over b} -1-k) \over \gamma({2\A\over b}) },
\fe
where $\widetilde \mu \equiv \pi \mu \gamma(b^2) b^{2-2b^2}$. 

By looking at the limit $y_1,y_2 \rightarrow z_1,z_2$ and using the OPE \eqref{degOPE}, the six-point function of interest in (\ref{yint}) reduces to four-point functions. 
Let us first consider the contribution to the six-point function in this limit from the second term in the OPE \eqref{degOPE}, of the form
\ie
C_-({ k-\ell\over2} b+\epsilon_1 b)C_-({k-\ell\over2}b+\epsilon_2 b) \left\langle V_{-b{\ell\over2}+{1\over b} +\epsilon_1 b} (z_1)V_{-b{\ell\over2}+{1\over b} +\epsilon_2 b} (z_2)V_{ {k-\ell\over2} b}(z_3)V_{ {k-\ell\over2} b}(z_4)\right\rangle\label{CC}.
\fe
We have previously asserted that the poles in $\epsilon_1,\epsilon_2$ will come from the integral in $y_1$ and $y_2$. One may worry about further poles coming from the OPE coefficients (\ref{opec}). Indeed, the coefficient $C_-({ k-\ell\over2} b+\epsilon b)$ has a pole in $\epsilon$ at $\epsilon=0$. However, this pole will be canceled by a zero in the four-point function $\left\langle V_{-b{\ell\over2}+{1\over b} +\epsilon_1 b} (z_1)V_{-b{\ell\over2}+{1\over b} +\epsilon_2 b} (z_2)V_{ {k-\ell\over2} b}(z_3)V_{ {k-\ell\over2} b}(z_4)\right\rangle$ at $\epsilon=0$. To see this, we make use of the reflection relation in the Liouville theory
\ie
V_\A = R^L (\A) V_{Q-\A},
\fe
where
\ie
R^L(\A) = -\left[ \pi \mu \gamma(b^2) \right]^{{Q-2\A\over b}}
{\Gamma\left( 1+b(2\A-Q) \right)\Gamma\left( 1+b^{-1}(2\A-Q) \right)\over \Gamma\left( 1-b(2\A-Q) \right)\Gamma\left( 1-b^{-1}(2\A-Q) \right)},
\fe
together with the relation between $C_-$ and reflection coefficients,
\ie
C_-(\A)=R^L (\A) R^L(Q-\A-{1\over 2b}).
\fe
Applying the reflection relation on $V_{-b{\ell\over 2}+{1\over b}+\epsilon b}$, $\epsilon=\epsilon_1, \epsilon_2$, (\ref{CC}) turns into a Liouville four-point function multiplied by two factors of the form
\ie
&C_-({ k -\ell\over 2} b +\epsilon b) R^L( -b{\ell\over 2} +b^{-1} +\epsilon b) \\
&= -[\pi \mu \gamma(b^2)]^{b^{-2} -1} b^{-4} 
{ \Gamma(1-k+\ell - 2\epsilon)\Gamma (- \ell-1 +2\epsilon) \over \Gamma(k-\ell) \Gamma(\ell+2) }\times
{\Gamma(2- 2b^2\ell -b^2) \Gamma( k-\ell)\over 
\Gamma( b^2\ell +b^2) \Gamma(2-k+\ell-2\epsilon)}\\
&= [\pi \mu \gamma(b^2)]^{b^{-2} -1} b^{-4} 
{ \Gamma (- \ell-1 +2\epsilon) \over \Gamma(\ell+2) }\times
{\Gamma(2- 2b^2\ell -b^2) \over 
\Gamma( b^2\ell +b^2)} {1\over 1-k+\ell}.
\fe 
In these manipulations, we have treated $k=1/b^2$ as a generic real number (as opposed to an integer), so that we do not have to worry about potential poles coming from $\Gamma(-\ell-1+2\epsilon)$. In the end when we take $k$ to be an integer, the pole from $\Gamma(-\ell-1+2\epsilon)$ will cancel against the zero from $1/\Gamma(-j_i-\bar m_i) = 1/\Gamma(-\ell-1+\epsilon_i)$ in (\ref{connection}) for $i=1,2$. On the other hand, the pole in $C_-({ k -\ell\over 2} b +\epsilon b) $ is canceled by the reflection coefficient $R( -b{\ell\over 2} +b^{-1} +\epsilon b)$ for general value of $b$. We are then left with a finite Liouville four-point function $\left\la V_{{\ell+2\over2}b}(z_1)V_{{\ell+2\over2}b}(z_2)V_{{k-\ell\over2} b}(z_3)V_{ { k-\ell\over2} b}(z_4) \right\ra_{Liouville}$.

Putting everything together, the residue of the $SL(2)$ correlator in $\epsilon_i$ with $i=1,\cdots,4$ is computed by
\ie
&\left\langle \prod_{i=1}^4 \Phi_{j_i,m_i,\bar m_i}(z_i) \right\rangle\Big|_{\epsilon_i\rightarrow 0}
= {\pi^{-6} b}\, {\rm vol}\cdot
{\Gamma(-1-\epsilon_3)\Gamma(-1-\epsilon_4)\Gamma(\ell+2)^2\over \Gamma(-\ell-1+\epsilon_1)\Gamma(-\ell-1+\epsilon_2)}\\
&
~~~~\times|z_{12}|^{2\ell +k +6}|z_{34}|^{-2\ell + k -2}|z_{13}|^{k+2}|z_{14}|^{k+2}|z_{23}|^{k+2}|z_{24}|^{k+2}
\\
&~~~~\times 
\int d^2y_1 d^2y_2 |y_{12}|^{k+2} |z_1-y_1|^{-2-2\epsilon_1} |z_1-y_2|^{- (\ell+ k +4)} |z_1-y_3|^{\ell-k}|z_1-y_4|^{\ell-k} \\
&~~~~~\times
|z_2-y_1|^{-(\ell+k+4)} |z_2-y_2|^{-2-2\epsilon_2} |z_2-y_3|^{\ell-k} |z_2-y_4|^{\ell-k}\\
&~~~~\times 
C_-({k-\ell\over 2} b + \epsilon_1 b)R(- b{\ell\over2}+b^{-1}+\epsilon_1 b)C_-({k-\ell\over 2} b + \epsilon_2 b)R(- b{\ell\over2}+b^{-1}+\epsilon_2 b)\\
&~~~~\times\la V_{{\ell+2 \over 2}b}(z_1)V_{{\ell+2 \over 2}b}(z_2)V_{{k-\ell \over 2}b}(z_3)V_{{k-\ell \over 2}b}(z_4) \ra_{Liouville}
\fe
In the limit $\epsilon_{1,2}\to 0$, there will be two identical contributions to the pole coming from the limiting regions $y_1\rightarrow z_1,y_2\rightarrow  z_2$ and $y_1\rightarrow z_2,y_2\rightarrow  z_1$,
\ie
&\int d^2y_1 d^2y_2 |y_{12}|^{k+2} |z_1-y_1|^{-2-2\epsilon_1} |z_1-y_2|^{- (\ell+ k +4)} |z_1-y_3|^{\ell-k}|z_1-y_4|^{\ell-k} \\
&~~~~~\times
|z_2-y_1|^{-(\ell+k+4)} |z_2-y_2|^{-2-2\epsilon_2} |z_2-y_3|^{\ell-k} |z_2-y_4|^{\ell-k}\\
&\sim { \pi^2\over \epsilon_1 \epsilon_2} |z_{12}|^{-2\ell - k -6}|z_{13}|^{\ell-k}  |z_{14}|^{\ell-k}|z_{23}|^{\ell-k} |z_{24}|^{\ell-k}.
\fe
We then arrive at the relation
\ie\label{relsl}
\left\langle \prod_{i=1}^4 \Phi_{j_i,m_i,\bar m_i}(z_i) \right\rangle\Big|_{\epsilon_i\rightarrow 0}
&\sim {1\over \epsilon_1\epsilon_2\epsilon_3\epsilon_4} \times|z_{34}|^{-2\ell+k-2} |z_{13}|^{\ell+2} |z_{14}|^{\ell+2} |z_{23}|^{\ell+2} |z_{24}|^{\ell+2}\\
&\times\la V_{{\ell+2 \over 2}b}(z_1)V_{{\ell+2 \over 2}b}(z_2)V_{{k-\ell \over 2}b}(z_3)V_{{k-\ell \over 2}b}(z_4) \ra_{Liouville}
\fe
Here we have dropped an immaterial overall constant factor, as well as the vol factor which will cancel against the $U(1)$ correlator in passing to the coset $SL(2)/U(1)$.

Recall the decomposition of the four-point function in terms of conformal blocks in Liouville theory, \footnote{These $\A_i$ are general and are not to be confused with the $\A_i$ identified in \eqref{alphaj}.}
\ie
&\left\langle V_{\A_1}(z_1) V_{\A_2}(z_2) V_{\A_3}(z_3) V_{\A_4}(z_4) \right\rangle \\
&~~~= |z_{14}|^{-4\Delta_1} |z_{24}|^{2(\Delta_1 - \Delta_2 + \Delta_3 - \Delta_4)} |z_{34}|^{2(\Delta_1+\Delta_2-\Delta_3-\Delta_4)} |z_{23}|^{2(\Delta_4-\Delta_1-\Delta_2-\Delta_3)} \\
&~~~\times 
\int_{0}^\infty {dP\over 2\pi} C(\A_1,\A_2,{Q\over 2}+iP) C(\A_3,\A_4,{Q\over 2}-iP) |F(\Delta_1,\Delta_2,\Delta_3,\Delta_4; \Delta_P; z)|^2,
\fe
where $z={z_{12}z_{34}\over z_{14}z_{32}}$ is a conformally invariant cross ratio and $\Delta_i = \A_i(Q-\A_i)$ is the conformal weight of $V_{\A_i}$. Here $\Delta_P = {Q^2\over 4} +P^2$. Importantly, in applying this formula, $\A_i$ are assumed to lie on the line ${Q\over 2} + i\mathbb{R}$. In order to go to real values of $\A_i$, for the application to $SL(2)/U(1)$ correlators, one performs an analytic continuation in $\A_i$'s, in which extra residue contribution need to be included whenever a pole of the three-point function coefficient in $P$ crosses the contour as we deform the $\A_i$'s. This significance of this phenomenon was explained in \cite{Maldacena:2001km}.

The Liouville three-point function coefficient $C(\A_1,\A_2,\A_3)$ are given by \cite{Zamolodchikov:1995aa,Teschner:2001gi}
\ie\label{structureconstant}
C(\A_1,\A_2,\A_3) = \widetilde\mu^{Q-\sum\A_i\over b} {\Upsilon_0 \prod_{i=1}^3\Upsilon(2\A_i)\over \Upsilon(\sum\A_i - Q) \Upsilon(\A_1+\A_2-\A_3)\Upsilon(\A_2+\A_3-\A_1)\Upsilon(\A_3+\A_1-\A_2)},
\fe
where $\Upsilon_0 \equiv \Upsilon'(0)$. The function $\Upsilon(x)$ is defined by
\ie
\Upsilon(-bj) = {b^{-{j(j+1)\over k}-j} \over k^{(j+1)(k-j)\over 2k} \Gamma_2(-j-1|1,k) \Gamma_2(k+j+2|1,k) },
\fe
where $\Gamma_2(x|1,\omega)$ is the Barnes double Gamma function. 
In particular $\Upsilon(x)$ has zeroes at $x=-nb-m/b$ and $x=(n+1)b+(m+1)/b$, for integers $n,m\geq 0$.

In our case of interest,  $\A_1=\A_2={\ell+2\over2}b$ and $\A_3=\A_4={k-\ell\over2}b$, hence
\ie\label{alphadelta}
&\Delta_1 = \Delta_2= {1\over 4k} (\ell+2)(2k-\ell),\\
&\Delta_3=\Delta_4 =  {1\over 4k} (k+\ell+2)(k-\ell).
\fe
We therefore obtain from \eqref{relsl}
\ie\label{phi4decomp}
&\left\langle \prod_{i=1}^4 \Phi_{j_i,m_i,\bar m_i}(z_i) \right\rangle\Big|_{\epsilon_i\rightarrow 0}
\sim {1\over \epsilon_1\epsilon_2\epsilon_3\epsilon_4}
 {\Gamma(2-2b^2\ell -b^2)^2 \over \Gamma(b^2\ell+b^2)^2}{1\over (1-k+\ell)^2} \\
 &~~
\times
 |z_{23}|^{ (\ell+2) ({\ell\over k}-1) } |z_{24}|^{\ell+2} |z_{13}|^{\ell+2} |z_{14}|^{ (\ell+2) ({\ell\over k}-1)}\\
&~~ \times\int_{0}^\infty {dP\over 2\pi}C(\A_1,\A_2,{Q\over 2}+iP) C(\A_3,\A_4,{Q\over 2}-iP)
|F(\Delta_1,\Delta_2,\Delta_3,\Delta_4; \Delta_P; z)|^2.
\fe
Once again, we have dropped an overall constant normalization factor that depends on $k$ only.
Note that since the crossing ratio $z$ is invariant under $(z_1,z_2)\leftrightarrow (z_3, z_4)$, the above expression is invariant under the exchange $(1,2)\leftrightarrow (3,4)$. Indeed since we have taken all $j_i$ to be equal to ${\ell\over 2}$, and $m_1=m_2=-m_3=-m_4$, this exchange simply flips the signs of all $m_i$'s, and the correlator remains invariant.

\subsection{Scattering amplitude in the double scaled little string theory}

Now let us put everything together including the $\bR^{1,5}$, $SL(2)_k/U(1)$, and $SU(2)_k/U(1)$ parts, and also the $bc$ and $\beta\gamma$ ghosts, to obtain the scattering amplitudes in the double scaled little string theory. We will expand the amplitudes in power series of the Mandelstam variables $s_{12}$ and $s_{13}$ and evaluate them numerically for $k=2,3,4,5$. 

Let us take $z_1=z,~z_2=0, ~z_3=1,~z_4=\infty$. The winding number conserving $SL(2)/U(1)$ correlators are related to the correlators in the bosonic $SL(2)_{k+2}$ WZW model by \eqref{winding}. The correlators involving $H$ and $X$ in \eqref{winding} can be evaluated straightforwardly,
\ie
&\la e^{ - i {1\over 2}H(z)} e^{ - i {1\over 2}H(0)} e^{ i {1\over 2}H(1)} e^{  i {1\over 2}H(\infty)}\ra
= z^{1\over 4}(1-z)^{-{1\over 4}},\\
&\la e^{ \sqrt{2\over k}{\ell+1 \over2} X(z)}e^{ \sqrt{2\over k}{\ell+1\over2} X(0)}e^{ -\sqrt{2\over k}{\ell+1\over2} X(1)}e^{ -\sqrt{2\over k}{\ell+1\over2} X(\infty)}\ra
= {\rm vol}~z^{-{ (\ell+1)^2\over 2k}}(1-z)^{{ (\ell+1)^2\over 2k}}.
\fe
We then have
\ie
& \left\langle V^{sl, - 1/2}_{j_1, m_1}(z_1)
V^{sl, - 1/2}_{j_2 ,m_2}(z_2)
 V^{sl, -1/2}_{j_3,m_3}(z_3)
   V^{sl, -1/2}_{j_4,m_4}(z_4)
 \right\rangle \\
& ={1\over{\rm vol }} \la \prod_{i=1}^4 \Phi_{j_i,m_i} (z_i)\ra \times
 z^{  {(\ell+1)^2\over 2k }   +{1\over 4}    }
 (1- z)^{  -{(\ell+1)^2\over 2k }   -{1\over 4}    }.
 \fe
An explicit expression for the bosonic $SL(2)_{k+2}$ four-point function $ \la \prod_{i=1}^4 \Phi_{j_i,m_i} (z_i)\ra$ was given in (\ref{phi4decomp}) in the last subsection.
For the $\bR^{1,5}$ part, the four-point functions are (up to some $z$-independent overall factors)
\ie
&\la c\tilde c e^{ip_1\cdot X}(z,\bar z) e^{ip_2\cdot X}(0,0)  c\tilde c e^{ip_3\cdot X}(1,1) c\tilde c e^{ip_4\cdot X}(\infty,\infty)\ra \sim\delta(\sum_{i=1}^4 p_i) |z|^{-{\A' \over 2} s_{12}}|1-z|^{-{\A' \over 2} s_{13}},\\
&\la e^{-\varphi(z)/2}
e^{-\varphi(0)/2}
e^{-\varphi(1)/2}
e^{-\varphi(\infty)/2}\ra = z^{-{1\over4}}(1-z)^{-{1\over4}} ,\\
&\la S_a(z)S_b(0)S_c(1)S_d(\infty)\ra = \epsilon_{abcd}z^{-{1\over4}}(1-z)^{-{1\over4}},
\fe
where $s_{ij} = -2 p_i \cdot p_j$.

The final expression for the four-gluon scattering amplitude with $\mathbb{Z}_k$ momenta $\ell+1,\,\ell+1,\,-\ell-1,\,-\ell-1$ is (up to an immaterial overall factor, the polarization factor, and the momentum conservation delta function $ \delta(\sum_{i=1}^4 p_i)$)
\ie\label{lstamplitude}
\cA_{DSLST}=&\underset{ j_i \rightarrow {\ell\over 2}}{\text{Res}} \, \int_\bC d^2z\ \la \mathcal{V}^+_{R,\ell}(z,\bar z)\mathcal{V}^+_{R,\ell}(0,0)\mathcal{V}^-_{R,\ell}(1,1)\mathcal{V}^-_{R,\ell}(\infty,\infty)\ra \\
& \sim \int_\bC d^2z\ \, |z|^{{(\ell+1)^2\over k}-s_{12} -{1\over2} } |1-z|^{\ell -{(\ell+1)^2\over k} -s_{13} +{1\over2}} \\
 ~~~~~~~~& \quad \times\,
 \la V^{su,(1/2,1/2)}_{ {\ell\over2} ,{\ell\over2},{\ell\over 2}}(z,\bar z)
 V^{su,(1/2,1/2)}_{ {\ell\over2} ,{\ell\over2},{\ell\over 2}}(0,0)
 V^{su,(-1/2,-1/2)}_{ {\ell\over2} ,-{\ell\over2},-{\ell\over 2}}(1,1)
  V^{su,(-1/2,-1/2)}_{ {\ell\over2} ,-{\ell\over2},-{\ell\over 2}}(\infty,\infty)\ra
  \\
~~~~~~~~& \quad \times\int_{0}^\infty {dP\over 2\pi}C(\A_1,\A_2,{Q\over 2}+iP) C(\A_3,\A_4,{Q\over 2}-iP)
|F(\Delta_1,\Delta_2,\Delta_3,\Delta_4; \Delta_P; z)|^2,
\fe
where $C(\A_1,\A_2,\A_3)$ and $F(\Delta_1,\Delta_2,\Delta_3,\Delta_4)$ are the structure constant \eqref{structureconstant} and the conformal block of the Liouville theory. Here $Q=b+1/b$ and $b^2=1/k$. $\A_i$ and $\Delta_i$ are given in \eqref{alphadelta}. The Liouville conformal block will be computed using Zamolodchikov's recurrence formula as reviewed in Appendix \ref{zrec}.  Once the  four-point function $\la V^{su}V^{su}V^{su}V^{su}\ra$ in the supersymmetric $SU(2)_k/U(1)$ coset model is obtained, the integral in \eqref{lstamplitude} can be computed numerically (see Appendix \ref{numerics}).

For the supersymmetric $SU(2)_k/U(1)$ coset model, the general four-point functions are known but rather complicated. We will instead look into a few lower-level examples, where correlators in the supersymmetric $SU(2)_k/U(1)$ theory can be computed from correlators of free bosons and parafermions.

\subsubsection{$k=2$}

The supersymmetric $SU(2)_2/U(1)$ has zero central charge and is a trivial theory. The $\bZ_k$ momentum $\ell+1$ can only be $1$ in the $k=2$ case, i.e., $\ell=0$. The complete scattering amplitude for the double scaled little string theory is (up to some immaterial overall factor and the polarization factor)
\ie\label{ampxa}
\cA_{DSLST}
&\sim
\int_\bC d^2z\ \,
\, |z|^{-{\A' \over 2} s_{12}} |1-z|^{-{\A' \over 2} s_{13}}\\
 &
 \quad \times \int_{0}^\infty {dP\over 2\pi}C({1\over \sqrt{2}},{1\over \sqrt{2}},{Q\over 2}+iP) C({1\over \sqrt{2}},{1\over \sqrt{2}},{Q\over 2}-iP)
|F(\Delta_1,\Delta_2,\Delta_3,\Delta_4; \Delta_P; z)|^2.
\fe
Here $Q=b+b^{-1}={3\sqrt {2}\over 2}.$


The amplitude can be computed numerically (Appendix \ref{numerics}) and we find
\ie
k=2, ~\ell=0:~~~\cA_{DSLST} \sim 27.92 +28.18 \left( {\A' \over 2} \right)^2 (s_{12}^2+s_{13}^2+s_{14}^2)+\mathcal{O} (\A'^3s^3),
\fe
where we have included the explicit $\A'$-dependence.
The linear term in the Mandelstam variables is absent in this case due to the symmetry in $s_{12}$, $s_{13}$, and $s_{14}$, so ${\cA^{(2)}_{DSLST}}$ is zero.



%
%

\subsubsection{$k=3$}

As we have seen in Section \ref{compact}, the supersymmetric $SU(2)_3/U(1)$ coset model is the compact boson $\phi'$ CFT with radius $R=1/\sqrt{3}$.
 The relevant primary operators for the gauge boson vertex operators  \eqref{gaugevertex-} and  \eqref{gaugevertex+} are
\ie
&V^{su, 1/2,1/2}_{ 0, 0 ,0 } = V^{su,-1/2,-1/2}_{{1\over2}, -{1\over 2} , -{1\over 2}},~~\Delta =\bar \Delta= {1\over 24} ,~~R= \bar R ={1\over 6},\\
&V^{su, 1/2,1/2}_{ {1\over 2}, {1\over 2} ,{1\over2} } = V^{su,-1/2,-1/2}_{ 0 , 0,0},~~\Delta =\bar \Delta= {1\over 24} ,~~R= \bar R =-{1\over 6}.
\fe
By comparing the dimensions and the $R$-charges, the above vertex operators can be written in terms of the compact boson $\phi'$ as
\ie
&V^{su, 1/2,1/2}_{ 0, 0 ,0 }(z,\bar z) = V^{su,-1/2,-1/2}_{{1\over2}, -{1\over 2} , -{1\over 2} }(z,\bar z) = 
\exp\left[  i   { 1 \over 2\sqrt{3} } \phi'(z)\right] \exp\left[ - i   { 1 \over 2\sqrt{3} } \bar \phi'(\bar z)\right] ,\\
&V^{su, 1/2,1/2}_{ {1\over 2}, {1\over 2} ,{1\over2} }(z,\bar z) = V^{su,-1/2,-1/2}_{ 0 , 0,0} (z,\bar z)= \exp\left[ - i   { 1 \over 2\sqrt{3} } \phi'(z)\right] \exp\left[ i   { 1 \over 2\sqrt{3} } \bar \phi'(\bar z)\right] .
\fe
The four-point functions of the supersymmetric $SU(2)_3/U(1)$ coset model can then be computed straightforwardly for each value of the $\bZ_k$ momentum $\ell = 1, 2$, 
\ie
& \la V^{su, 1/2,1/2}_{ 0, 0 ,0 } (z,\bar z) V^{su, 1/2,1/2}_{ 0, 0 ,0 } (0,0) V^{su, -1/2,-1/2}_{ 0, 0 ,0 }  (1,1) V^{su, -1/2,-1/2}_{ 0, 0 ,0 } (\infty,\infty) \ra \\
&= \la V^{su, 1/2,1/2}_{ {1\over 2}, {1\over 2} ,{1\over 2} } (0,0) V^{su, 1/2,1/2}_{ {1\over 2}, {1\over 2} ,{1\over 2} } (z,\bar z) V^{su, -1/2,-1/2}_{ {1\over 2},-{1\over 2} ,-{1\over 2} }  (1,1) V^{su, -1/2,-1/2}_{ {1\over 2}, -{1\over 2} ,-{1\over 2} } (\infty,\infty) \ra\\
 &= | z| ^{ {1\over 6} }  |1-z|^{ -{1\over 6}}.
\fe

Combining with the correlators in the $\bR^{1,5}$ and $SL(2)_3/U(1)$ parts, we obtain the following scattering amplitude,
\ie\label{ampxb}
\cA_{DSLST}&\sim   \int_\bC d^2z\ |z|^{  {(\ell+1)^2 \over 3 }   -{\A' \over 2} s_{12}  -{1\over3}  }
  |1-z|^{\ell  - {( \ell+1)^2\over 3} - {\A' \over 2} s_{13} +{1\over3} } \\
& \times  \int_{0}^\infty {dP\over 2\pi}C(\A_1,\A_2,{Q\over 2}+iP) C(\A_3,\A_4,{Q\over 2}-iP)
|F(\Delta_1,\Delta_2,\Delta_3,\Delta_4; \Delta_P; z)|^2,
 \fe
 where
 \ie
 &\A_1 =\A_2 = {\ell+2\over 2\sqrt{3}},~~~\A_3 = \A_4= {3-\ell\over 2\sqrt{3}},
 \fe
and $Q  = b+b^{-1} = 1/\sqrt{3} + \sqrt{3}$.

The amplitudes are the same for $\ell=0$ and $\ell=1$ as expected. They are\ie
k=3,~\ell=0,1:~~~\cA_{DSLST} \sim 51.28 - 60.11\, \times{\A'\over2} s_{12}+\mathcal{O} (\A'^2s^2),
\fe
The ratio between the subleading and the leading order terms in the $\A'$ expansion is
\ie
k=3,~\ell=0,1: ~~~{\cA^{(2)}_{DSLST} \over s_{12} \cA^{(1)}_{DSLST}}=-1.172.
\fe

\subsubsection{$k=4$}

 The relevant operators in the supersymmetric $SU(2)_4/U(1)$ coset model for the gauge boson vertex operators  \eqref{gaugevertex-} and  \eqref{gaugevertex+} are
\ie
&V^{su, 1/2,1/2}_{ 0, 0 ,0 } = V^{su,-1/2,-1/2}_{{1}, -{1} , -{1}},~~\Delta =\bar \Delta= {1\over 16} ,~~R= \bar R ={1\over 4},\\
&V^{su, 1/2,1/2}_{ {1\over2}, {1\over2} ,{1\over2} } = V^{su,-1/2,-1/2}_{ {1\over2} , -{1\over2},-{1\over2}},~~\Delta =\bar \Delta= {1\over 16} ,~~R= \bar R =0,\\
&V^{su, 1/2,1/2}_{ {1}, {1} ,{1} } = V^{su,-1/2,-1/2}_{ 0 , 0,0},~~\Delta =\bar \Delta= {1\over 16} ,~~R= \bar R =-{1\over 4}.
\fe
These three operators with dimension 1/16 can be realized as the disorder field $\sigma$ in the Ising model and a free boson $\varphi'$ \cite{Mussardo:1988av}:
\ie
&V^{su, 1/2,1/2}_{ 0, 0 ,0 }(z,\bar z) = V^{su,-1/2,-1/2}_{{1}, -{1} , -{1}}(z,\bar z)
= \exp \left[ i {1\over 2\sqrt{2}} \varphi ' (z)\right] 
\exp \left[- i {1\over 2\sqrt{2}} \bar\varphi '(\bar z) \right],\\
&V^{su, 1/2,1/2}_{ {1\over2}, {1\over2} ,{1\over2} } (z,\bar z)= V^{su,-1/2,-1/2}_{ {1\over2} , -{1\over2},-{1\over2}}(z,\bar z)=\sigma(z,\bar z),\\
&V^{su, 1/2,1/2}_{ {1}, {1} ,{1} }(z,\bar z) = V^{su,-1/2,-1/2}_{ 0 , 0,0}(z,\bar z)= \exp \left[ -i {1\over 2\sqrt{2}} \varphi '(z) \right] 
\exp \left[i {1\over 2\sqrt{2}} \bar\varphi ' (\bar z)\right].
\fe
The four-point functions of the $SU(2)_4/U(1)$ coset model can then be computed straightforwardly \cite{Ginsparg:1988ui} for each value of  $\ell$, which takes three possible values, $0,1,2$:
\ie
\ell=0, 2 \quad &\la V^{su, 1/2,1/2}_{ 0, 0 ,0 } (z,\bar z) V^{su, 1/2,1/2}_{ 0, 0 ,0 } (0,0) V^{su, -1/2,-1/2}_{ 0, 0 ,0 }  (1,1) V^{su, -1/2,-1/2}_{ 0, 0 ,0 } (\infty,\infty) \ra \\
&= \la V^{su, 1/2,1/2}_{ {1}, {1} ,{1} } (0,0) V^{su, 1/2,1/2}_{ {1}, {1} ,{1} } (z,\bar z) V^{su, -1/2,-1/2}_{ {1},-{1} ,-{1} }  (1,1) V^{su, -1/2,-1/2}_{ {1}, -{1} ,-{1} } (\infty,\infty) \ra\\
&= | z| ^{ {1\over 4} }  |1-z|^{ -{1\over 4}}\\
 \ell=1,~~&\la V^{su, 1/2,1/2}_{ {1\over2}, {1\over2} ,{1\over2} } (z,\bar z) V^{su, 1/2,1/2}_{ {1\over2}, {1\over2} ,{1\over2} } (0,0) V^{su, -1/2,-1/2}_{ {1\over2},-{1\over2} ,-{1\over2} }  (1,1) V^{su, -1/2,-1/2}_{ {1\over2}, -{1\over2} ,-{1\over2} } (\infty,\infty) \ra\\
& = {1\over2} |z|^{-{1\over4}} |1-z|^{-{1\over4}} \left( 
\Big|  1+ \sqrt{1-z} \Big| + \Big| 1-\sqrt{1-z} \Big|
\right),
\fe

Combining with the correlators in the $\bR^{1,5}$ and $SL(2)_4/U(1)$ parts, we obtain the following scattering amplitude,
\ie\label{ampxc}
 \cA_{DSLST}&\sim  \int_\bC d^2z\ |z|^{  {(\ell+1)^2 \over 4 }   -{\A' \over 2}s_{12}  -{1\over4}  }
  |1-z|^{\ell  - {( \ell+1)^2\over 4} -{\A' \over 2}s_{13}  +{1\over4} } \\
  &
 \quad \times  \left[  1-\delta_{\ell,1}  +{ \delta_{\ell,1} \over2} |z|^{-{1 \over 2}} \left( 
\Big|  1+ \sqrt{1-z} \Big| + \Big| 1-\sqrt{1-z} \Big|
\right)      \right]
  \\
 & \quad \times \int_{0}^\infty {dP\over 2\pi}C(\A_1,\A_2,{Q\over 2}+iP) C(\A_3,\A_4,{Q\over 2}-iP)
|F(\Delta_1,\Delta_2,\Delta_3,\Delta_4; \Delta_P; z)|^2,
 \fe
 where
 \ie
 &\A_1 =\A_2 = {\ell+2\over 4}, \quad \A_3 = \A_4= {4-\ell\over 4},
 \fe
and $Q  = b+b^{-1} = 5/2$.

The amplitudes are the same for $\ell=0$ and $\ell=2$. They are
\ie
k=4,~\ell=0,2:~~~\cA_{DSLST} \sim 78.96 - 144.6\, \times{\A'\over2} s_{12}+\mathcal{O} (\A'^2s^2).
\fe
The ratio between the subleading and the leading order terms in the $\A'$ expansion is
\ie
k=4,~\ell=0,2: ~~~{\cA^{(2)}_{DSLST} \over s_{12} \cA^{(1)}_{DSLST}}=-1.831.
\fe
As for the $\ell=1$ case, the linear term in the Mandelstam variables is absent in this case due to the symmetry in $s_{12}$, $s_{13}$, and $s_{14}$, so ${\cA^{(2)}_{DSLST}}$ is zero.


%

\subsubsection{$k=5$}
The relevant operators in the supersymmetric $SU(2)_5/U(1)$ coset model for the gauge boson vertex operators  \eqref{gaugevertex-} and  \eqref{gaugevertex+} are
\ie
&V^{su, 1/2,1/2}_{ 0, 0 ,0 }(z,\bar z) = V^{su,-1/2,-1/2}_{{3\over 2}, -{3\over 2} , -{3\over 2}}(z,\bar z),~~\Delta =\bar \Delta= {3\over 40} ,~~R= \bar R ={3\over 10},\\
&V^{su, 1/2,1/2}_{ {1\over2}, {1\over2} ,{1\over2} } (z,\bar z)= V^{su,-1/2,-1/2}_{ {1} , -{1},-{1}}(z,\bar z),~~\Delta =\bar \Delta= {3\over 40} ,~~R= \bar R ={1\over 10}\\
&V^{su, 1/2,1/2}_{ {1}, {1} ,{1} }(z,\bar z) = V^{su,-1/2,-1/2}_{{1\over2}, -{1\over2} ,-{1\over2}}(z,\bar z),~~\Delta =\bar \Delta= {3\over 40} ,~~R= \bar R =-{1\over 10}\\
&V^{su, 1/2,1/2}_{ {3\over 2}, {3\over 2} ,{3\over 2} }(z,\bar z) = V^{su,-1/2,-1/2}_{ 0 , 0,0}(z,\bar z)~~\Delta =\bar \Delta= {3\over 40} ,~~R= \bar R =-{3\over 10}.
\fe
These four operators with dimension 3/40 can be realized as a free boson $\varphi'$ and the complex order parameter $\sigma$ of dimension 1/15 in the three-state Potts model\footnote{The three-state Potts model is reviewed in Appendix~\ref{Three-state Potts Model}.} and also its complex conjugate $\sigma^*$,
\ie
&V^{su, 1/2,1/2}_{ 0, 0 ,0 }(z,\bar z) = V^{su,-1/2,-1/2}_{{3\over 2}, -{3\over 2} , -{3\over 2}}(z,\bar z)
= \exp \left[ i \sqrt{3\over 20} \varphi ' (z)\right] 
\exp \left[- i \sqrt{3\over 20}  \bar\varphi '(\bar z) \right],\\
&V^{su, 1/2,1/2}_{ {1\over2}, {1\over2} ,{1\over2} } (z,\bar z)= V^{su,-1/2,-1/2}_{ {1} , -{1},-{1}}(z,\bar z)=\sigma(z,\bar z)\exp \left[ i {1\over \sqrt{60}} \varphi '(z) \right] 
\exp \left[-i {1\over \sqrt{60}} \bar\varphi ' (\bar z)\right],\\
&V^{su, 1/2,1/2}_{ {1}, {1} ,{1} }(z,\bar z) = V^{su,-1/2,-1/2}_{{1\over2}, -{1\over2} ,-{1\over2}}(z,\bar z)=\sigma^*(z,\bar z) \exp \left[ -i {1\over \sqrt{60}} \varphi '(z) \right] 
\exp \left[i {1\over \sqrt{60}} \bar\varphi ' (\bar z)\right],\\
&V^{su, 1/2,1/2}_{ {3\over 2}, {3\over 2} ,{3\over 2} }(z,\bar z) = V^{su,-1/2,-1/2}_{ 0 , 0,0}(z,\bar z)= \exp \left[ -i \sqrt{3\over 20}  \varphi '(z) \right] 
\exp \left[i\sqrt{3\over 20}  \bar\varphi ' (\bar z)\right].
\fe

The four-point functions of the $SU(2)_5/U(1)$ coset model for each value of $\ell$, which takes three possible values, $0,1,2,3$, are then
\ie
\ell=0, 3 \quad &\la V^{su, 1/2,1/2}_{ 0, 0 ,0 } (z,\bar z) V^{su, 1/2,1/2}_{ 0, 0 ,0 } (0,0) V^{su, -1/2,-1/2}_{ 0, 0 ,0 }  (1,1) V^{su, -1/2,-1/2}_{ 0, 0 ,0 } (\infty,\infty) \ra \\
&= \la V^{su, 1/2,1/2}_{ {3 \over 2}, {3 \over 2} ,{3 \over 2} } (z, \bar z) V^{su, 1/2,1/2}_{  {3 \over 2}, {3 \over 2} ,{3 \over 2} } (0,0) V^{su, -1/2,-1/2}_{  {3 \over 2}, -{3 \over 2} ,-{3 \over 2}  }  (1,1) V^{su, -1/2,-1/2}_{  {3 \over 2}, -{3 \over 2} ,-{3 \over 2}  } (\infty,\infty) \ra\\
 &= | z| ^{ {3\over 10} }  |1-z|^{ -{3 \over 10}}\\
 \ell=1,2 \quad &\la V^{su, 1/2,1/2}_{ {1\over2}, {1\over2} ,{1\over2} } (z,\bar z) V^{su, 1/2,1/2}_{ {1\over2}, {1\over2} ,{1\over2} } (0,0) V^{su, -1/2,-1/2}_{ {1\over2},-{1\over2} ,-{1\over2} }  (1,1) V^{su, -1/2,-1/2}_{ {1\over2}, -{1\over2} ,-{1\over2} } (\infty,\infty) \ra\\
&= \la V^{su, 1/2,1/2}_{ {1}, {1} ,{1} } (0,0) V^{su, 1/2,1/2}_{ {1}, {1} ,{1} } (z,\bar z) V^{su, -1/2,-1/2}_{ {1},-{1} ,-{1} }  (1,1) V^{su, -1/2,-1/2}_{ {1}, -{1} ,-{1} } (\infty,\infty) \ra\\
& =  | z| ^{ {1\over 30} }  |1-z|^{ -{1\over 30}} G^{Potts}_{\sigma\sigma\sigma^*\sigma^*}(z, \bar z).
\fe

For $\ell = 0,3$, the scattering amplitude is
\ie\label{ampxe}
\cA_{DSLST}&\sim  \int_\bC d^2z\ |z|^{ {(\ell+1)^2 \over 5}-{\A' \over 2}s_{12} - {1 \over 5} }
  |1-z|^{ \ell - {(\ell+1)^2 \over 5} -{\A' \over 2}s_{13} +{1 \over 5}} \\
& \times  \int_{0}^\infty {dP\over 2\pi}C(\A_1,\A_2,{Q\over 2}+iP) C(\A_3,\A_4,{Q\over 2}-iP)
|F(\Delta_1,\Delta_2,\Delta_3,\Delta_4; \Delta_P; z)|^2,
 \fe
where
\ie
\A_1 = \A_2 = {1 \over \sqrt 5}, ~~~\A_3 = \A_4 = {5 \over 2 \sqrt 5}, \\
\fe
and $Q = b+b^{-1} = 1/\sqrt5 + \sqrt5$. The amplitude is numerically computed to be
\ie
k=5,~\ell=0,3:~ \cA_{DSLST} \sim 110.4-264.5 \times {\A'\over 2} s_{12} +\mathcal{O}(\A'^2s^2).
\fe
The ratio between the second order and the first order $\A'$ expansion is
\ie
k=5,~\ell=0,3: ~~~{\cA^{(2)}_{DSLST} \over s_{12} \cA^{(1)}_{DSLST}}=-2.397.
\fe

%

For $\ell = 1,2$, the scattering amplitude is
\ie\label{ampxd}
\cA_{DSLST}&\sim \int_\bC d^2z\  |z|^{  {(\ell+1)^2 \over 5 }   -{\A' \over 2}s_{12}  -{7\over15}  }
  |1-z|^{\ell  - {( \ell+1)^2\over 5} -{\A' \over 2}s_{13}  +{7\over15} } \times G_{\sigma\sigma\sigma^*\sigma^*}^{Potts}(z,\bar z)
  \\
 \times &\int_{0}^\infty {dP\over 2\pi}C(\A_1,\A_2,{Q\over 2}+iP) C(\A_3,\A_4,{Q\over 2}-iP)
 | F(\Delta_1, \Delta_2, \Delta_3, \Delta_4; \Delta_P | z) |^2,
\fe
where
\ie
\A_1 = \A_2 = {3 \over 2\sqrt 5}, ~~~\A_3 = \A_4 = {2 \over \sqrt 5}, \\
\fe
and $G_{\sigma\sigma\sigma^*\sigma^*}^{Potts}(z,\bar z)$ is the four-point function in the three-state Potts model reviewed in Appendix \ref{Three-state Potts Model}.  The amplitude is numerically computed to be
\ie
k=5,~\ell=1,2:~ \cA_{DSLST} \sim 220.7 - 304.6 \times {\A'\over 2} s_{12} +\mathcal{O}(\A'^2s^2).
\fe
The ratio between the subleading and the leading order terms in the $\A'$ expansion is
\ie
k=5,~\ell=1,2: ~~~{\cA^{(2)}_{DSLST} \over s_{12} \cA^{(1)}_{DSLST}}=-1.380.
\fe

We summarize the DSLST amplitudes to second order in the $\A'$ expansion in Table~\ref{tab:lstresults}.  The overall normalization of the amplitudes (which depends on $k$ and $\ell$) are fixed here, so only the relative coefficients at different orders in the $\A'$ expansion, or equivalently, the expansion in the Mandelstam variables, of a particular amplitude are computed unambiguously. These exactly match with the 1-loop and 2-loop 6D $SU(k)$ SYM amplitudes listed in Table~\ref{tab:2loop}.
It is straightforward to carry out the $\A'$-expansion of the DSLST amplitudes to higher order, and they should be compared to higher than 2-loop amplitudes on the SYM side. We will comment on this later.

\begin{table}[h!]
\centering
\begin{tabular}{c|c|c|c}
$k$ & $\ell$ & ${\cal A}_{DSLST}$ & ${\cal A}_{DSLST}^{(2)} / s_{12}{\cal A}_{DSLST}^{(1)}$ \\\hline\hline
2 & 0 & $27.92 +  0\times s_{12}$ & 0\\\hline
3 & 0, 1 & $51.28 - 60.11 s_{12}$ & $-1.172$ \\\hline
4 & 0, 2 & $78.96-144.6 s_{12}$ & $-1.831$ \\ 
& 1 & 157.9 + $0\times s_{12}$& 0 \\\hline
5 & 0, 3 & $110.4-264.5s_{12}$ & $-2.397$ \\ 
& 1, 2 & $220.7-304.6s_{12}$ & $-1.380$ \\ 
\end{tabular}
\caption{DSLST amplitudes for different $k$ (number of NS5-branes) and $\ell+1$ ($\bZ_k$ momentum) to subleading order in the $\alpha'$ expansion (up to a $k$- and $\ell$-dependent overall normalization factor).  The last column records the ratios between the subleading and leading order terms in the $\A'$ expansion, i.e. the coefficient of $s_{12}$ divided by the  $s_{12}$-independent term in the full amplitudes. These ratios exactly match with the ratios of the 6D $SU(k)$ SYM amplitudes listed in the last column of Table~\ref{tab:2loop}. Here we set $\A'=2$.}
\label{tab:lstresults}
\end{table}

\section{Comparison to 6D super-Yang-Mills amplitudes}

The strong coupling limit of DSLST, namely LST, reduces to 6D $(1,1)$ $SU(k)$ SYM in the low energy limit.
It is conceivable that the full dynamics of the massless degrees of freedom at the origin of the Coulomb branch is described by a Wilsonian effective action, which is that of $SU(k)$ SYM deformed by an infinite set of higher dimensional operators, with a floating cutoff $\Lambda$.\footnote{A fully supersymmetric regulator  is assumed.} It is reasonable to assume that the coefficient of the higher dimensional operators involve non-negative powers of $g_{YM}$ (and arbitrary functions of $\Lambda$).

The DSLST corresponds to a point away from the origin on the Coulomb branch of this theory, at which a $\mathbb{Z}_k\times U(1)$ subgroup of the $SO(4)$ R-symmetry is preserved. In the gauge theory description, scalar vevs of order $\langle\phi\rangle\sim m_W/g_{YM}$ are turned on. 
A perturbative scattering amplitude of gluons in the Cartan $U(1)^{k-1}$ at energy $E$ schematically takes the form
\ie
{\cal A}(E) = \sum_{L=0}^\infty (g_{YM} E)^{2L+2} {\cal A}_L({m_W/E}).
\fe
A priori, ${\cal A}_L$ is not quite the same as the $L$-loop amplitude. As we expand the scalars around their vevs, we obtain couplings that involve positive powers of $g_{YM}$ and non-negative powers of $\langle\phi\rangle$. It seems reasonable to assume that the Wilsonian effective Lagrangian on the Coulomb branch at fixed $m_W$ is non-singular at $g_{YM}=0$, thus there should always be more powers of $g_{YM}$ than $\langle\phi\rangle$ in the coefficients of the operators in the Lagrangian.
With such vertices, ${\cal A}_L$ generally receives contributions from diagrams of {\it no more than} $L$ loops. By construction, the $\Lambda$ dependence drops out of ${\cal A}(E)$.

The perturbative amplitude in DSLST, on the other hand, has the structure (after identifying $g_s$ and $\A'$ with gauge theory parameters, as explained in the introduction)
\ie
{\cal A}(E) = \sum_{h=0}^\infty (g_{YM} m_W)^{-2h-2} {\cal A}^{lst}_h \quad \left(\A'E^2={g_{YM}^2E^2\over 16\pi^3}\right)
\fe
where $h$ labels the genus (which is entirely unrelated to the loop order in the SYM theory).

While ${\cal A}_L(m_W/g_{YM})$ is naturally defined by an analytic expansion in $m_W/g_{YM}$ (as the theory is free of infrared divergences), the duality with DSLST suggests that ${\cal A}_L(m_W/g_{YM})$ also has an analytic expansion in $1/m_W^2$ at large $m_W$, namely
\ie
{\cal A}_L(m_W/E) = \sum_{n=1}^\infty \left( {E\over m_W} \right)^{2n} {\cal A}_L^{(n)},
\fe
and in particular, we expect the tree level DSLST amplitude to agree with the $1/m_W^2$ part of the gauge theory amplitude,
\ie
{\cal A}_0^{lst}\left(\A'E^2={g_{YM}^2E^2\over 16\pi^3}\right) = \sum_{L=0}^\infty (g_{YM} E)^{2L+4} {\cal A}_L^{(1)} .
\fe

If we take the cutoff $\Lambda$ to infinity, the $L$-loop amplitude in the undeformed 6D SYM generally diverges for $L\geq 3$. However, as already argued (and verified explicitly in Appendix \ref{3ldiv}), the 3-loop divergence is absent when the external legs are restricted to the Cartan subalgebra. It turns out that the first UV divergence of the Cartan gluon four-point amplitude arises at four-loop.\footnote{We thank the authors of \cite{Bern:2012uf} for pointing this out. The relevant four-loop divergence can be extracted from \cite{Bern:2012uf}.} In any case, one can ask whether the perturbative $L$-loop amplitude in SYM captures (a part of) ${\cal A}_L(m_W/E)$. At $L=1$ the answer is known to be yes, and in fact the finite 1-loop amplitude in SYM, when expanded to first order in $1/m_W^2$, precisely agrees with the leading low energy term in the tree level DSLST amplitude \cite{Aharony:2003vk}. Since the 6D SYM by itself is UV finite at 2-loop as well, one might suspect that ${\cal A}_2(m_W/E)$ is also given entirely by the 2-loop 6D SYM amplitude (with $\Lambda\to \infty$ and no contribution from higher dimensional operators). Remarkably, we will find that this is indeed the case.



\subsection{Structure of perturbative amplitudes}

We will consider four-point amplitudes in $SU(k)$ maximally supersymmetric Yang-Mills theory obtained from unitarity cut methods \cite{Dixon:1996wi,Dixon:2010gz,Bern:1996je,Brandhuber:2010mm,Bern:1997nh,Bern:2010qa}. While such amplitudes are mostly studied in four-dimensional gauge theories, where the $L$-loop result can be expressed in terms of the tree level amplitude together with scalar loop integrals, such formulae admit straightforward generalizations to higher dimensions. It is known that up to 3-loop order in any spacetime dimension $D$, and at 4-loop for $D\leq 6$, the formula derived in four dimensions can be extended to $D$ dimensions by simply replacing the relevant scalar loop integrals by the $D$-dimensional loop integrals~\cite{Dixon:1996wi,Dixon:2010gz}.

We will express the amplitudes in 6D SYM in terms of 6-dimensional spinor helicity variables $\lambda_i^{Aa}$ and $\tilde \lambda_{iB\dot b}$ \cite{Cheung:2009dc,Dennen:2009vk,Bern:2010qa}, where $i=1,2,3,4$ label the external particle legs, $a,\dot b$ are $SU(2)\times SU(2)$ little group indices, and $A,B$ are spinor indices of $SO(6)$ or $SO(5,1)$ Lorentz group. The amplitudes involving various particles in a supermultiplet will be contracted with Grassmann polarization variables $\eta_{ia}$ and $\tilde \eta_i^{\dot b}$ to form a superamplitude.
It is convenient to express the latter in terms of the supermomenta
\ie
q_i^A = \lambda_i^{Aa}\eta_{ia},~~~~ \tilde q_{iB} = \tilde\lambda_{iB\dot b} \tilde\eta_i^{\dot b}.
\fe
The {\it color-ordered} four-point tree-level superamplitude can be written as
\ie
{\cal A}^{tree}(1,2,3,4) = -{i\over s_{12}s_{23}} \delta^8(\sum_{i=1}^4 {\bf q}_i).
\fe
Here $s_{ij}\equiv -(p_i+p_j)^2$ are the Mandelstam variables, and the color factor has been stripped off.
The delta function is the one in Grassmann variables. Explicitly, it can be expanded as
\ie
&\delta^8(\sum_{i=1}^4 {\bf q}_i) = \delta^4(\sum_{i=1}^4 { q}_i^A)\delta^4(\sum_{i=1}^4 {\tilde q}_{iB})
\\
&= {1\over (4!)^2} \sum_{i,j,k,\ell,m,n,r,s=1}^4 \epsilon_{ABCD} q_i^A q_j^B q_k^C q_\ell^D \epsilon^{EFGH} \tilde q_{mE} \tilde q_{nF} \tilde q_{rG} \tilde q_{sH}.
\fe
Suppose we are interested in the 4-scalar amplitude. Then each set of $(\eta_i,\tilde \eta_i)$ should appear in the superamplitude in the combination $1,\, \eta_i^2( = \epsilon^{ab}\eta_{ia}\eta_{ib}),\,\tilde\eta_i^2, \,\eta_i^2\tilde\eta_i^2$. With respect to the $SU(2)\times SU(2)$ R-symmetry (not to be confused with the little group), or rather, $U(1)\times U(1)$ Cartan generators of the R-symmetry group, these scalars have charges $(-,-)$, $(+,-)$, $(-,+)$, $(+,+)$. In the 5-brane description, the two rotations in the transverse $\mathbb{R}^2\times \mathbb{R}^2$ of the 5-brane are linear combinations of these two Cartan generators. So the scalar labeled by $1,\eta^2\tilde\eta^2$ are the collective coordinates in one $\mathbb{R}^2$ (denote by $\phi^{1,2}$), while $\eta^2$ and $\tilde\eta^2$ give scalars parameterizing the other $\mathbb{R}^2$ (denote by $\phi^{3,4}$). For instance, there are terms in the tree superamplitude proportional to $\eta_1^2 \eta_2^2 \tilde\eta_1^2\tilde\eta_2^2$, or $\eta_1^2 \eta_2^2 \tilde\eta_3^2\tilde\eta_4^2$. The former describes a four-point scattering amplitude of the scalars $\phi^1$ and $\phi^2$, while the latter describes a four-point amplitude of $\phi^3$ and $\phi^4$.

In this paper we are interested in the scattering amplitudes of the massless gluons in the Cartan $U(1)^{k-1}$ on the Coulomb branch of $SU(k)$ gauge theory, where the remaining $k^2-k$ $W$-bosons are massive. In SYM there is no tree level amplitude of the Cartan gluons due to the vanishing color factor. The nonvanishing loop amplitudes of the Cartan gluons contain $W$-bosons in the loops (as well as possibly massless gluon propagators at two loops and higher).

\begin{figure}[htb]
\begin{center}
\includegraphics[scale=1.1]{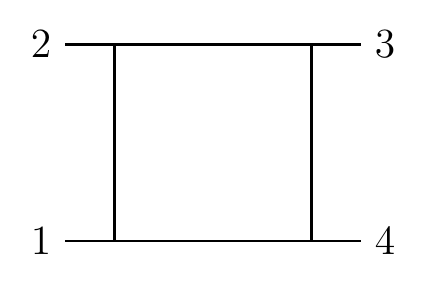}
\end{center}
\caption{The 1-loop scalar integral $I_4^{1-loop}(s_{12},s_{14})$.}\label{fig:1loop}
\end{figure}

The {\it full} 1-loop amplitude in SYM is given by\footnote{While ${\cal A}^{tree}(1,2,3,4)$ refers to the color-ordered partial amplitude (the full amplitude is obtained by summing over $s_{12}, s_{13}, s_{14}$ channels), ${\cal A}^{1-loop}(1,2,3,4)$ and ${\cal A}^{2-loop}(1,2,3,4)$ are full amplitudes.} \cite{Bern:1996je} (see also \cite{Brandhuber:2010mm})
\ie
{\cal A}^{1-loop}(1,2,3,4) = - s_{12} s_{23} {\cal A}^{tree}(1,2,3,4) \left[ C_{1234} I_4^{1-loop}(s_{12},s_{23}) + (2\leftrightarrow 3) + (3\leftrightarrow 4) \right],
\fe
where $I_4^{1-loop}(s_{12},s_{14})$ in Figure \ref{fig:1loop} is the 1-loop massless scalar box integral. $C_{1234}$ is the color factor associated with the box diagram.  This relation holds in any $D$, which we now take to be $D=6$, and is easily generalized to the case of a $W$-boson loop, where we simply need to replace $I_4^{1-loop}(s_{12},s_{14})$ by\footnote{The generalization to massive propagators in the loop is justified by consideration of unitarity cuts.}
\ie
I_4^{1-loop}&(s_{12},s_{14})\\
&= \int {d^6\ell\over (2\pi)^6} {1\over (\ell^2+m_W^2)((\ell+p_1)^2+m_W^2)((\ell+p_1+p_2)^2+m_W^2)((\ell-p_4)^2+m_W^2)}
\\
&= {1\over m_W^2} \int {d^6\ell\over (2\pi)^6} {1\over (\ell^2+1)^4} + {\cal O}({1\over m_W^4}) .
\fe
In the last line we have expanded the result in $1/m_W^2$. As explained, it is the order $1/m_W^2$ result of the SYM amplitude that will be compared with the {\it genus zero} amplitude of double scaled little string theory.

In the end, the 1-loop amplitude of Cartan gluons on the Coulomb branch can be written in the form 
\ie
{\cal A}^{1-loop}(1,2,3,4) = -i \delta^8(\sum_{i=1}^4 {\bf q}_i) \left[ {C_{1234}+C_{1324}+C_{1243}\over m_W^2} {1\over 384\pi^3} + {\cal O}(m_W^{-4}) \right],
\fe
summed over the species of $W$-bosons if $k > 2$. 

The {\it full} 2-loop amplitude is given by the tree-level amplitude multiplied by 2-loop scalar integrals \cite{Bern:1997nh}
\ie
\label{2loopamp}
&{\cal A}^{2-loop}(1,2,3,4) = -s_{12} s_{23} {\cal A}^{tree}(1,2,3,4) \\
&~~~~~~\times\bigg[ s_{12} ( {\cal A}_{1234}^{2-loop,P} + {\cal A}_{1234}^{2-loop,NP} + {\cal A}_{3421}^{2-loop,P} + {\cal A}_{3421}^{2-loop,NP})
+({\rm cyclic~in~2,3,4})\bigg],
\fe
Here ${\cal A}^{2-loop,P}_{abcd}$ and ${\cal A}^{2-loop,NP}_{abcd}$ are the color-weighted 2-loop scalar integrals given in Figure~\ref{fig:2loop}.  Once again, the propagators in the loops will be replaced by the appropriate massive $W$-boson or massless gluon propagators in the amplitudes on the Coulomb branch of the theory.

\begin{figure}[htb]
\begin{center}
\includegraphics[scale=0.9]{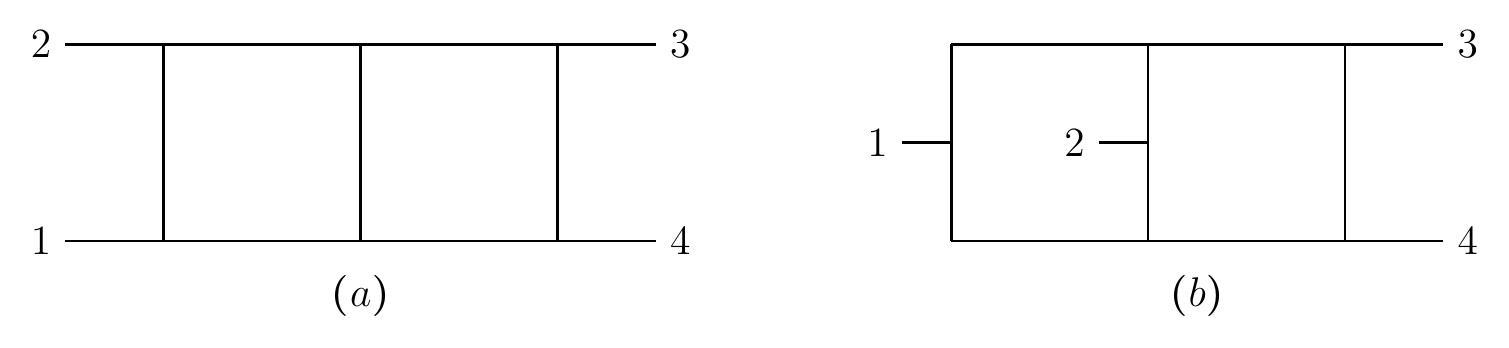}
\end{center}
\caption{In (a), the planar 2-loop scalar integral. In (b), the non-planar 2-loop scalar integral.}\label{fig:2loop}
\end{figure}

The 3- and higher-loop amplitudes generally contain logarithmic divergences. It is likely that they still contain nontrivial information that captures the DSLST amplitudes expanded to the corresponding order in $\A'$, but this is beyond the scope of the current paper.

\subsection{Evaluation of color factors and box integrals}

We will label the $W$-bosons in the loops by a pair of gauge indices $(ij)$ ($i,j=1,\cdots,k$, and $i\not=j$). There can also be massless Cartan gluons in the loops, labeled by $(ii)$ (we will not need to impose the traceless condition by hand in this case, as the overall $U(1)$ decouples due to the interaction vertices). The external massless gluons will be labeled by vectors $\vec v_1,\cdots,\vec v_4$ in the Cartan subalgebra of $su(k)$. The mass of the $(ij)$-$W$-boson is
\ie
m_{ij} = r_0 |\omega^i -\omega^j| = 2r_0 \left| \sin {\pi (i-j)\over k} \right|.
\fe
Here $\omega=e^{2\pi i/k}$ is the primitive $k$-th root of unity, $r_0$ is a (radial) Coulomb branch parameter that will be related to the inverse string coupling of DSLST.

Expanding around the point in Coulomb branch with $\mathbb{Z}_k$ symmetry, corresponding to the NS5-branes spreading out on the circle in a transverse $\mathbb{R}^2$, it is convenient to take $\vec v_a$ to be $\mathbb{Z}_k$ charge eigenstates,
\ie
v_a^j = \omega^{(j-1) n_a},~~~j=1,\cdots,k,
\fe
where 
the $\mathbb{Z}_k$ momentum $n_a$ is an integer ranging from 1 to $k-1$. We also need $\mathbb{Z}_k$ charge conservation,
\ie
\sum_{a=1}^4 n_a \equiv 0 {\rm~mod~} k.
\fe

As discussed before, the gluon vertex operator $\cV^\pm_{R,\ell}$ in DSLST has $\mathbb{Z}_k$ momentum $\pm (\ell+1)$. Therefore, we see that in order to compare with the DSLST scattering amplitude $\la \cV^+_{R,\ell} (z_1,\bar z_1)\cV^+_{R,\ell} (z_2,\bar z_2)\cV^-_{R,\ell} (z_3,\bar z_3)\cV^-_{R,\ell} (z_4,\bar z_4)\ra$ computed in the last section, the $\mathbb{Z}_k$ charges $n_a$ of interest are\footnote{We shift $n_3$ and $n_4$ by $k$ for later convenience.}
\ie
n_1=n_2 = \ell+1,~~n_3=n_4=k -(\ell+1),
\fe
for $\ell=0,1,\cdots,k-2$.

The 1-loop amplitude, expanded to order $1/m_W^2$, is of the form
\ie
{\cal A}^{1-loop} = 
\sum_{i\not=j} {\prod_{a=1}^4 (v_a^i-v_a^j)\over m_{ij}^2} {\cal A}(s_{12},s_{14}) + {\cal O}(m_W^{-4}),
\fe
where
\ie
{\cal A}(s_{12}, s_{14}) = 
- { s_{12} s_{13} {\cal A}^{tree}(s_{12},s_{14}) \over 128 \pi^3}.
\fe
Plugging in the explicit expression for $v_a^i$, we can further write
\ie
{\cal A}^{1-loop} = {4k\over r_0^2} {\cal A}(s_{12},s_{14}) \sum_{\ell=1}^{k-1} {\prod_{a=1}^4 \sin{\pi n_a \ell\over k}\over \sin^2({\pi \ell\over k})} + {\cal O}(r_0^{-4})
\fe
As was shown in \cite{Aharony:2003vk}, the sum collapses into a curiously simple answer,
\ie
{\cal A}^{1-loop} = {2k^2\over r_0^2} {\cal A}(s_{12},s_{14}) {\rm min}\{n_a, k-n_a\} + {\cal O}(r_0^{-4}).
\fe


Now consider the 2-loop amplitude.  In the planar case, let us label the $W$-boson running through vertices 1,2 by $(ij)$, the $W$-boson running through 3,4 by $(\ell m)$, and the $W$-boson in the middle line $(nr)$.  Then ${\cal A}_{1234}^{2-loop,P}$ of (\ref{2loopamp}) is given by
\ie
& \sum_{i,j,\ell,m,n,r} I_4^{2-loop, P}(m_{ij},m_{\ell m}, m_{nr}) \big(  \delta_{jn}\delta_{r\ell}\delta_{m i} -  \delta_{j\ell}\delta_{mn}\delta_{r i} \big)^2 \prod_{a=1,2} (v_a^i-v_a^j) \prod_{a=3,4} (v_a^\ell-v_a^m) \\
&= 2 \sum_{i\not=j, i\not=\ell} I_4^{2-loop, P} (m_{ij},m_{\ell i}, m_{j\ell }) \prod_{a=1,2} (v_a^i-v_a^j) \prod_{a=3,4} (v_a^\ell-v_a^i),
\fe
where the scalar loop integral is
\ie
& I_4^{2-loop, P} = \int {d^6\ell_1\over (2\pi)^6} {d^6\ell_2\over (2\pi)^6} {1\over (\ell_1^2+m_{ij}^2)
((\ell_1+p_2)^2+m_{ij}^2)((\ell_1+p_1+p_2)^2+m_{ij}^2)}
\\
&~~~~ \times {1\over (\ell_2^2+m_{\ell i}^2)((\ell_2+p_4)^2+m_{\ell i}^2)((\ell_2-p_1-p_2)^2+m_{\ell i}^2) ((\ell_1+\ell_2)^2 + m_{j\ell}^2)} 
\\
& = {1\over 4r_0^2}  \int {d^6\ell_1\over (2\pi)^6}{d^6\ell_2\over (2\pi)^6} {1\over (\ell_1^2+\sin^2{\pi(i-j)\over k})^3(\ell_2^2+\sin^2{\pi(i-\ell)\over k})^3 ((\ell_1+\ell_2)^2+\sin^2{\pi(j-\ell)\over k})} + {\cal O}(r_0^{-4})
\fe
This gives the color-weighted planar amplitude
\ie
& {\cal A}_{1234}^{2-loop,P} \\
&
= {8 \over r_0^2} \sum_{i\not=j, i\not=\ell} \int {d^6\ell_1\over (2\pi)^6}{d^6\ell_2\over (2\pi)^6} {\prod_{a=1,2}e^{{\pi i n_a\over k}(j-\ell)} \sin({\pi n_a(i-j)\over k}) \prod_{a=3,4}\sin({\pi n_a(i-\ell)\over k}) \over (\ell_1^2+\sin^2{\pi(i-j)\over k})^3(\ell_2^2+\sin^2{\pi(i-\ell)\over k})^3 ((\ell_1+\ell_2)^2+\sin^2{\pi(j-\ell)\over k})} + {\cal O}(r_0^{-4}) \\
&
= {8k \over r_0^2} \sum_{m,r=0}^{k-1} \int {d^6\ell_1\over (2\pi)^6}{d^6\ell_2\over (2\pi)^6} {\prod_{a=1,2}e^{{\pi i n_a\over k}(r-m)} \sin({\pi n_a m \over k}) \prod_{a=3,4}\sin({\pi n_a r\over k}) \over (\ell_1^2+\sin^2{\pi m\over k})^3(\ell_2^2+\sin^2{\pi r\over k})^3 ((\ell_1+\ell_2)^2+\sin^2{\pi(m-r)\over k})} + {\cal O}(r_0^{-4}).
\fe
Similarly,
\ie
& {\cal A}_{1234}^{2-loop,NP} \\
&
=-{8 \over r_0^2} \sum_{i\not=j\not=\ell} \int {d^6\ell_1\over (2\pi)^6}{d^6\ell_2\over (2\pi)^6} {e^{{\pi i n_1\over k}(j-\ell) + {\pi i n_2\over k}(j-i)} \sin({\pi n_1(i-j)\over k}) \sin({\pi n_2(j-\ell)\over k}) \prod_{a=3,4}\sin({\pi n_a(i-\ell)\over k}) \over (\ell_1^2+\sin^2{\pi(i-j)\over k})^2 (\ell_2^2+\sin^2{\pi(i-\ell)\over k})^3 ((\ell_1+\ell_2)^2+\sin^2{\pi(j-\ell)\over k})^2} + {\cal O}(r_0^{-4}) \\
&
= -{8k \over r_0^2} \sum_{0\leq m\not=-r\leq k-1} \int {d^6\ell_1\over (2\pi)^6}{d^6\ell_2\over (2\pi)^6} {e^{{\pi i n_1\over k}r - {\pi i n_2\over k}m} \sin({\pi n_1 m\over k}) \sin({\pi n_2 r\over k})\prod_{a=3,4}\sin({\pi n_a (m+r)\over k}) \over (\ell_1^2+\sin^2{\pi m\over k})^2 (\ell_2^2+\sin^2{\pi r\over k})^2 ((\ell_1+\ell_2)^2+\sin^2{\pi (m+r)\over k})^3 } + {\cal O}(r_0^{-4}).
\fe

These convergent integrals and sums over color factors can be computed numerically.  The results for the planar and non-planar contributions to the two-loop amplitude are given in Table~\ref{tab:PvsNP}, and the full two-loop amplitudes, whose expression is given by (\ref{2loopamp}), are listed in Table~\ref{tab:2loop}.  We see that the ratios listed in the last column remarkably match with the ratios computed from DSLST that are listed in Table~\ref{tab:lstresults}, to the numerical precision of the conformal block integration.  (The matching of ${\cal A}^{1-loop}$ with DSLST amplitudes, where the overall normalization is needed, was already demonstrated in \cite{Aharony:2003vk}.)

\begin{table}[h!]
\centering
\begin{tabular}{c|c|c|c}
$k$ & $\ell$ & $ { g_{YM}^2} {\cal A}^{2-loop}_P / s_{12}$ & ${ g_{YM}^2} {\cal A}^{2-loop}_{NP} / s_{12}$ \\\hline\hline
2 & 0 & 0 & 0  \\\hline
3 & 0, 1 & $-6.048$ & $-4.500$ \\\hline 
4 & 0, 2 & $-16.876$ &  $-12.435$  \\ 
& 1 & 0 & 0  \\\hline
5 & 0, 3 & $-34.594$ & $-25.327$  \\ 
& 1, 2 & $-39.883$ &  $-29.136$ \\ 
\end{tabular}
\caption{The planar and non-planar contributions to the 2-loop amplitudes in leading orders of $1/m_W^2$ expansion for four-gluon scattering in 6D $SU(k)$ SYM. Here we choose the $\mathbb{Z}_k$ charges for the external gluons to be $ n_1 = n_2 =\ell+1$ and $n_3 = n_4=k-(\ell+1)$ with $\ell=0,1,\cdots,k-2$.   The numbers are in units of $-s_{12} s_{23} {\cal A}^{tree} / 64 \pi^3 r_0^2$.  In order to compare with the DSLST results in Table~\ref{tab:lstresults}, here we set $g_{YM}^2 = 32 \pi^3 $ (see \eqref{idaa} with $\alpha'=2$).}
\label{tab:PvsNP}
\end{table}

\begin{table}[h!]
\centering
\begin{tabular}{c|c|c|c|c}
$k$ & $\ell$ & ${\cal A}^{1-loop}$ & ${ g_{YM}^2 } {\cal A}^{2-loop} / s_{12}$ & ${ g_{YM}^2 } {\cal A}^{2-loop} / s_{12} {\cal A}^{1-loop}$ \\\hline\hline
2 & 0 & 4 & 0 & 0 \\\hline
3 & 0, 1 & 9 & $-10.548$ & $-1.1720$  \\\hline 
4 & 0, 2 & 16 & $-29.311$ &  $-1.8319$ \\ 
& 1 & 32 & 0 & 0 \\\hline
5 & 0, 3 & 25 & $-59.922$ & $-2.3969$ \\ 
& 1, 2 & 50 & $-69.019$ &  $-1.3804$ \\ 
\end{tabular}
\caption{The 1 and 2-loop amplitudes in leading orders of $1/m_W^2$ expansion and their ratios for four-gluon scattering in 6D $SU(k)$ SYM. Here we choose the $\mathbb{Z}_k$ charges for the external gluons to be $ n_1 = n_2 =\ell+1$ and $n_3 = n_4=k-(\ell+1)$ with $\ell=0,1,\cdots,k-2$.   ${\cal A}^{1-loop}$ and ${\cal A}^{2-loop}$ are both in units of $-s_{12} s_{23} {\cal A}^{tree} / 64 \pi^3 r_0^2$.  In order to compare with the DSLST results in Table~\ref{tab:lstresults}, here we set $g_{YM}^2 = 32 \pi^3 $ (see \eqref{idaa} with $\alpha'=2$). The loop amplitudes ratios exactly match the  ratios between different  $\A'$ expansion orders of the DSLST amplitudes in  Table~\ref{tab:lstresults}. }
\label{tab:2loop}
\end{table}

%

\section{Discussion}

The tree DSLST amplitudes provides all order results $g_{YM}^2$ and first order in $1/m_W^2$ of the UV completed 6D gauge theory on its Coulomb branch. While the agreement of 6D SYM amplitudes at 2-loop with DSLST at next to leading order in $\A'$ is already striking, a burning question is whether the SYM 3-loop amplitude, which as discussed is finite when the external lines are restricted to the Cartan subalgebra, agrees with the DSLST at next-to-next-to leading order in $\A'$. As the relevant 3-loop superamplitudes have already been reduced to scalar integrals, it is merely a matter of evaluating these scalar integrals to answer the question. We hope to report on the result in the near future.

One may also try to carry out the DSLST amplitude computation to higher genus, and compare with the higher order terms in the $1/m_W^2$ expansion of the SYM amplitude at each loop order. This is not easy as the relevant genus one four-point function in the cigar coset CFT is not yet known, but would nonetheless be interesting.

From the point of view of the Abelian effective action on the Coulomb branch, the 2-loop amplitude of order $1/m_W^2$ comes from the $1/4$ BPS dimension 10 operator of the form $m_W^{-2}D^2F^4+\cdots$. Presumably, our finding suggests a non-renormalization theorem of this term in the Coulomb effective action, with respect to higher dimensional non-BPS operator corrections to the non-Abelian SYM theory. If so, then the 3-loop test will be particularly important, and an agreement with the DSLST tree amplitude at the next order in $\A'$ would be more surprising.\footnote{We thank Ofer Aharony for pointing this out.}

In any case, the big question here is, to what extent will the agreement between the massless amplitudes of pure 6D SYM on the Coulomb branch and DSLST hold, and why do they agree? It so happens that the Cartan gluon amplitude becomes divergent at four-loop \cite{Bern:2012uf}. Therefore we will definitely see some nontrivial disagreement with the LST amplitude at ${\mathcal O}(\A'^3)$. At five-loop, the scalar integrand for the four-point amplitude is known but the UV divergence with external legs in the Cartan subalgebra has yet to be extracted \cite{Bern:2012uc}.\footnote{
We thank the authors of \cite{Bern:2012uf} for explaining to us the results of \cite{Bern:2012uf,Bern:2012uc,Bern:2010tq}.
}  
A priori, there could be all sorts of higher dimensional operators that enter the Wilsonian effective action of the 6D gauge theory and correct the amplitudes of the SYM theory itself. After all, we do expect the presence of the dimension 10 non-BPS operator (see for instance \cite{Movshev:2009ba}) as the counter term that cancels the general 3-loop divergence, even though this operator vanishes when the fields are restricted to the Cartan subalgebra. A systematic investigation of the higher dimensional counter terms and their effect on the Cartan gluon scattering amplitude is left to future work.

Finally, let us mention that the W-bosons in the 6D SYM are dual to D1-branes stretched between the NS5-branes. The scatterings of strings with the D1-branes correspond to the scatterings of the Cartan gluons with the W-bosons, and also the scatterings of the D1-branes with themselves are dual to the scatterings of W-bosons. Some aspects of open strings and D-branes in DSLST are studied in \cite{Israel:2005fn} (also see \cite{Israel:2004jt} for the D-branes in $\cN=2$ Liouville theory). It would be interesting to extend their analysis, for example, to the closed string two-point amplitudes on a disc ending on stretched D1-branes, and compare with the scattering amplitudes of two Cartan gluons and two W-bosons. 


\bigskip

\section*{Acknowledgments}

We would like to thank Ofer Aharony, Clay C\'ordova, Lance Dixon, Thomas Dumitrescu, Yu-tin Huang, Daniel Jafferis, Ingo Kirsch, Soo-Jong Rey, David Simmons-Duffin, and Andy Strominger for helpful conversations and correspondences at various stages of this project. We would like to thank the Kavli Institute for Theoretical Physics and Aspen Center for Physics during the course of this work. 
C.M.C. has been supported in part by a KITP Graduate Fellowship. C.M.C. would like to thank the Physics Department of National Taiwan University for hospitality during the final stage of the work. 
S.H.S. is supported by the Kao Fellowship and the An Wang Fellowship at Harvard University. 
X.Y. is supported by a Sloan Fellowship and a Simons Investigator Award from the Simons Foundation. This work is also supported by NSF Award PHY-0847457, and by the Fundamental Laws Initiative 
Fund at Harvard University.


\appendix

\section{Normalizable vertex operators}

\subsection{NS-sector}
\label{n1}
We consider $F_L=0$ and concentrate on the internal CFT. The on-shell condition we want to solve is
\ie
2(j'-j)(j'+j+1)= k(1-\eta^2-{\eta'}  ^2).
\fe
Meanwhile the GSO condition demands that
\ie
\eta^2+\eta'^2, \quad |\eta-\eta'|\in 2{\mathbb Z}+1,
\fe
and therefore,
\ie
j'-j\in\mathbb Z_{\leq 0}.
\fe
For normalizable vertex operators we have
\ie
|m|>j,~j'\geq|m'|,
\fe
and it follows that
\ie
 j-j' <|m|-|m'|\leq |m-m'|=|\eta-{\eta'} |.
\fe
On the other hand, we have the restriction $0\leq j,~j'\leq{k/2-1}$, or $j+j'\leq k-2<k$, which implies that
\ie
2|j'-j|>|\eta^2+{\eta'}  ^2-1|.
\fe
Assuming $j>j'$, we can combine the two equalities above to get
\ie
|\eta^2+{\eta'}  ^2-1|<2(j-j')<2|\eta-\eta'|.
\fe
This further implies,
\ie
 2|\eta-\eta'|\geq|\eta^2+{\eta'}  ^2-1|+4\geq \eta^2+\eta'^2+3 \geq {1\over 2} |\eta-\eta'|^2+3   
\fe
which is equivalent to
\ie
4|\eta-\eta'|\geq |\eta-\eta'|^2+6.
\fe
This is impossible. 
%

Therefore the only normalizable solutions that survive the GSO projection satisfy $j=j'$ with $\eta^2+{\eta'}  ^2=1$.

\subsection{R-sector}
\label{n2}
The on-shell condition in the R sector is
\ie
4(j'-j)(j'+j+1)= k(1-2\eta^2-2{\eta'}  ^2),
\fe
and the GSO condition becomes
\ie
\eta+\eta'\in 2{\mathbb Z}.
\fe
Since we are looking at half-integer spectral flows, this implies that
\ie
\eta-\eta'\in 2{\mathbb Z}+1,
\fe
and therefore,
\ie
 2\eta^2+2\eta'^2,|\eta-\eta'|^2\in 4{\mathbb Z}+1.
\fe
Then the mass-shell condition requires that
\ie
j'-j\in \mathbb Z_{\leq 0}.
\fe
Assuming $j>j'$, we have the following inequality from the normalizability condition
\ie
|2\eta^2+2{\eta'}  ^2-1|<4(j-j')<4|\eta-\eta'|,
\fe
which demands
\ie
4|\eta-\eta'| \geq |2\eta^2+2{\eta'}  ^2-1|+8 \geq |\eta-\eta'| ^2+7.
\fe
This is impossible.

Hence the only possibility is $j=j', \eta^2+\eta'^2={1\over 2}$, which are solved by
\ie
\eta=\pm{1\over 2}, \quad \eta'=\mp{1\over 2 }.
\fe

\section{Three-state Potts model}\label{Three-state Potts Model}

The three-state Potts Model is the non-diagonal modular invariant of the $c={4\over 5}$ minimal model. The scalar primary operators in this theory are
\ie
1,~~\epsilon,~~X,~~Y,~~\sigma,~~\sigma^*,~~Z,~~Z^*.
\fe
Their dimensions are
\ie
\Delta_{\epsilon}={2\over 5},~~\Delta_X={7\over 5},~~\Delta_Y=3,~~\Delta_{\sigma}=\Delta_{\sigma^*}={1\over 15},~~\Delta_Z=\Delta_{Z^*}={2\over 3}.
\fe
In addition to the scalar primaries, there are spin 1 primaries
\ie
&\Phi_{{2\over 5},{7\over 5}},~~~\Delta=\bar\Delta-1={2\over 5},
\\
&\Phi_{{7\over 5},{2\over 5}},~~~\Delta=\bar\Delta+1={7\over 5},
\fe
and also spin 3 primaries,
\ie
&\Phi_{0,3},~~~\Delta=\bar\Delta-3=0,
\\
&\Phi_{3,0},~~~\Delta=\bar\Delta+3=3.
\fe

The fusion rules of the primaries are given in \cite{McCabe:1995uq}. Here, we only present the part that is relevant to us,
\ie
&\sigma \times \sigma=\sigma^*+Z^*,
\\
&\sigma\times\sigma^*=1+\epsilon+X+Y+\Phi_{{2\over 5},{7\over 5}}+\Phi_{{7\over 5},{2\over 5}}+\Phi_{0,3}+\Phi_{3,0}.
\fe
We are interested in the 4-point function $\vev{\sigma(z_1)\sigma(z_2)\sigma^*(z_3)\sigma^*(z_4)}$, and by fusion rules it can be written as
\ie
&\vev{\sigma(z_1)\sigma(z_2)\sigma^*(z_3)\sigma^*(z_4)} \equiv{1\over |z_{14}z_{23}|^{4\over 15}}G_{\sigma\sigma\sigma^*\sigma^*}^{Potts}(z,\bar z) \\
& = {1\over |z_{14}z_{23}|^{4\over 15}}\left(\left|C_{\sigma\sigma\sigma}F_{1\over 15}(z)\right|^2+\left|C_{\sigma\sigma Z}F_{2\over 3}(z)\right|^2\right),
\fe
where $C_{\sigma\sigma\sigma},C_{\sigma\sigma Z}$ are structure constants, and $F_{\Delta}(z)\equiv{\cal F}({1\over 15},{1\over 15},{1\over 15},{1\over 15};\Delta;z)$ is the conformal block. Similarly, the 4-point functions $\vev{\sigma(z_1)\sigma^*(z_2)\sigma(z_3)\sigma^*(z_4)}$ and $\vev{\sigma(z_1)\sigma^*(z_2)\sigma^*(z_3)\sigma(z_4)}$ can be written as
\ie
&\vev{\sigma(z_1)\sigma^*(z_2)\sigma(z_3)\sigma^*(z_4)} \equiv {1\over |z_{14}z_{23}|^{4\over 15}} G_{\sigma\sigma^*\sigma\sigma^*}^{Potts}(z,\bar z) \\
& = {1\over |z_{14}z_{23}|^{4\over 15}}\left(\left|F_0(z)\right|^2+\left|C_{\sigma\sigma^* \epsilon}F_{2\over 5}(z)\right|^2+\left|C_{\sigma\sigma^* X}F_{7\over 5}(z)\right|^2+\left|C_{\sigma\sigma^* Y}F_3(z)\right|^2\right.
\\
&~~~-\left|C_{\sigma\sigma^* \Phi_{{2\over 5},{7\over 5}}}\right|^2F_{2\over 5}(z)\overline{F_{7\over 5}(z)}-\left|C_{\sigma\sigma^* \Phi_{{7\over 5},{2\over 5}}}\right|^2F_{7\over 5}(z)\overline{F_{2\over 5}(z)}
\\
&~~~\left.-\left|C_{\sigma\sigma^* \Phi_{0,3}}\right|^2F_{0}(z)\overline{F_{3}(z)}-\left|C_{\sigma\sigma^* \Phi_{3,0}}\right|^2F_{3}(z)\overline{F_{0}(z)}\right),
\\
&\vev{\sigma(z_1)\sigma^*(z_2)\sigma^*(z_3)\sigma(z_4)} \equiv {1\over |z_{14}z_{23}|^{4\over 15}} G_{\sigma\sigma^*\sigma^*\sigma}^{Potts}(z,\bar z) \\
& = {1\over |z_{14}z_{23}|^{4\over 15}}\left(\left|F_0(z)\right|^2+\left|C_{\sigma\sigma^* \epsilon}F_{2\over 5}(z)\right|^2+\left|C_{\sigma\sigma^* X}F_{7\over 5}(z)\right|^2+\left|C_{\sigma\sigma^* Y}F_3(z)\right|^2\right.
\\
&~~~+\left|C_{\sigma\sigma^* \Phi_{{2\over 5},{7\over 5}}}\right|^2F_{2\over 5}(z)\overline{F_{7\over 5}(z)}+\left|C_{\sigma\sigma^* \Phi_{{7\over 5},{2\over 5}}}\right|^2F_{7\over 5}(z)\overline{F_{2\over 5}(z)}
\\
&~~~\left.+\left|C_{\sigma\sigma^* \Phi_{0,3}}\right|^2F_{0}(z)\overline{F_{3}(z)}+\left|C_{\sigma\sigma^* \Phi_{3,0}}\right|^2F_{3}(z)\overline{F_{0}(z)}\right),
\fe
where $\left|C_{\sigma\sigma^* \Phi_{{2\over 5},{7\over 5}}}\right| = \left|C_{\sigma\sigma^* \Phi_{{7\over 5},{2\over 5}}}\right|$ and $\left|C_{\sigma\sigma^* \Phi_{0,3}}\right| = \left|C_{\sigma\sigma^* \Phi_{3,0}}\right|$.  The structure constants can be solved by bootstrap. $C_{\sigma\sigma\sigma}$ and $C_{\sigma\sigma Z}$ are given in closed form in \cite{McCabe:1995uq}
\ie
& |C_{\sigma\sigma\sigma}| 
={\sqrt{{1\over 2}(1+\sqrt{5})}\Gamma({3\over 5})^2\over\Gamma({2\over 5})\Gamma({4\over 5})}\approx 1.09236,
\\
& |C_{\sigma\sigma Z}| 
 = {1\over 3},
\fe
and the rest are found to be
\ie
& |C_{\sigma\sigma^*\epsilon}| = 0.546178, \\
& |C_{\sigma\sigma^*X}| = 0.0260085, \\
& |C_{\sigma\sigma^*Y}| = 0.000474834, \\
& \left|C_{\sigma\sigma^*\Phi_{{2 \over 5}, {7 \over 5}}}\right| = 0.0217907, \\
& \left|C_{\sigma\sigma^*\Phi_{0,3}}\right| = 0.119186.
\fe

\section{Numerical Methods}\label{numerics}

\subsection{Zamolodchikov recurrence formula for conformal blocks}\label{zrec}

Consider a CFT of central charge $c=1+6Q^2$, where $Q = b+{1 \over b}$. The conformal dimension of an operator will be labeled $\Delta_i = \A_i(Q-\A_i) = {Q^2 \over 4} - \lambda_i^2$, and $\Delta = {Q^2\over 4} + P^2$.  The correlation function of four scalars is expressed in terms of the three-point functions and the conformal blocks as
\ie\label{phi4}
& \langle \phi_1(z_1) \phi_2(z_2) \phi_3(z_3) \phi_4 (z_4)\rangle \\
&= |z_{14}|^{-4\Delta_1} |z_{24}|^{2(\Delta_1 - \Delta_2 + \Delta_3 - \Delta_4)} |z_{34}|^{2(\Delta_1+\Delta_2-\Delta_3-\Delta_4)} |z_{23}|^{2(\Delta_4-\Delta_1-\Delta_2-\Delta_3)} \\
& \hspace{1.5in} \times \sum_\Delta C_{12\Delta} C_{34\Delta} |{F}(\Delta_{i}; \Delta | z)|^2, \\
\fe
The conformal block of 4 primary operators of weight $\Delta_{\A_1},\cdots,\Delta_{\A_4}$ through a channel given by the conformal family of a primary of weight $\Delta$ has the following expression \cite{Zamolodchikov:1985ie,Zamolodchikov:1987}
\ie
F(\Delta_{i}; \Delta| z) &= (16 q)^{P^2} z^{{Q^2\over 4} - \Delta_{\A_1} - \Delta_{\A_2}} (1-z)^{{Q^2\over 4} - \Delta_{\A_1} - \Delta_{\A_3}} \\
& \hspace{.5in} \times \theta_3(q)^{3Q^2 - 4 (\Delta_{\A_1}+\Delta_{\A_2}+\Delta_{\A_3}+\Delta_{\A_4}) } H(\lambda_i^2; \Delta|q),
\fe
where $z$ is the cross ratio
\ie
z = {z_{12} z_{34}\over z_{14} z_{32}},
\fe
$q$ is defined by
\ie
q(z)=e^{\pi i\tau(z)} ,\quad \tau(z) = i {K(1-z)\over K(z)},~~~~ K(z) = {1\over 2} \int_0^1 {dt\over \sqrt{t(1-t) (1-z t) }},
\fe
and $\theta_3$ is the Jacobi theta function
\ie
\theta_3(p) = \sum_{n=-\infty}^\infty p^{n^2}.
\fe
Finally, the function $H$ is determined by the following recurrence relation 
\ie\label{recH}
H(\lambda_i^2, \Delta| q) = 1 + \sum_{m,n\geq 1} {q^{mn} R_{m,n}(\{\lambda_i\}) \over \Delta - \Delta_{m,n} }
H(\lambda_i^2, \Delta_{m,n} + mn|q)
\fe
where $\Delta_{m,n}$ are the conformal weights of degenerate representations of the Virasoro algebra,
\ie
\Delta_{m,n} = {Q^2\over 4} - \lambda_{m,n}^2, \quad \lambda_{m,n} = {1\over 2} ({m\over b} + nb).
\fe
and $R_{m,n}(\{\lambda_i \})$ are given by
\ie
R_{m,n}(\{\lambda_i \}) = 2 {\prod_{r,s} (\lambda_1+\lambda_2 - \lambda_{r,s}) (\lambda_1-\lambda_2 - \lambda_{r,s}) (\lambda_3+\lambda_4 - \lambda_{r,s}) (\lambda_3-\lambda_4 - \lambda_{r,s}) \over \prod_{k,\ell}' \lambda_{k,\ell} }.
\fe
The product of $(r,s)$ is taken over
\ie\label{rsrange}
& r = -m+1, -m+3, \cdots, m-1,
\\
& s = -n+1, -n+3, \cdots, n-1,
\fe
and the product of $(k,\ell)$ is taken over
\ie
& k = -m+1, -m+2, \cdots, m,
\\
& \ell = -n+1, -n+2, \cdots, n,
\fe
{\it excluding} $(k,\ell)=(0,0)$ and $(k,\ell)=(m,n)$.

Because $|q(z)|$ is always less than 1 except at $z=1, \infty$ where it is equal to 1, the series expansion of $H$ in $q$ converges except at $z=1, \infty$.  In fact, a series in $q$ converges much faster than the corresponding series in $x$; this can be seen, for example, for small values of $z$ if one notes $q = {z \over 16} + {\cal O}(z^2)$.

For fixed values of external weights $\lambda_i$, $H(\lambda_i^2, \Delta| q)$ for general $\Delta$ is determined via the recurrence relation (\ref{recH}) once we know $H$ for degenerate values of $\Delta = \Delta_{m,n}$.  In practice, we can consider the set of degenerate weights $\Delta_{m,n}$ satisfying $mn \leq N$, and solve the matrix equation (\ref{recH}) formed by having $\Delta=\Delta_{m,n}+mn$ on the left-hand side with $m,n$ in this set.  As a simple example, consider $N=2$ where the set of $(m, n)$ are $(1,1), (1,2), (2,1)$, then the matrix equation is
\ie
\begin{pmatrix}
H(\lambda_i^2, 1 | q) \\
H(\lambda_i^2, \Delta_{1,2} + 2 | q) \\
H(\lambda_i^2, \Delta_{2,1} + 2 | q)
\end{pmatrix}
=
\begin{pmatrix}
1 \\
1 \\
1
\end{pmatrix}
+
\begin{pmatrix}
 & {q^2 R_{1,2} \over 1-\Delta_{1,2}} & {q^2 R_{2,1} \over 1-\Delta_{2,1}} \\
{ q R_{1,1} \over \Delta_{1,2}+2 } & & {q^2 R_{2,1} \over \Delta_{1,2}+2-\Delta_{2,1}} \\
{ q R_{1,1} \over \Delta_{2,1}+2 } & {q^2 R_{1,2} \over \Delta_{2,1}+2-\Delta_{1,2}}
\end{pmatrix}
\begin{pmatrix}
H(\lambda_i^2, 1 | q) \\
H(\lambda_i^2, \Delta_{1,2} + 2 | q) \\
H(\lambda_i^2, \Delta_{2,1} + 2 | q)
\end{pmatrix}
\fe
After solving this equation, we have a series expansion for $H(\lambda_i^2, \Delta| q)$ for general $\Delta$ accurate to order $q^N$.

We note some caveats in the implementation of this method.  For special values of the central charge, for example when $c$ equals the central charge of a minimal model, or when $b^2$ is an integer, $\Delta_{r,s} + rs - \Delta_{m,n}$ or the denominator of $R_{m,n}$ can become zero, and therefore certain coefficients appearing in the recurrence relation diverge.  Nonetheless, we can deform the value of the central charge from $c$ to $c+\epsilon$, and as it must all the poles in $\epsilon$ cancel.  Therefore, with a small $\epsilon$ and high enough numerical precision (high enough so that the divergences cancel properly on the computer), we can still compute the conformal blocks for these seemingly pathological values of the central charge.

\subsection{Crossing symmetry}

Consider the four-point function (\ref{phi4}), and let us define
\ie
G(1,2,3,4| z) \equiv \sum_\Delta C_{12\Delta} C_{34\Delta} |{F}(\Delta_1,\Delta_2,\Delta_3,\Delta_4; \Delta| z)|^2,
\fe
which satisfies the following crossing relations
\ie
\label{xingrel}
G(1,2,3,4|z) = G(1,3,2,4|1-z) = |z|^{-4\Delta_1} G(1,4,3,2|{1/z}).
\fe
The two transformations $T \equiv 2 \leftrightarrow 3$ ($z \to 1-z$) and $S \equiv 2 \leftrightarrow 4$ ($z \to 1/z$) generate the permutation group $S_3$.  We can think of this $S_3$ as permuting 0, 1, and $\infty$.  The relations are
\ie
T^2 = S^2 = 1, \quad (TS)^3 = 1.
\fe
The complete set of transformations are
\ie
& 1: z \to z, \quad (1234) \\
& S: z \to {1 \over z}, \quad (1432) \\
& TS: z \to {z-1 \over z}, \quad (1342) \\
& STS: z \to {z \over z-1}, \quad (1243) \\
& ST: z \to {1 \over 1-z}, \quad (1423) \\
& T: z \to 1-z, \quad (1324).
\fe
We can divide the complex plane into six fundamental regions:
\ie
& I: {\rm Re}\,z \leq {1 \over 2}, |z-1| \leq 1, \\
& II: |z| \leq 1, |z-1| \geq 1, \\
& III: {\rm Re}\,z \leq {1 \over 2}, |z| \geq 1,
\fe
and $IV, V, VI$ their mirror images under $z \to 1-z$.  Regions $II$-$VI$ can be obtained from region $I$ by the $STS, TS, T, ST, S$ transformations, respectively.  An integral involving conformal blocks over the entire complex plane can be rewritten as an integral over only region $I$.  This is useful for doing numerical integration because, first, region $I$ is bounded, and second, in this region $|q|$ is bounded above by $0.0658287$, which means that the Zamolodchikov recurrence formula (\ref{recH}) converges very quickly.

By repeated use of (\ref{xingrel}) and noting that $\Delta_1 = \Delta_2$, $\Delta_3 = \Delta_4$, we rewrite the LST amplitude for $k = 2, 3, 4, 5$ and $\ell = 0$ as
\ie
& \int_\mathbb{C} d^2z |z|^{- s_{12}} |1-z|^{- s_{13}} G(1,2,3,4|z) \\
= & \int_I d^2z \Big( |z|^{- s_{12}} ( |1-z|^{- s_{13}} + |1-z|^{- s_{14}} ) G(1,2,3,4|z) \\
& \hspace{.5in} ( |z|^{- s_{13}} |1-z|^{- s_{14}} + |z|^{- s_{14}} |1-z|^{- s_{13}} ) G(1,4,3,2|z) \\
& \hspace{.5in} |1-z|^{-s_{12}} (|z|^{- s_{13}} + |z|^{- s_{14}}) G(1,3,2,4|z) \Big), \\
\fe
and for $k = 5$, $\ell = 1$ as
\ie
& \int_\mathbb{C} d^2z |z|^{{1 \over 3} - s_{12}} |1-z|^{{2 \over 3} - s_{13}} G(1,2,3,4|z) G_{\sigma\sigma\sigma^*\sigma^*}^{Potts}(z) \\
= & \int_I d^2z \left( |z|^{- s_{12}} ( |1-z|^{- s_{13}} + |1-z|^{- s_{14}} ) |z|^{{1 \over 3}} |1-z|^{{2 \over 3}} G(1,2,3,4|z) G_{\sigma\sigma\sigma^*\sigma^*}^{Potts}(z) \right. \\
& \hspace{.5in} ( |z|^{- s_{13}} |1-z|^{- s_{14}} + |z|^{- s_{14}} |1-z|^{- s_{13}} ) |z|^{{2 \over 3}} |1-z|^{{2 \over 3}} G(1,4,3,2|z) G_{\sigma\sigma^*\sigma^*\sigma}^{Potts}({z}) \\
& \hspace{.5in} \left. |1-z|^{-s_{12}} (|z|^{- s_{13}} + |z|^{- s_{14}}) |z|^{{2 \over 3}} |1-z|^{{1 \over 3}} G(1,3,2,4|z) G_{\sigma\sigma^*\sigma\sigma^*}^{Potts}({z}) \right),
\fe



\section{3-Loop UV finiteness of the four-point amplitude for the Cartan gluons}
\label{3ldiv}

In this appendix we verify that the 3-loop 4-point amplitudes of the $U(1)$ Cartan gluons (``photon") in the $SU(k)$ SYM is UV finite (see also \cite{Bern:2010tq}). This is indeed expected both from the DSLST amplitude (corresponding to the third order term in the $\alpha'$ expansion) and from the inspection on the possible counter terms mentioned in the introduction.

The amplitude is reduced to scalar 3-loop integrals, summarized in Figure 2 of \cite{Bern:2007hh}. The four potentially logarithmic divergent diagrams are shown in Figure \ref{fig:3loop} (ignoring the signs on vertices for now). We will compute the divergent parts of these diagrams with color factors included and show that they cancel among themselves.

\begin{figure}[h]
\begin{center}
\includegraphics[scale=1.9]{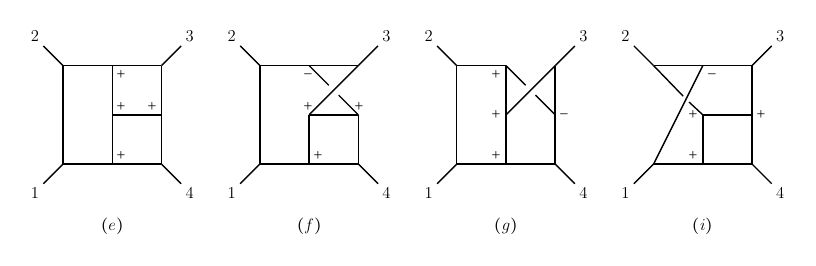}
\end{center}
\caption{The four UV divergent scalar integrals for the 3-loop 4-gluon scattering amplitude in 6D SYM. The signs for the internal vertices denote the two index structures in the double line notation; plus for the left vertex and minus for the right vertex in Figure \ref{fig:DL}. We label the diagrams following the notation in Figure 2 of \cite{Bern:2007hh}. The above sign assignments together with the other four obtained by flipping all the $+/-$ are the only eight diagrams that contribute to the UV divergence of the scattering amplitudes of four Cartan gluons.  }\label{fig:3loop}
\end{figure}

\begin{figure}[h]
\begin{center}
\includegraphics[scale=1.75]{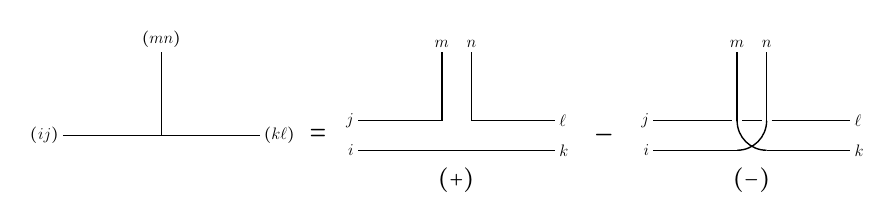}
\end{center}
\caption{The two index structures for the 3-gluon vertex in the double line notation.}\label{fig:DL}
\end{figure}

Let us start with the $SU(2)$ case where no actual calculation is needed to show the cancellation of UV divergence. 
The key fact here is that since there is only one species of Cartan gluon,  the amplitudes are invariant under permutation of all four external legs. It follows that the logarithmic divergent part of the amplitude is proportional to $(s_{12}+s_{13}+s_{14}) \ln \Lambda=0$ (see Table 1 of \cite{Bern:2007hh}) and vanishes in the end.

Moving on to the general $SU(k)$ 6D SYM, it suffices to show that the UV divergent part is invariant under cyclic permutations on the external legs $2,3,4$, from which it again follows that the logarithmic divergence is proportional to $(s_{12}+s_{13}+s_{14})\ln \Lambda=0$.

In the double line notation, each 3-point vertex can be written as the difference of two vertices shown in Figure \ref{fig:DL} with different index structures. Each diagram in Figure \ref{fig:3loop} then becomes a sum of $2^4=16$ diagrams with different sign assignments on the four vertices. It is easy to see that the only diagrams that give noncyclic invariant amplitudes are the four shown in Figure \ref{fig:3loop} together with the other four with all the $+/-$ flipped.

The color factors for the four external Cartan gluons will be labeled by $\vec{v}_1,\cdots,\vec{v}_4$ in the Cartan subalgebra of $su(k)$ as in the previous section. The four diagrams in Figure \ref{fig:3loop} can be expressed as\footnote{The full 3-loop amplitude will be $\cA^{3-loop, (e,f,g,h)}$ multiplied by some overall factors including $\cA^{tree}$ \cite{Bern:2007hh}.}
\ie
&\mathcal{A}^{3-loop, (e)}_{1234} = 2N^2 s_{12}(\vec{v_1}\cdot \vec{v_2}) (\vec{v_3}\cdot \vec{v_4}) \, I_2^{log}+ \cdots,\\ 
&\mathcal{A}^{3-loop, (f)}_{1234} = -2N^2s_{12}\left[ (\vec{v_1}\cdot \vec{v_3}) (\vec{v_2}\cdot \vec{v_4}) +
(\vec{v_1}\cdot \vec{v_4}) (\vec{v_2}\cdot \vec{v_3}) \right] \, I_1^{log}+ \cdots,\\ 
&\mathcal{A}^{3-loop, (g)}_{1234} = -2N^2 s_{12}\times 2(\vec{v_1}\cdot \vec{v_2}) (\vec{v_3}\cdot \vec{v_4})  \, I_1^{log}+ \cdots,\\ 
&\mathcal{A}^{3-loop, (i)}_{1234} = -2N^2(s_{12} -s_{13})\left[ -(\vec{v_1}\cdot \vec{v_2}) (\vec{v_3}\cdot \vec{v_4}) +
(\vec{v_1}\cdot \vec{v_4}) (\vec{v_2}\cdot \vec{v_3}) \right] \,\left(I_1^{log}-{1\over 3}I_2^{log} \right)+ \cdots,\\ 
\fe
where the factor 2 comes from the other diagrams obtained by flipping all the $+/-$ for the vertices. The $\cdots$ stands for the finite as well as the cyclic invariant terms. The two UV-divergent scalar integrals $I_1^{log}$ and $I_2^{log}$ are defined in Figure \ref{fig:circle}.

\begin{figure}[htb]
\begin{center}
\includegraphics[scale=1.6]{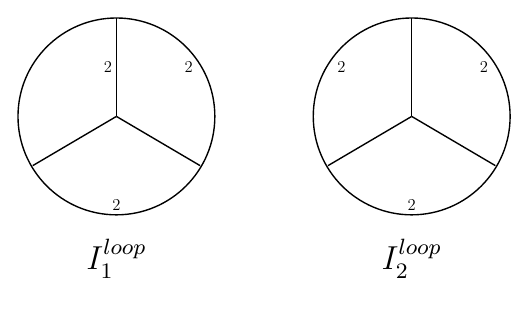}
\end{center}
\caption{The two UV-divergent scalar integrals in the 3-loop 4-point amplitude. Since we are only interested in the divergent part, we have set the external momenta to be zero.The number indicates the propagator should be raised to the corresponding power.}\label{fig:circle}
\end{figure}

Next, we need to sum over all the permutations on the external legs. After taking the symmetry factors for each diagram appropriately, the noncyclic invariant part of the divergent amplitude is proportional to
\ie
2N^2 s_{12}(\vec{v_1}\cdot \vec{v_2}) (\vec{v_3}\cdot \vec{v_4})\left[  
{1\over 2} I_2^{log}
+{1\over 2}I_1^{log}
- 2I_1^{log}
-{1\over2}\times 2\times \left( -{3\over2}\right) \left(I_1^{log}-{1\over 3}I_2^{log}\right)
\right]=0.
\fe
Note that we have grouped $(\vec{v_1}\cdot \vec{v_3}) (\vec{v_2}\cdot \vec{v_4})+(\vec{v_1}\cdot \vec{v_4}) (\vec{v_2}\cdot \vec{v_3})$ with $(\vec{v_1}\cdot \vec{v_2}) (\vec{v_3}\cdot \vec{v_4})$ to the cyclic invariant part, which in the end is proportional to $(s_{12}+s_{13}+s_{14})\ln \Lambda=0$. 

In summary, in this appendix we have showed that the 3-loop 4-point amplitudes for gluons in the Cartan subalgebra is free from divergence and we are then left with a finite amplitude
. The comparison with the DSLST amplitude at this order will be left for future analysis.

\bibliography{lstrefs,6dsymrefs} 

\providecommand{\href}[2]{#2}\begingroup\raggedright\begin{thebibliography}{10}

\bibitem{Berkooz:1997cq}
M.~Berkooz, M.~Rozali, and N.~Seiberg, {\it {Matrix description of M theory on
  T**4 and T**5}},  {\em Phys.Lett.} {\bf B408} (1997) 105--110,
  [\href{http://xxx.lanl.gov/abs/hep-th/9704089}{{\tt hep-th/9704089}}].

\bibitem{Seiberg:1997zk}
N.~Seiberg, {\it {New theories in six-dimensions and matrix description of M
  theory on T**5 and T**5 / Z(2)}},  {\em Phys.Lett.} {\bf B408} (1997)
  98--104, [\href{http://xxx.lanl.gov/abs/hep-th/9705221}{{\tt
  hep-th/9705221}}].

\bibitem{Aharony:1998ub}
O.~Aharony, M.~Berkooz, D.~Kutasov, and N.~Seiberg, {\it {Linear dilatons, NS
  five-branes and holography}},  {\em JHEP} {\bf 9810} (1998) 004,
  [\href{http://xxx.lanl.gov/abs/hep-th/9808149}{{\tt hep-th/9808149}}].

\bibitem{Giveon:1999zm}
A.~Giveon, D.~Kutasov, and O.~Pelc, {\it {Holography for noncritical
  superstrings}},  {\em JHEP} {\bf 9910} (1999) 035,
  [\href{http://xxx.lanl.gov/abs/hep-th/9907178}{{\tt hep-th/9907178}}].

\bibitem{Aharony:1999ks}
O.~Aharony, {\it {A Brief review of 'little string theories'}},  {\em
  Class.Quant.Grav.} {\bf 17} (2000) 929--938,
  [\href{http://xxx.lanl.gov/abs/hep-th/9911147}{{\tt hep-th/9911147}}].

\bibitem{Kutasov:lecturenote}
Kutasov, {\it Introduction to little string theory},  Prepared for ICTP Spring
  School on Superstrings and Related Matters, Trieste, Italy, 2-10 Apr 2001.

\bibitem{Giveon:1999px}
A.~Giveon and D.~Kutasov, {\it {Little string theory in a double scaling
  limit}},  {\em JHEP} {\bf 9910} (1999) 034,
  [\href{http://xxx.lanl.gov/abs/hep-th/9909110}{{\tt hep-th/9909110}}].

\bibitem{Giveon:1999tq}
A.~Giveon and D.~Kutasov, {\it {Comments on double scaled little string
  theory}},  {\em JHEP} {\bf 0001} (2000) 023,
  [\href{http://xxx.lanl.gov/abs/hep-th/9911039}{{\tt hep-th/9911039}}].

\bibitem{Bossard:2010pk}
G.~Bossard, P.~Howe, U.~Lindstrom, K.~Stelle, and L.~Wulff, {\it {Integral
  invariants in maximally supersymmetric Yang-Mills theories}},  {\em JHEP}
  {\bf 1105} (2011) 021, [\href{http://xxx.lanl.gov/abs/1012.3142}{{\tt
  arXiv:1012.3142}}].

\bibitem{Dixon:1996wi}
L.~J. Dixon, {\it {Calculating scattering amplitudes efficiently}},
  \href{http://xxx.lanl.gov/abs/hep-ph/9601359}{{\tt hep-ph/9601359}}.

\bibitem{Dixon:2010gz}
L.~J. Dixon, {\it {Ultraviolet Behavior of N = 8 Supergravity}},
  \href{http://xxx.lanl.gov/abs/1005.2703}{{\tt arXiv:1005.2703}}.

\bibitem{Bern:1996je}
Z.~Bern, L.~J. Dixon, and D.~A. Kosower, {\it {Progress in one loop QCD
  computations}},  {\em Ann.Rev.Nucl.Part.Sci.} {\bf 46} (1996) 109--148,
  [\href{http://xxx.lanl.gov/abs/hep-ph/9602280}{{\tt hep-ph/9602280}}].

\bibitem{Brandhuber:2010mm}
A.~Brandhuber, D.~Korres, D.~Koschade, and G.~Travaglini, {\it {One-loop
  Amplitudes in Six-Dimensional (1,1) Theories from Generalised Unitarity}},
  {\em JHEP} {\bf 1102} (2011) 077,
  [\href{http://xxx.lanl.gov/abs/1010.1515}{{\tt arXiv:1010.1515}}].

\bibitem{Bern:1997nh}
Z.~Bern, J.~Rozowsky, and B.~Yan, {\it {Two loop four gluon amplitudes in N=4
  superYang-Mills}},  {\em Phys.Lett.} {\bf B401} (1997) 273--282,
  [\href{http://xxx.lanl.gov/abs/hep-ph/9702424}{{\tt hep-ph/9702424}}].

\bibitem{Bern:2010qa}
Z.~Bern, J.~J. Carrasco, T.~Dennen, Y.-t. Huang, and H.~Ita, {\it {Generalized
  Unitarity and Six-Dimensional Helicity}},  {\em Phys.Rev.} {\bf D83} (2011)
  085022, [\href{http://xxx.lanl.gov/abs/1010.0494}{{\tt arXiv:1010.0494}}].

\bibitem{Ribault:2005wp}
S.~Ribault and J.~Teschner, {\it {H+(3)-WZNW correlators from Liouville
  theory}},  {\em JHEP} {\bf 0506} (2005) 014,
  [\href{http://xxx.lanl.gov/abs/hep-th/0502048}{{\tt hep-th/0502048}}].

\bibitem{Aharony:2003vk}
O.~Aharony, B.~Fiol, D.~Kutasov, and D.~A. Sahakyan, {\it {Little string theory
  and heterotic / type II duality}},  {\em Nucl.Phys.} {\bf B679} (2004) 3--65,
  [\href{http://xxx.lanl.gov/abs/hep-th/0310197}{{\tt hep-th/0310197}}].

\bibitem{Bern:2007hh}
Z.~Bern, J.~Carrasco, L.~J. Dixon, H.~Johansson, D.~Kosower, {\em et.~al.},
  {\it {Three-Loop Superfiniteness of N=8 Supergravity}},  {\em Phys.Rev.Lett.}
  {\bf 98} (2007) 161303, [\href{http://xxx.lanl.gov/abs/hep-th/0702112}{{\tt
  hep-th/0702112}}].

\bibitem{Callan:1991at}
J.~Callan, Curtis~G., J.~A. Harvey, and A.~Strominger, {\it {Supersymmetric
  string solitons}},  \href{http://xxx.lanl.gov/abs/hep-th/9112030}{{\tt
  hep-th/9112030}}.

\bibitem{Ooguri:1995wj}
H.~Ooguri and C.~Vafa, {\it {Two-dimensional black hole and singularities of CY
  manifolds}},  {\em Nucl.Phys.} {\bf B463} (1996) 55--72,
  [\href{http://xxx.lanl.gov/abs/hep-th/9511164}{{\tt hep-th/9511164}}].

\bibitem{Kutasov:1995te}
D.~Kutasov, {\it {Orbifolds and solitons}},  {\em Phys.Lett.} {\bf B383} (1996)
  48--53, [\href{http://xxx.lanl.gov/abs/hep-th/9512145}{{\tt
  hep-th/9512145}}].

\bibitem{Sfetsos:1998xd}
K.~Sfetsos, {\it {Branes for Higgs phases and exact conformal field theories}},
   {\em JHEP} {\bf 9901} (1999) 015,
  [\href{http://xxx.lanl.gov/abs/hep-th/9811167}{{\tt hep-th/9811167}}].

\bibitem{Giveon:2003wn}
A.~Giveon, A.~Konechny, A.~Pakman, and A.~Sever, {\it {Type 0 strings in a 2-d
  black hole}},  {\em JHEP} {\bf 0310} (2003) 025,
  [\href{http://xxx.lanl.gov/abs/hep-th/0309056}{{\tt hep-th/0309056}}].

\bibitem{Aharony:2004xn}
O.~Aharony, A.~Giveon, and D.~Kutasov, {\it {LSZ in LST}},  {\em Nucl.Phys.}
  {\bf B691} (2004) 3--78, [\href{http://xxx.lanl.gov/abs/hep-th/0404016}{{\tt
  hep-th/0404016}}].

\bibitem{Evans:1998wq}
J.~Evans, M.~Gaberdiel, and M.~Perry, {\it {The No - ghost theorem and strings
  on AdS(3)}},  \href{http://xxx.lanl.gov/abs/hep-th/9812252}{{\tt
  hep-th/9812252}}.

\bibitem{Maldacena:2000hw}
J.~M. Maldacena and H.~Ooguri, {\it {Strings in AdS(3) and SL(2,R) WZW model
  1.: The Spectrum}},  {\em J.Math.Phys.} {\bf 42} (2001) 2929--2960,
  [\href{http://xxx.lanl.gov/abs/hep-th/0001053}{{\tt hep-th/0001053}}].

\bibitem{Lerche:1989uy}
W.~Lerche, C.~Vafa, and N.~P. Warner, {\it {Chiral Rings in N=2 Superconformal
  Theories}},  {\em Nucl.Phys.} {\bf B324} (1989) 427.

\bibitem{Schwimmer:1986mf}
A.~Schwimmer and N.~Seiberg, {\it {Comments on the N=2, N=3, N=4 Superconformal
  Algebras in Two-Dimensions}},  {\em Phys.Lett.} {\bf B184} (1987) 191.

\bibitem{DiFrancesco:1988xz}
P.~Di~Francesco, H.~Saleur, and J.~Zuber, {\it {Generalized Coulomb Gas
  Formalism for Two-dimensional Critical Models Based on SU(2) Coset
  Construction}},  {\em Nucl.Phys.} {\bf B300} (1988) 393.

\bibitem{Fateev:1985mm}
V.~Fateev and A.~Zamolodchikov, {\it {Parafermionic Currents in the
  Two-Dimensional Conformal Quantum Field Theory and Selfdual Critical Points
  in Z(n) Invariant Statistical Systems}},  {\em Sov.Phys.JETP} {\bf 62} (1985)
  215--225.

\bibitem{Parnachev:2001gw}
A.~Parnachev and D.~A. Sahakyan, {\it {Some remarks on D-branes in AdS(3)}},
  {\em JHEP} {\bf 0110} (2001) 022,
  [\href{http://xxx.lanl.gov/abs/hep-th/0109150}{{\tt hep-th/0109150}}].

\bibitem{Maldacena:2001km}
J.~M. Maldacena and H.~Ooguri, {\it {Strings in AdS(3) and the SL(2,R) WZW
  model. Part 3. Correlation functions}},  {\em Phys.Rev.} {\bf D65} (2002)
  106006, [\href{http://xxx.lanl.gov/abs/hep-th/0111180}{{\tt
  hep-th/0111180}}].

\bibitem{Zamolodchikov:1995aa}
A.~B. Zamolodchikov and A.~B. Zamolodchikov, {\it {Structure constants and
  conformal bootstrap in Liouville field theory}},  {\em Nucl.Phys.} {\bf B477}
  (1996) 577--605, [\href{http://xxx.lanl.gov/abs/hep-th/9506136}{{\tt
  hep-th/9506136}}].

\bibitem{Teschner:2001gi}
J.~Teschner, {\it {Crossing symmetry in the H(3)+ WZNW model}},  {\em
  Phys.Lett.} {\bf B521} (2001) 127--132,
  [\href{http://xxx.lanl.gov/abs/hep-th/0108121}{{\tt hep-th/0108121}}].

\bibitem{Mussardo:1988av}
G.~Mussardo, G.~Sotkov, and M.~Stanishkov, {\it {N=2 SUPERCONFORMAL MINIMAL
  MODELS}},  {\em Int.J.Mod.Phys.} {\bf A4} (1989) 1135.

\bibitem{Ginsparg:1988ui}
P.~H. Ginsparg, {\it {APPLIED CONFORMAL FIELD THEORY}},
  \href{http://xxx.lanl.gov/abs/hep-th/9108028}{{\tt hep-th/9108028}}.

\bibitem{Bern:2012uf}
Z.~Bern, J.~Carrasco, L.~Dixon, H.~Johansson, and R.~Roiban, {\it {Simplifying
  Multiloop Integrands and Ultraviolet Divergences of Gauge Theory and Gravity
  Amplitudes}},  {\em Phys.Rev.} {\bf D85} (2012) 105014,
  [\href{http://xxx.lanl.gov/abs/1201.5366}{{\tt arXiv:1201.5366}}].

\bibitem{Cheung:2009dc}
C.~Cheung and D.~O'Connell, {\it {Amplitudes and Spinor-Helicity in Six
  Dimensions}},  {\em JHEP} {\bf 0907} (2009) 075,
  [\href{http://xxx.lanl.gov/abs/0902.0981}{{\tt arXiv:0902.0981}}].

\bibitem{Dennen:2009vk}
T.~Dennen, Y.-t. Huang, and W.~Siegel, {\it {Supertwistor space for 6D maximal
  super Yang-Mills}},  {\em JHEP} {\bf 1004} (2010) 127,
  [\href{http://xxx.lanl.gov/abs/0910.2688}{{\tt arXiv:0910.2688}}].

\bibitem{Bern:2012uc}
Z.~Bern, J.~Carrasco, H.~Johansson, and R.~Roiban, {\it {The Five-Loop
  Four-Point Amplitude of N=4 super-Yang-Mills Theory}},  {\em Phys.Rev.Lett.}
  {\bf 109} (2012) 241602, [\href{http://xxx.lanl.gov/abs/1207.6666}{{\tt
  arXiv:1207.6666}}].

\bibitem{Bern:2010tq}
Z.~Bern, J.~Carrasco, L.~J. Dixon, H.~Johansson, and R.~Roiban, {\it {The
  Complete Four-Loop Four-Point Amplitude in N=4 Super-Yang-Mills Theory}},
  {\em Phys.Rev.} {\bf D82} (2010) 125040,
  [\href{http://xxx.lanl.gov/abs/1008.3327}{{\tt arXiv:1008.3327}}].

\bibitem{Movshev:2009ba}
M.~Movshev and A.~Schwarz, {\it {Supersymmetric Deformations of Maximally
  Supersymmetric Gauge Theories}},  {\em JHEP} {\bf 1209} (2012) 136,
  [\href{http://xxx.lanl.gov/abs/0910.0620}{{\tt arXiv:0910.0620}}].

\bibitem{Israel:2005fn}
D.~Israel, A.~Pakman, and J.~Troost, {\it {D-branes in little string theory}},
  {\em Nucl.Phys.} {\bf B722} (2005) 3--64,
  [\href{http://xxx.lanl.gov/abs/hep-th/0502073}{{\tt hep-th/0502073}}].

\bibitem{Israel:2004jt}
D.~Israel, A.~Pakman, and J.~Troost, {\it {D-branes in N=2 Liouville theory and
  its mirror}},  {\em Nucl.Phys.} {\bf B710} (2005) 529--576,
  [\href{http://xxx.lanl.gov/abs/hep-th/0405259}{{\tt hep-th/0405259}}].

\bibitem{McCabe:1995uq}
J.~McCabe and T.~Wydro, {\it {Critical Correlation Functions of the
  2-Dimensional, 3-State Potts Model}},
  \href{http://xxx.lanl.gov/abs/cond-mat/9507033}{{\tt cond-mat/9507033}}.

\bibitem{Zamolodchikov:1985ie}
A.~Zamolodchikov, {\it {CONFORMAL SYMMETRY IN TWO-DIMENSIONS: AN EXPLICIT
  RECURRENCE FORMULA FOR THE CONFORMAL PARTIAL WAVE AMPLITUDE}},  {\em
  Commun.Math.Phys.} {\bf 96} (1984) 419--422.

\bibitem{Zamolodchikov:1987}
A.~Zamolodchikov, {\it {Conformal symmetry in two-dimensional space: Recursion
  representation of conformal block}},  {\em Theoretical and Mathematical
  Physics} {\bf 73} (1987), no.~1 1088--1093.

\end{thebibliography}\endgroup
\bibliographystyle{JHEP}

\end{document}